\documentclass{JHEP3}
\usepackage{amsmath,epsfig}
\usepackage{mathrsfs}
\usepackage{amssymb}
\usepackage{mathtools}
\usepackage{graphics}
\usepackage{cancel}
\usepackage{bbm}
\usepackage{slashed}


\newcommand{\be}{\begin{equation}}
\newcommand{\ee}{\end{equation}}
\newcommand{\bea}{\begin{eqnarray}}
\newcommand{\eea}{\end{eqnarray}}

\title{Leptogenesis in GeV-scale seesaw models}

\preprint{IFIC/15-44\\ SISSA 35/2015/FISI}

\author{ P. ~Hern\'andez$^a$, M.~Kekic$^a$, J. L\'opez-Pav\'on$^b$, J. Racker$^a$ and N. Rius$^a$
\\
$^a$Instituto de F\'{\i}sica Corpuscular, Universidad de Valencia and CSIC,\\
 Edificio Institutos Investigaci\'on, Apt.\ 22085, E-46071 Valencia, Spain\\
$^b$ SISSA and INFN Sezione di Trieste, via Bonomea 265, 34136 Trieste  Italy.}

\abstract{We revisit the production of leptonic asymmetries  in minimal extensions of the Standard
Model  that can explain neutrino masses, involving extra  singlets with Majorana masses in the GeV scale. 
We study the quantum kinetic equations
both analytically, via a perturbative expansion up to third order in the mixing angles, and numerically. The analytical 
solution allows us to identify the relevant CP invariants, and simplifies the exploration of the parameter space. 
We find that sizeable lepton asymmetries are compatible with non-degenerate neutrino masses and measurable active-sterile
mixings. }

\begin{document}

\section{Introduction}

One of the interesting potential implications of (Majorana) neutrino masses is the generation of a matter-antimatter asymmetry in the Universe. It has been demonstrated that 
the generation of sizeable leptonic asymmetries, leptogenesis,  is generic  in  extensions of the Standard Model that can account for neutrino masses \cite{Fukugita:1986hr}. In particular 
two new ingredients are essential for this mechanism to work:  the existence of new weakly interacting particles that are not in thermal equilibrium sometime before the electroweak phase transition and the existence of new sources of CP violation. 

Leptogenesis from the out-of-equilibrium decay of heavy Majorana fermions that appear in type I seesaw models \cite{Fukugita:1986hr} has been extensively studied (for a comprehensive review see e.g.~\cite{davidson08}). The simplest version 
requires however relatively large Majorana masses $> 10^8$ GeV~\cite{davidson02,hambye03} (or $>10^6$ if flavour effects are included \cite{racker12}), which imply that this scenario would be very difficult to test experimentally. It is possible to have sizeable asymmetries for smaller masses if a large degeneracy exists, through resonant leptogenesis \cite{pilaftsis03}.

On the other hand, for Majorana masses in the GeV range, when the neutrino Yukawa couplings are small, another mechanism 
might be at work. In particular, the non-equilibrium condition takes place not in the decay, but in the production of 
the heavy sterile neutrinos. The small Yukawa couplings imply that some of the species might never reach thermal equilibrium
and a lepton asymmetry can be generated at production  and seed the baryon asymmetry in the Universe. This mechanism 
was first proposed by Akhmedov, Rubakov and Smirnov  (ARS) in their pioneering work \cite{Akhmedov:1998qx} and pursued, with important refinements in refs.~ \cite{Asaka:2005pn,Shaposhnikov:2008pf}. For a recent review and further references see \cite{Canetti:2012kh}.  In most of these works, the case of just two extra sterile species is considered, which is also the limiting case of the so-called 
$\nu$MSM where there are three 
species, but one of them plays the role of warm dark matter  (WDM) and is almost decoupled, having no impact in the generation of the lepton asymmetry. 
When the mechanism involves just two species,  it has been found that the observed baryon asymmetry is only possible if the two states are highly degenerate in mass. This however was not the conclusion of the ARS paper. 

The purpose of this paper is to explore systematically the parameter space in the case of three sterile species (which encompass the one with two neutrinos) as regards the baryon asymmetry, in particular  we do not want
to restrict the parameter space to have a WDM candidate. The model has many free parameters 
(only 5 out of the 18 parameters are fixed by the measured light neutrino masses and mixings) and the exploration 
of the full parameter space  is challenging. Only with the help of approximate analytical solutions to 
the kinetic equations this task  is feasible. The analytical solutions furthermore allows us to identify the relevant CP invariants and  to reach regions of parameter space where the equations become stiff and very difficult to deal with numerically.

The paper is organised as follows. In section 2 we present the model, which is essentially a generic type I seesaw model,  establish the notation and discuss on general grounds what are the CP reparametrization and flavour invariants we expect to find in computing any CP violating quantity such as any putative lepton asymmetry. In section 3 we present the kinetic equations that describe the production of sterile neutrinos and solve them analytically via a perturbative expansion in the mixing angles up to the third order. In section 4 we compare
the analytical and numerical solutions for several choices of the parameters, and identify the region of parameter space where the analytical solution accurately describes the numerical one. In section 5 we use the analytical solutions and perform a Monte Carlo scan (using the software package MultiNest \cite{Feroz:2007kg,Feroz:2008xx}) to find regions of parameter space that can reproduce the observed baryon asymmetry, and that are compatible with the measured neutrino masses and mixings. In section 6 we conclude. 

\section{Minimal Model of neutrino masses}
We will concentrate on the arguably simplest model of neutrino masses that includes three right-handed singlets. The Lagrangian is given by:
   \begin{eqnarray}
{\cal L} = {\cal L}_{SM}- \sum_{\alpha,i} \bar L^\alpha Y^{\alpha i} \tilde\Phi N^i_R - \sum_{i,j=1}^3 {1\over 2} \bar{N}^{ic}_R M^{ij} N_R^j+ h.c., \nonumber
\label{eq:lag}
\end{eqnarray}
where $Y$ is a $3\times 3$ complex matrix and $M$ a diagonal real matrix. The spectrum of this theory has six massive Majorana neutrinos, and the mixing is described in terms of six angles and six CP phases generically. One convenient parametrization for the problem at hand is in terms of the eigenvalues of the yukawa  and  majorana mass matrices together with two unitary matrices, $V$ and $W$. In the basis where the Majorana mass is diagonal, $M = {\rm Diag}(M_1,M_2,M_3)$, the neutrino Yukawa matrix is given by:
\begin{eqnarray}
Y \equiv V^\dagger {\rm Diag}(y_1,y_2,y_3) W. 
\label{eq:yuk}
\end{eqnarray} 
Without loss of generality, using rephasing invariance, we can reduce the unitary matrices to the  form\footnote{Although we use the same notation for the mixing angles and phases of $W$ as those in the usual PMNS matrix, they should not be confused.}:
\begin{eqnarray}
W = U(\theta_{12},\theta_{13}, \theta_{23},\delta)^\dagger {\rm Diag}(1, e^{i \alpha_1}, e^{i \alpha_2}),\nonumber\\
V = {\rm Diag}(1, e^{i \phi_1}, e^{i \phi_2}) U(\bar{\theta}_{12},\bar{\theta}_{13}, \bar{\theta}_{23},\bar{\delta}), 
\end{eqnarray}
where\footnote{Note the unconventional ordering of the 2$\times$2 rotation matrices in $U$.}
\begin{footnotesize}
\begin{eqnarray}
U(\theta_1,\theta_2,\theta_3,\delta) \equiv \left( \begin{array}{lll}
\cos\theta_1 & \sin\theta_1 & 0 \\
-\sin\theta_1 & \cos\theta_1 & 0 \\
0& 0 &1\end{array}\right)
\left( \begin{array}{lll}
\cos\theta_2 & 0& \sin\theta_2 e^{-i\delta}\\ 
0 & 1 & 0 \\
-\sin\theta_2 e^{i�\delta} & 0 &\cos\theta_2\\
 \end{array}\right) 
\left( \begin{array}{lll}
1 & 0 & 0 \\
0& \cos\theta_3 & \sin\theta_3 \\
0& -\sin\theta_3  &\cos\theta_3\\
\end{array}\right).
\end{eqnarray}
\end{footnotesize}
Obviously not all the parameters are free, since this model must reproduce the light neutrino masses, which approximately implies the seesaw relation:
\begin{eqnarray}
m_\nu \simeq -{v^2\over 2} Y {1\over M} Y^T ,
\end{eqnarray}
 where $v=246$~GeV is the vev of the Higgs. On the other hand, the known neutrino masses and mixings do not give us enough information to determine the Majorana spectrum, not even the absolute scale. Very strong constraints can be derived
from neutrino oscillation experiments for masses below the eV range \cite{deGouvea:2009fp,deGouvea:2011zz,Donini:2011jh,Donini:2012tt}. Cosmology can exclude a huge window below 100~MeV \cite{Dolgov:2000pj,Dolgov:2000jw,Ruchayskiy:2012si,Hernandez:2013lza,Hernandez:2014fha,Drewes:2015iva}, except maybe for one species 
that could be lighter provided the lightest active neutrino mass is below $\lesssim 3 \times 10^{-3}$eV \cite{Hernandez:2013lza,Hernandez:2014fha}. The GeV range is interesting because an alternative mechanism for lepton asymmetry generation could be at work \cite{Akhmedov:1998qx,Asaka:2005pn,Shaposhnikov:2008pf}. Majorana neutrinos in this range are heavy enough to safely decay before Big Bang Nucleosynthesis, while they are light enough that they might have
not reached thermal equilibrium by the time of the electroweak phase transition (EWPT), behaving as reservoirs of a putative lepton asymmetry. 

Our goal in this paper is to explore the full parameter space of this model allowed by neutrino masses, as regards leptogenesis. An essential condition will be that at least one of the sterile neutrinos does not reach thermal equilibrium before the EWPT. This can be ensured assuming a large hierarchy in the yukawas  \cite{Akhmedov:1998qx}:
\begin{eqnarray}
y_3 \ll y_1,y_2.
\end{eqnarray}
It is mandatory, however, 
to have an accurate analytical description, since the unconstrained parameter space is huge. We will solve the quantum kinetic equations in  a perturbative expansion in the mixings in the next section. Since the lepton asymmetry is necessarily a CP-odd observable, on general grounds we can derive what are the expectations in terms of weak-basis CP invariants. 

\subsection{CP invariants}

In  \cite{Branco:2001pq}, weak basis (WB)  invariants sensitive to the CP violating phases which appear in leptogenesis,  within the  type I seesaw model, were derived.
 All of them should vanish if CP 
is conserved, and conversely the non-vanishing of any of these invariants signals CP violation.
They must be invariant under the basis transformations:
\bea
\label{eq:basis}
\ell_L & \rightarrow &W_L \ell_L, 
\nonumber \\
N_R & \rightarrow &W_R N_R \ .
\eea
Defining $h\equiv Y^\dagger Y$, and $H_M\equiv M^\dagger M$, a subset of the invariants 
can be written as:
\bea
\label{eq:inv1}
I_1 &\equiv& {\rm Im Tr} [h H_M M^* h^* M],  \\
I_2 &\equiv& {\rm Im Tr} [h H_M^2 M^* h^* M], \\
I_3 &\equiv& {\rm Im Tr} [h H_M^2 M^* h^* M H_M]. 
\eea
Since the $I_i$ are WB invariants, we can evaluate them in any basis. In  the 
WB where the sterile neutrino mass matrix $M$ is real and diagonal, one obtains:
\bea
I_1 &= &M_1 M_2 \Delta M^2_{21} {\rm Im}(h_{12}^2) +  M_1 M_3 \Delta M^2_{31} {\rm Im}(h_{13}^2) +  M_2 M_3 \Delta M^2_{32} {\rm Im}(h_{23}^2),  \\
I_2 &=& M_1 M_2 (M_2^4-M_1^4) {\rm Im}(h_{12}^2) +  
M_1 M_3 (M_3^4-M_1^4) {\rm Im}(h_{13}^2) \nonumber \\
&+& 
 M_2 M_3 (M_3^4-M_2^4) {\rm Im}(h_{23}^2),  \\
I_3 &=& M_1^3 M_2^3 \Delta M^2_{21} {\rm Im}(h_{12}^2) +  
M_1^3 M_3^3 \Delta M^2_{31} {\rm Im}(h_{13}^2) + 
 M_2^3 M_3^3 \Delta M^2_{32} {\rm Im}(h_{23}^2),  
 \eea
where $\Delta M^2_{ij} \equiv  M_i^2 - M_j^2$ and, using the parametrization of eq.~(\ref{eq:yuk}) 
\bea
{\rm Im} (h_{ij}^2) =  {\rm Im} [(Y^\dagger Y)_{ij}^2] = \sum_{\alpha, \beta}
   y_\alpha^2 y_\beta^2 \, {\rm Im} [ W^*_{\alpha i} W^*_{\beta i}  W_{\alpha j}  W_{\beta j} ].
 \eea
It is explicit in the above expression that such unflavoured 
invariants depend only on the CP phases of the 
sterile neutrino sector, which are encoded in the unitary matrix 
$W$:
one Dirac-type phase, $\delta$ and two Majorana-type phases $\alpha_1,  \alpha_2$.
Not surprisingly, these invariants are the relevant ones in unflavoured leptogenesis, i.e., in the conventional computation of the CP asymmetry generated by  heavy Majorana neutrino decay neglecting flavour effects.

The combinations of $W$ matrix elements which appear in 
${\rm Im} (h_{ij}^2)$
 can be expressed in terms of 
the rephasing invariants defined in \cite{Jenkins:2007ip} as follows: 
\be 
{\rm Im} [W^*_{\alpha i} W^*_{\beta i}  W_{\alpha j}  W_{\beta j}]
=  \frac{ {\rm Im} [W_{\alpha i} W^*_{\beta i}  W^*_{\alpha j}  W_{\beta j}
(W_{\alpha j}  W^*_{\alpha i} )^2  ]}
{|W_{\alpha i}  W_{\alpha j}|^2} \ .
\ee

Notice that $ J_W \equiv \pm {\rm Im} [W_{\alpha i} W^*_{\beta i}  W^*_{\alpha j}  W_{\beta j}]$ is the 
Jarlskog invariant for the matrix $W$, while the quantities 
${\rm Im} [ (W_{\alpha j}  W^*_{\alpha i} )^2]$ determine the Majorana phases, $\alpha_{1,2}$.
When considering
processes, such as heavy neutrino oscillations, where the Majorana nature does not
play a role, only the Dirac phase $\delta$ will be relevant and therefore we expect to find just the Jarlskog invariant of the matrix W.

Since there are six independent CP-violating phases, it is possible to construct three more 
independent WB invariants, which would complete the description of CP violation in the 
leptonic sector. One simple choice are those invariants obtained from $I_i$ under the 
change of the matrix $h$ by $\bar h \equiv Y^\dagger h_{\ell} Y$, with 
$h_\ell = \lambda_\ell \lambda_\ell^\dagger$, being $\lambda_\ell$ the  charged lepton Yukawa couplings, 
i.e., 
\be 
\label{eq:inv2}
\bar{I}_1 =  {\rm Im Tr} [Y^\dagger h_\ell Y  H_M M^* Y^T h_\ell^* Y^* M]  \ , 
\ee
and analogously for $\bar{I}_2,\bar{I}_3$. 
The corresponding CP odd invariants are ${\rm Im} (\bar{h}_{ij}^2)$, which 
in the basis where also the charged lepton Yukawa matrix is real and diagonal 
can be written as:
\be
\label{eq:CPLV}
{\rm Im} (\bar{h}_{ij}^2) = \sum_{\alpha, \beta} \lambda_\alpha^2 \lambda_\beta^2 \, 
{\rm Im} [ Y^*_{\alpha i}  Y_{\alpha j}  Y_{\beta j} Y^*_{\beta i} ] \ .
\ee

The lepton number (L) violating part of the flavoured CP asymmetries in leptogenesis 
depends on the above combinations \cite{Covi:1996wh}:
\be
\epsilon^{{\not L}}_{i \alpha} = \sum_{\beta,j} 
{\rm Im} [ Y^*_{\alpha i}  Y_{\alpha j}  Y_{\beta j} Y^*_{\beta i}]  \tilde{f}(M_i, M_j) \ ,   
\ee
where $\tilde{f}$ is an arbitrary function. Upon substitution of 
 the neutrino Yukawa couplings as given in eq.~(\ref{eq:yuk}) 
can be written as:
\be
\epsilon^{ {\not L}}_{i \alpha} = \sum_j
\sum_{\beta,\delta,\sigma} y_\beta \, y_\delta \, y_\sigma^2  \, {\rm Im} 
[W^*_{\beta i} V_{\beta \alpha} V^*_{\delta \alpha}  W_{\delta j} W^*_{\sigma i}   W_{\sigma j}  ]
\tilde{f}(M_i, M_j) \ .
\ee
These asymmetries contain the additional rephasing invariants of the form 
${\rm Im}[W^*_{\beta i} V_{\beta \alpha} V^*_{\delta \alpha}  W_{\delta j}]$, which depend on 
the phases in the matrix $V$($\bar{\delta}, \phi_1, \phi_2$), showing that the flavoured CP asymmetries of leptogenesis are also sensitive to the CP phases in the $V$ 
leptonic mixing matrix, besides those in $W$.

Alternatively,  we  choose  to construct  the WB invariants which  will appear when the Majorana character of the sterile neutrinos is not  relevant, i.e., L-conserving 
ones.
These are given by: 
\bea
\label{eq:inv3}
\bar{I}_1' &\equiv& {\rm Im Tr} [h H_M^2 \bar h H_M] 
\nonumber  \\
&=& 
M_1^2 M_2^2  \Delta M^2_{21} {\rm Im}(h_{12} \bar{h}_{21} )
 +  M_1^2 M_3^2 \Delta M^2_{31} {\rm Im}(h_{13}  \bar{h}_{31} )  \nonumber \\
 &+& 
 M_2^2 M_3^2  \Delta M^2_{32} {\rm Im}(h_{23}  \bar{h}_{32})\ ,  \\
\bar{I}_2' &\equiv& {\rm Im Tr} [h H_M^3 \bar h H_M ] 
\nonumber \\
&=&M_1^2 M_2^2 (M_2^4-M_1^4) {\rm Im}(h_{12} \bar{h}_{21}  ) +  
M_1^2 M_3^2 (M_3^4-M_1^4) {\rm Im}(h_{13} \bar{h}_{31} ) \nonumber \\
&+& 
 M_2^2 M_3^2 (M_3^4-M_2^4) {\rm Im}(h_{23} \bar{h}_{32} )\ ,  \\
\bar{I}_3' &\equiv& {\rm Im Tr} [h H_M^3 \bar h  H_M^2]  
\nonumber \\
&=& M_1^4 M_2^4  \Delta M^2_{21} {\rm Im}(h_{12} \bar{h}_{21} )
 +  M_1^4 M_3^4 \Delta M^2_{31} {\rm Im}(h_{13}  \bar{h}_{31} )  \nonumber \\
 &+& 
 M_2^4 M_3^4  \Delta M^2_{32} {\rm Im}(h_{23}  \bar{h}_{32})   \ , 
\eea
where
\be 
\label{eq:CPLC}
 {\rm Im}(h_{ij} \bar{h}_{ji} ) = 
 \sum_{\alpha, \beta} \lambda_\alpha^2 \, {\rm Im} [Y_{\alpha i} Y^*_{\alpha j}  Y_{\beta j} Y^*_{\beta i}] \ .
\ee

The L-conserving CP asymmetry in leptogenesis via heavy 
 neutrino decay, as well as the CP asymmetries encountered in leptogenesis through sterile 
neutrino oscillations, are sensitive to the above combinations of Yukawa couplings \cite{Covi:1996wh}:
\be
\epsilon^{L}_{i \alpha} = \sum_{j,\beta} 
{\rm Im} [Y_{\alpha i} Y^*_{\alpha j}  Y_{\beta j} Y^*_{\beta i}] \, f(M_i, M_j) \ , 
\ee
where $f$ is an arbitrary function, and  can be written in terms of the rephasing invariants as:
\be
\label{eq:epsLC}
\epsilon^{L}_{i \alpha} = - \sum_j
\sum_{\beta,\delta,\sigma} y_\beta y_\delta y_\sigma^2  \, {\rm Im} 
[W^*_{\beta i} V_{\beta \alpha} V^*_{\delta \alpha}  W_{\delta j} W_{\sigma i}   W^*_{\sigma j}  ] 
\, f(M_i, M_j)\ .
\ee
Notice that the crucial difference between 
the $L$-violating and the $L$-conserving CP asymmetries is that in $\epsilon^{L}_{i \alpha}$
the combination of $W$ matrix elements is such that all dependence on the Majorana 
phases $\alpha_{1,2}$  disappears, as expected. 

In the approximation of neglecting $y_3 \ll y_1,y_2$, 
we obtain that 
$ {\rm Im} [ Y_{\alpha i}  Y^*_{\alpha j} (Y^\dagger Y)_{ij} ] =
\sum_{\beta}{\rm Im} [ Y_{\alpha i}  Y^*_{\alpha j}  Y_{\beta j} Y^*_{\beta i}] $
reduces to
\bea
 {\rm Im}   [ Y_{\alpha i}  Y^*_{\alpha j} (Y^\dagger Y)_{ij} ]
&=& y_1^2 y_2^2  (|V_{2\alpha}|^2 - |V_{1\alpha}|^2)
{\rm Im} [ W^*_{1i} W_{1j} W^*_{2j} W_{2i}] 
\nonumber \\
 &+& y_1 y_2 
 \left\{ 
 \left [ y_2^2 |W_{2i}|^2  - y_1^2 |W_{1i}|^2 \right]  
 {\rm Im} [W^*_{1 j} V_{1 \alpha} V^*_{2 \alpha}  W_{2 j} ] \right.
 \nonumber \\
 \label{eq:invLC}
&+& \left. \left[  y_1^2 |W_{1j}|^2 \ - y_2^2 |W_{2j}|^2  \right] 
{\rm Im} [W^*_{1 i} V_{1 \alpha} V^*_{2 \alpha}  W_{2 i } ] \right \} \ ,
\eea
so in principle we expect that the lepton asymmetry will depend on ten CP invariants, 
 namely ${\rm Im} [W^*_{1 i} V_{1 \alpha} V^*_{2 \alpha}  W_{2 i } ] $, 
with $i=1,2,3$  and $\alpha=1,2,3$ and $J_W$.

However, they are not all independent. In ref.~\cite{Jenkins:2007ip} it has been shown that 
in the minimal seesaw 
there are only six independent  CP invariants that can be made out of the matrices $V, W$. 
Two of them correspond to the Majorana phases of $W$, $\alpha_{1,2}$, which as we 
have argued before will not contribute in the limit of small sterile neutrino Majorana masses 
that we are considering. Other two are the equivalent of the Jarlskog invariants for the 
matrices $V,W$ and therefore determine the Dirac phases, $\bar{\delta}, \delta$, respectively.
The last two are of the form 
${\rm Im} [W^*_{1 i} V_{1 \alpha} V^*_{2 \alpha}  W_{2 i } ] $, 
for two reference values of $i, \alpha$, that fix the additional phases $\phi_{1,2}$. 
Moreover, it can be shown that
since we are neglecting the Yukawa coupling $y_3$, the phase 
$\phi_2$ of the matrix $V$ does not appear in 
eq.~(\ref{eq:invLC}), thus we are left with only three independent 
invariants.

The unitarity of the mixing matrices 
$V, W$ implies that 
\bea
\sum_\alpha V_{1 \alpha} V^*_{2 \alpha} & =& 0 \ , 
\\
\sum_i W^*_{1 i}   W_{2 i } &=& 0 \ ,
\eea
which allows to write  the invariants 
${\rm Im} [W^*_{1 i} V_{1 \alpha} V^*_{2 \alpha}  W_{2 i } ] $ for $\alpha=2$ in terms 
of those with $\alpha=1,3$, and the invariants  for $i=2$
in terms of the corresponding ones with $i=1,3$. 
By exploiting the identities 
\be
  {\rm Im} [W^*_{1 i} V_{1 \beta} V^*_{2 \beta}  W_{2 i } ] 
= \frac{{\rm Im} [(W^*_{1 i} V_{1 \alpha} V^*_{2 \alpha}  W_{2 i } )
(V^*_{2 \beta} V_{2 \alpha} V^*_{1 \alpha}  V_{1 \beta } )]  }
{|V_{1 \alpha} V_{2 \alpha} | ^2}.
\ee
we can
 write for instance  one of the invariants with $\beta =3$ in terms of the 
invariant with $\alpha=1$ and the Jarlskog invariant for $V$, 
${\rm Im} [V^*_{2 \beta} V_{2 \alpha} V^*_{1 \alpha}  V_{1 \beta } ]  = \pm J_V $.

 It is simpler, though, to write the results in terms of   the following 
 four  invariants,  even if only three are  independent, expanded up to  
3rd order in the small mixing angles $\theta_{ij}, \bar{\theta}_{ij}$:
\bea 
  I_1^{(2)}& =&  -{\rm Im} [W^*_{1 2} V_{1 1} V^*_{2 1}  W_{2 2 } ] \simeq
 \theta_{12} \bar{\theta}_{12} \sin \phi_1,\nonumber
\\
I_1^{(3)} &=&   {\rm Im} [W^*_{12} V_{13} V^*_{2 3}  W_{2 2 } ] 
\simeq
    \theta_{12} \bar{\theta}_{13}  \bar{\theta}_{23} \sin(\bar{\delta} + \phi_1),\nonumber
\\
I_2^{(3)} &=&  {\rm Im} [W^*_{13} V_{12} V^*_{2 2}  W_{2 3 } ] \simeq 
  \bar{\theta}_{12}   \theta_{13}  \theta_{23} \sin (\delta -\phi_1),\nonumber\\
  J_W  &=& - {\rm Im}[W^*_{23} W_{22} W^*_{32} W_{33} ] \simeq {\theta}_{12}   \theta_{13}  \theta_{23} \sin\delta .
  \label{eq:cpinvs}
 \eea

A generic expectation for the CP-asymmetry relevant for leptogenesis is 
\be 
 \Delta_{CP}= \sum_{\alpha, k} |Y_{\alpha k} |^2 \,  \Delta_\alpha,
\ee
with
\be 
\Delta_\alpha = \sum_{i} \epsilon^L_{i \alpha}
=
\sum_{i,j}  {\rm Im} [ Y_{\alpha i}  Y^*_{\alpha j} (Y^\dagger Y)_{ij} ] f(M_i,M_j).
\ee
Since the CP rephasing invariants are at least second order in the angles, 
we just need to take the diagonal elements in $\Delta_{CP}$, to keep the result up to 3rd order.
Then, in the limit $y_3 =0$, we get:
\bea 
\Delta_{CP} &=& y_1^2 y_2^2 (y_2^2 - y_1^2) \sum_{i,j} 
 {\rm Im} [ W^*_{1 i}  W_{1 j} W^*_{2 j}  W_{2 i} ]  f(M_i,M_j)
 \nonumber \\
 &+&y_1 y_2 \left ( (y_2^2 -y_1^2) \left\{ I_1^{(2)} [g(M_1) - g(M_2)] +  I_2^{(3)}
 [g(M_1) - g(M_3)]  \right\} \right. \\
&-& \left. y_2^2  I_1^{(3)} [g(M_1) - g(M_2)] \right),
\nonumber 
\eea
where
\be 
g(M_i) \equiv y_1^2 [f(M_1,M_i) - f(M_i,M_1)] - y_2^2 [f(M_2,M_i) - f(M_i,M_2)] \ .
\ee

From the above definition of $g(M_i)$, it immediately follows that
$g(M_1) - g(M_2) = (y_2^2 - y_1^2) [f(M_1,M_2) - f(M_2,M_1)] $, so 
$\Delta_{CP}$ simplifies to
\bea 
\Delta_{CP} &=& y_1^2 y_2^2 (y_2^2 - y_1^2) \sum_{i,j} 
 {\rm Im} [ W^*_{1 i}  W_{1 j} W^*_{2 j}  W_{2 i} ]  f(M_i,M_j)
\nonumber  \\
 &+&y_1 y_2 (y_2^2 - y_1^2) \left \{ \left [ (y_2^2 -y_1^2)  I_1^{(2)}  - y_2^2  I_1^{(3)}\right]
  [f(M_1,M_2) - f(M_2,M_1)] \right.
  \nonumber \\
  &+& \left.
    I_2^{(3)} [g(M_1) - g(M_3)]  \right\}  \ .
\label{deltaCPN}
\eea

We will see in the next section that this is precisely the yukawa and mixing angle dependence we will find when solving 
the kinetic equations, which is a strong crosscheck of the result.

\section{Perturbative Solution of the Raffelt-Sigl equation}

\subsection{Sterile neutrino production}

Our starting point is the Raffelt-Sigl formulation \cite{Sigl:1992fn} of the kinetic equations that describe the production of sterile neutrinos in the early Universe. The density matrix is the expectation value of the one-particle number operator for momentum $k$: $\rho_N(k)$ for neutrinos,  and $\bar{\rho}_N(k)$ for antineutrinos. We will assume that only sterile neutrinos and  the lepton doublets are out of chemical equilibrium, but assume that all the particles are in kinetic equilibrium, using Maxwell-Boltzmann statistics:
\begin{eqnarray}
\rho_a(k) = A_a \rho_{\rm eq}(k), \;\;\; A_a = e^{\mu_a}; \;\; \rho_{\bar{a}}(k) = A_{\bar{a}} \rho_{\rm eq}(k), \;\;\; A_{\bar{a}} = e^{-{\mu_a}}, 
\end{eqnarray}
where $\rho_{\rm eq}(k) \equiv e^{-k_0/T}$, with $k_0= |\mathbf{k}|$, and $\mu_a$ denotes the chemical potential normalised by the temperature. 
 We will furthermore neglect spectator processes and the washout induced by the asymmetries in all the fields other than the sterile neutrinos and lepton doublets. We expect this approximation to give uncertainties of ${\mathcal O}(1)$ which for our purpose is good enough~\cite{Nardi:2005hs}.

In \cite{Akhmedov:1998qx}, only the asymmetry in the sterile sector was considered, neglecting the feedback of the leptonic chemical potentials. In this case, the equations get the standard form
\begin{eqnarray}
\dot{\rho}_N = -i \big[ H, \rho_N\big] -{1 \over 2} \big\{ \Gamma, \rho_N - \rho_{\rm eq}\big\},
\label{eq:ars}
\end{eqnarray}
and the analogous for $\bar{\rho}_N$ with $H \rightarrow H^*$, 
where $H$ is the Hamiltonian (we neglect matter potentials for the time being but we will include them later on)
\begin{eqnarray}
H \equiv  W \Delta W^\dagger, \;\;\; \Delta \equiv {\rm Diag}\Big(0, {\Delta M^2_{12}\over 2 k_0}, {\Delta M^2_{13}\over 2 k_0}\Big).
\end{eqnarray}
$\Gamma$ is the rate of production/annihilation of sterile neutrinos in the plasma, which is diagonal in the basis that diagonalises the neutrino Yukawa's:
\begin{eqnarray}
\Gamma= {\rm Diag}(\Gamma_1, \Gamma_2, 0),\;\;\;\Gamma_i \propto y_i^2,
\end{eqnarray}
where we assume $y_3=0$.
 In deriving eq.(\ref{eq:ars}) it is assumed that the particles involved in the production/annihilation of the sterile neutrinos are in full equilibrium (all chemical potentials vanish), and that kinematical effects of neutrino masses are negligible. 

Note that only the matrix $W$ appears in these equations and therefore any CP asymmetry generated can only be proportional to the invariant $J_{W}$  which 
depends at third order on the mixing angles of  $W$. 

In \cite{Asaka:2005pn} it was correctly pointed out that the asymmetries in the sterile sector will be modified by the leptonic chemical potentials that will be generated as soon as sterile neutrinos start to be produced. Including the evolution of the leptonic chemical potentials has two important consequences:  new sources of CP violation become relevant and washout effects are effective. Leptons are fastly interacting through
 electroweak interactions in the plasma and therefore it is a good approximation to assume they are in kinetic equilibrium. 

An important question is what is the flavour structure of these chemical potentials.  For $T \lesssim 10^9$~GeV the Yukawa interactions of the tau and muon are very fast,
which implies that $\mu$ will be diagonal in the basis that diagonalises the charged lepton Yukawa matrix, since no other interaction changing flavour is in equilibrium before the heavy neutrinos are produced. Note however that this is not the basis where the neutrino Yukawas are diagonal, the two are related by the mixing matrix $V$. As a result, when 
the evolution of the lepton chemical potentials is taken into account, the CP phases of the matrix $V$ become relevant. 

\begin{figure}
 \begin{center}
\includegraphics[scale=0.7]{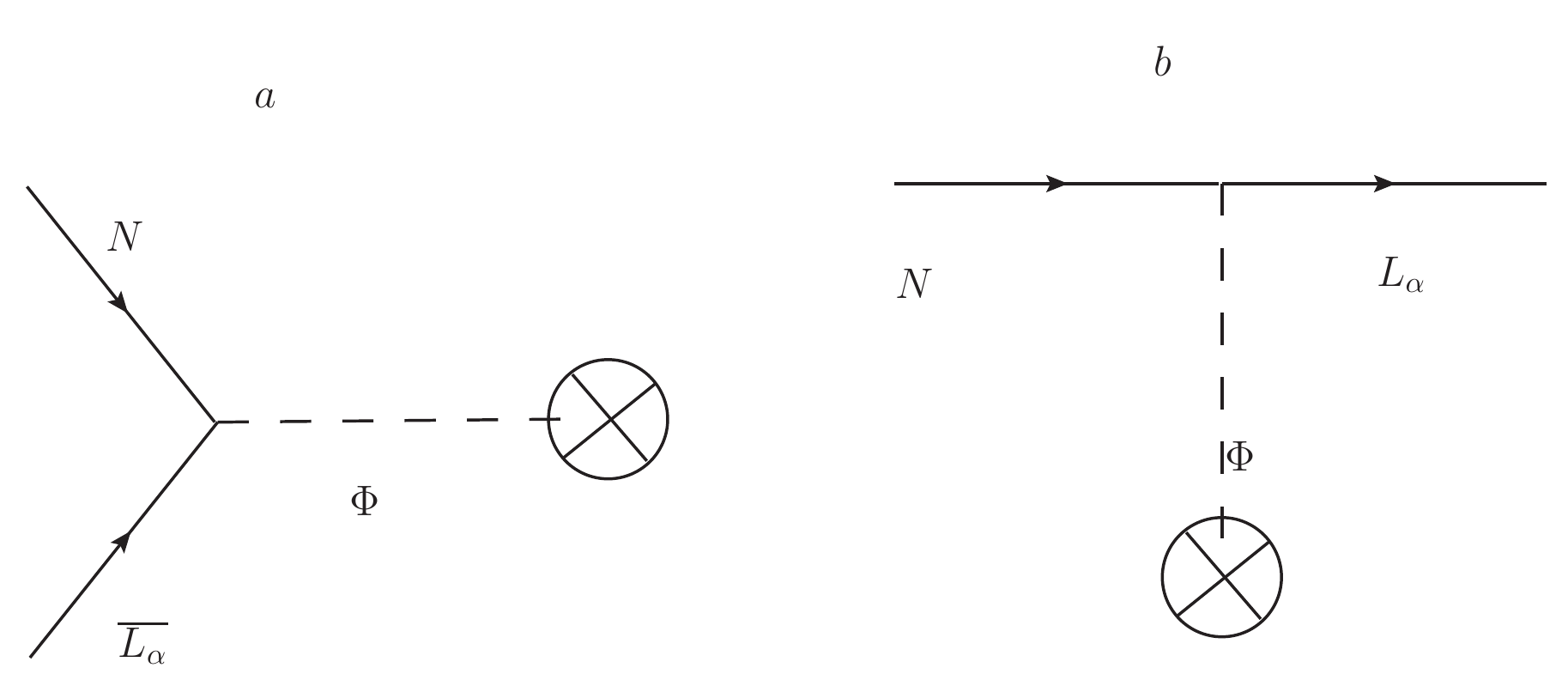} 
\caption{\label{fig:diagr2} $a,b$ topologies for annihilation/production of sterile neutrinos }
\end{center}
\end{figure}

Adapting the derivation of \cite{Sigl:1992fn} to this situation, we find that the evolution of the CP-even and CP-odd parts of the neutrino densities: $\rho_\pm \equiv {\rho_N\pm {\bar \rho}_N\over 2}$ and the lepton chemical potentials
, $\mu_\alpha$,   to linear order in $\mu_\alpha, \rho_-$, satisfy in this case:
\begin{eqnarray}
\dot{\rho}_+ &=& -i [H_{\rm re}, \rho_+] +  [ H_{\rm im}, \rho_-] -{\gamma_N^a+\gamma_N^b\over 2} \{Y^\dagger Y, \rho_+-\rho_{\rm eq}\} \nonumber\\
&&+ i \gamma_N^b {\rm Im}[Y^\dagger \mu Y]  \rho_{\rm eq}  + i {\gamma_N^a \over 2}  \big\{{\rm Im}[Y^\dagger \mu Y],\rho_+\big\},  \nonumber\\
\dot{\rho}_- &=& -i [H_{\rm re}, \rho_-] +  [ H_{\rm im}, \rho_+] -{\gamma_N^a+\gamma_N^b\over 2} \big\{Y^\dagger Y, \rho_-\big\}\nonumber\\
&&+  \gamma_N^b {\rm Re}[Y^\dagger \mu Y]  \rho_{\rm eq}  +  {\gamma^a_N \over 2}  \big\{{\rm Re}[Y^\dagger \mu Y],\rho_+\big\},  \nonumber\\
\dot{\mu}_\alpha &=&   -  \mu_\alpha  \left( \gamma_\nu^b {\rm Tr}[YY^\dagger I_\alpha]+ \gamma_\nu^a{\rm Tr}\big[{\rm Re}[Y^\dagger I_\alpha Y] r_+ \big] \right)\nonumber\\
&&+ (\gamma_\nu^a+ \gamma_\nu^b)\left( {\rm Tr}\big[ {\rm Re}[Y^\dagger I_\alpha Y] r_- \big]+i {\rm Tr}\big[ {\rm Im}[Y^\dagger I_\alpha Y] r_+ \big]\right)\ ,
\end{eqnarray}
where $H_{\rm re} \equiv {\rm Re}[H]$, $H_{\rm im} \equiv {\rm Im}[H]$, $I_\alpha$ is the projector on flavour $\alpha$ and  $\gamma^{a,b}_N, \gamma_\nu^{a,b}$ are the rates of production/annihilation of a sterile neutrino or a lepton doublet neglecting all masses, after factorizing the flavour structure in the Yukawas, 
\begin{eqnarray}
\gamma_{N(\nu)}^{a(b)} \equiv {1\over 2 k_0}
\sum_i \int_{\mathbf{ p_1,p_2,p_3}}  \rho_{\rm eq}(p_1)  |{\mathcal M_{N(\nu),i}^{(a(b))}}|^2  (2 \pi)^4 \delta(k+p_1-p_2-p_3) ,
\end{eqnarray}
where $k$ is the momentum of the $N$ or $\nu$ and 
 $a(b)$ refer to the s-channel (t,u-channels) depicted in figure~\ref{fig:diagr2}. In topology $a$ the lepton and sterile neutrino are {\it both} in the initial or final state, while topology $b$ corresponds to those
diagrams where one is in the initial and other in the final state. Finally 
\begin{eqnarray}
r_{\pm} \equiv  {\sum_i \int_{\mathbf{p_1,p_2,p_3}}  \rho_\pm(\mathbf{p_1})  |{\mathcal M_{\nu i}^{(a)}}|^2  (2 \pi)^4 \delta(k+p_1-p_2-p_3)   \over \sum_i \int_{\mathbf{ p_1,p_2,p_3}}  \rho_{\rm eq}(\mathbf{p_1})  |{\mathcal M_{\nu i}^{(a)}}|^2  (2 \pi)^4 \delta(k+p_1-p_2-p_3)} \  .
\end{eqnarray}
 A similar derivation can be found in \cite{Asaka:2011wq} and we agree with their findings.  

These equations reduce to those in eq.~(\ref{eq:ars}) in the limit $\mu \rightarrow 0$ with $\Gamma_i =y_i^2 (\gamma_N^a+ \gamma_N^b)$. 
 
Most previous studies have assumed that the rates are dominated by the top quark scatterings. In this case, the rates are given (in the Boltzman approximation) by the well-known result \cite{Luty:1992un,Besak:2012qm}
\begin{eqnarray}
\gamma_{N,Q}^b = 2 \gamma_{N,Q}^a = 2 \gamma_{\nu,Q}^b = 4 \gamma_{\nu,Q}^a = {3 \over 16 \pi^3} {y_t^2 T^2\over k_0}. 
\end{eqnarray}
The factor of 2 difference between the rates of the $N$ and the $\nu$ is due to the fact that the lepton is a doublet and the sterile neutrino is a singlet. Note that there is a non-linear term of the form ${\mathcal O}(\mu \rho_+)$, as first noted  in \cite{Asaka:2011wq}.  More recently in \cite{Shuve:2014zua}, the equations have been written in terms of the $\mu_{B- L_\alpha/3}$ chemical potentials, however not all the chemical potentials (e.g. higgs and top quark) have been included. A full treatment including all chemical potentials  will be postponed for a future work, but we expect that including these spectator effects will change the results by  factors of ${\mathcal O}(1)$. 
 
In \cite{Besak:2012qm,Ghisoiu:2014ena}, it has been pointed out that the scattering processes $\bar{L} N \leftrightarrow W H$ get a strong enhancement from hard thermal loops and are actually the dominant scatterings. The results of  \cite{Besak:2012qm,Ghisoiu:2014ena} however do not include the chemical potentials of spectators, so it is not clear
how to include them  consistently in the above equations. We will neglect these effects in the following. Note however that the lepton flavour structure of these and of the top quark scatterings is the same. 

It is easy to see also that total lepton number is conserved as it should:
\begin{eqnarray}
2 \sum_\alpha \dot{\mu}_\alpha + {\rm Tr}[\dot{r}_- ]=0.
\end{eqnarray}

Two approximations are  often used in solving these equations: 1) assume that the momentum dependence of $\rho_\pm$ follows that of $\rho_{\rm eq}$, i.e. kinetic equilibrium for the sterile states, which implies $r_\pm = \rho_\pm/\rho_{\rm eq}$ are constants  and the integro-differential equations become just differential equations,  2) neglect the $k_0$ dependence of the rates by approximating
\begin{eqnarray}
\langle k^{-1}_0 \rangle 
\simeq {T^{-1}\over 2}.
\end{eqnarray}
The effect of these approximations has been studied numerically in \cite{Asaka:2011wq} and the results do not differ too much. We will therefore adopt both approximations that simplify considerably the perturbative treatment. 

\subsection{Lepton asymmetries in the sterile sector }

We are going to solve these equations   perturbing in the mixing angles up to third order. We first consider the simpler case, neglecting leptonic chemical potentials and considering in turn the evolution in a static Universe and in the expanding case. 

\subsubsection{Static Universe}
\label{static}	

 We start with eq.~(\ref{eq:ars}) and assume $y_3=0$. In this case, neither $H$ nor $\Gamma$ depend on time.  Defining 
 $\rho_{N ij}/\rho_{\rm eq} \equiv a_{ij} + i b_{ij}$ and taking into account the hermiticity of $\rho_N$ we change the matrix equation into 
 a vector equation:
\begin{eqnarray}
r \equiv (a_{11}, a_{22}, a_{12}, b_{12}, a_{13}, b_{13}, a_{23}, b_{23}, a_{33}).
\end{eqnarray}
At 0-th order  the system of equations of eq.~(\ref{eq:ars}) can be rewritten as
\begin{eqnarray}
\dot{r}^{(0)} = A_0 r^{(0)} + h_0, 
\end{eqnarray}
with 
\begin{eqnarray}
h_0 \equiv (\Gamma_1 \rho_{\rm eq}, \Gamma_2 \rho_{\rm eq},0,....0), 
\end{eqnarray}
and the matrix $A_0$ is constant and has a block structure:
\begin{eqnarray}
A_0\equiv \left(\begin{array}{lll} (A_I)_{4\times 4} & 0 & 0\\
0 & (A_{II})_{4\times 4} & 0 \\
0 & 0  & 0\end{array}\right), 
\end{eqnarray}
\begin{footnotesize}
\begin{eqnarray}A_I\equiv \left(\begin{array}{llll} -\Gamma_1 & 0 & 0 &  0\\
0 & -\Gamma_2 & 0 & 0  \\
0 & 0  & -{\Gamma_1 + \Gamma_2 \over 2} & -\Delta_{12} \\
& &\Delta_{12} & -{\Gamma_1 + \Gamma_2 \over 2}  \end{array}\right), \; A_{II}\equiv \left(\begin{array}{llll} -\Gamma_1/2 &  -\Delta_{13}& 0 &  0 \\
 \Delta_{13} & -\Gamma_1/2 & 0&0   \\
0 &  0 & -{\Gamma_2 \over 2} & \Delta_{12} -\Delta_{13} \\
0& 0 & -(\Delta_{12}-\Delta_{13})& -{ \Gamma_2 \over 2}  \end{array}\right).
\end{eqnarray}
\end{footnotesize}
The matrix can be easily diagonalised and exponentiated so the general solution to the equation is
\begin{eqnarray}
r^{(0)}(t) = e^{A_0 t} \int_0^t d x~ e^{-A_0 x} h_0.\;\;
\end{eqnarray}

At the next order we have to keep ${\mathcal O}(\theta_{ij})$ in the Hamiltonian and translate the matrix form into the vector form:
\begin{eqnarray}
- i [H^{(1)}, \rho^{(0)}(t)] \rightarrow A_1 r^{(0)}.
\end{eqnarray}
The equation for the first order correction to the density is
\begin{eqnarray}
\dot{r}^{(1)} = A_0 r^{(1)} + A_1  r^{(0)}(t).
\end{eqnarray}
The solution at this order is therefore
\begin{eqnarray}
r^{(1)}(t) = e^{A_0 t} \int_0^t d x e^{-A_0 x} A_{1} r^{(0)}(x).
\end{eqnarray}
We can iterate this procedure to get the correction at order $n$: 
\begin{eqnarray}
\dot{r}^{(n)}(t) = A_0 r^{(n)}(t) + \sum_{i=1}^{n-1} A_i  r^{(n-i)}(t),
\end{eqnarray}
with solution
\begin{eqnarray}
r^{(n)}(t) = e^{A_0 t} \int_0^t d x e^{-A_0 x} \sum_{i=1}^{n-1} A_i r^{(n-i)}(x).
\label{eq:nthorder}
\end{eqnarray}
We can define the evolution operator 
\begin{eqnarray}
U_0(t,x) \equiv e^{A_0 t}  e^{-A_0 x}, 
\end{eqnarray}
so that the solution can be written as
\begin{eqnarray}
r^{(n)}(t) =  \int_0^t d x~ U_0(t,x) \sum_{i=1}^{n-1} A_i r^{(n-i)}(x).
\label{eq:nthorderU}
\end{eqnarray}
As a first estimate of the leptonic asymmetry that can be generated, we are interested in 
$\Delta\rho_{33}$ since this is the sector that will never reach equilibrium (in the absence of mixing) and therefore can act as reservoir of the leptonic asymmetry until the electroweak phase transition \cite{Akhmedov:1998qx}.

\begin{figure}
 \begin{center}
\includegraphics[scale=0.14]{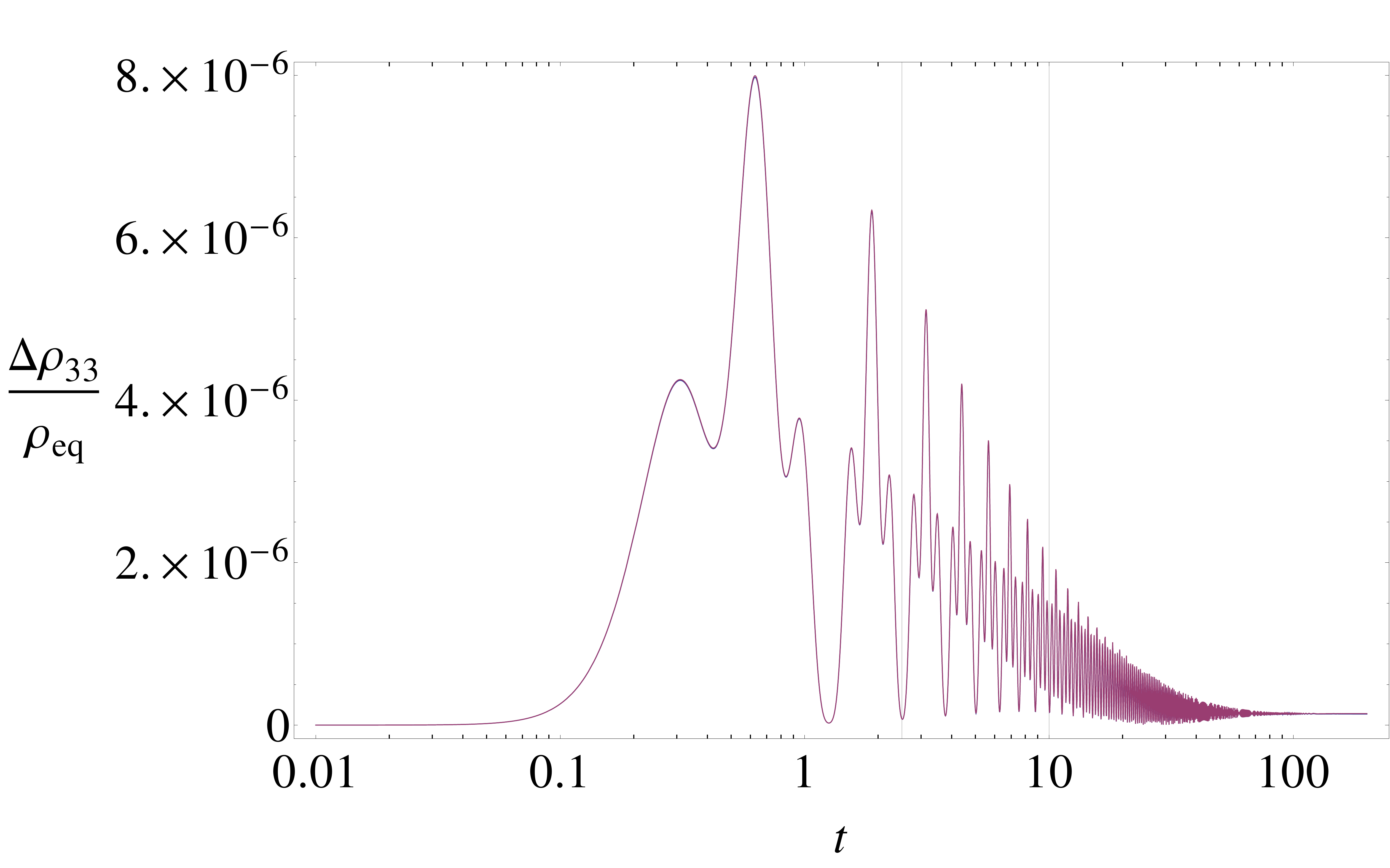} \includegraphics[scale=0.17]{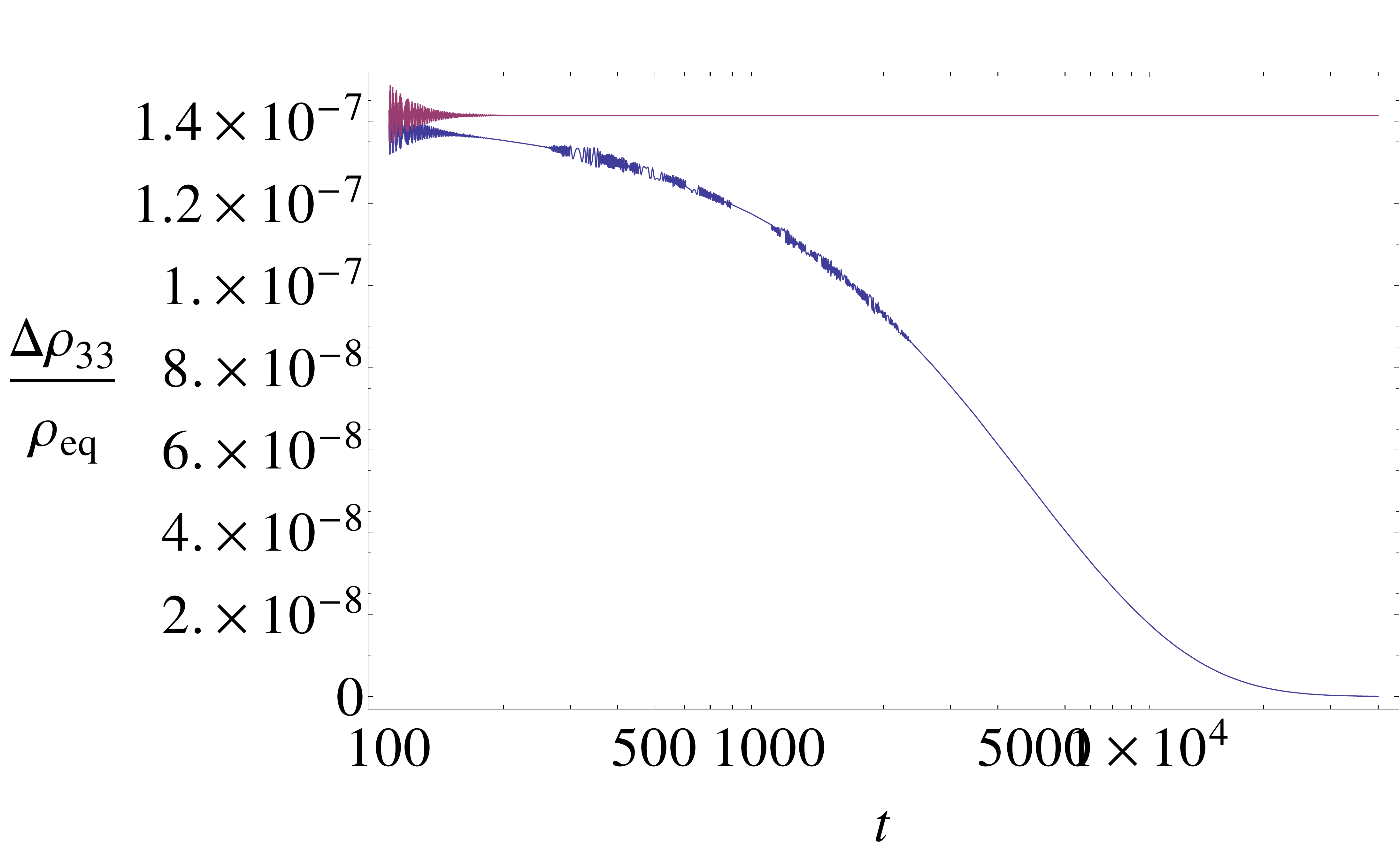}
\caption{\label{fig:nohubble} Comparison of numerical (blue) and perturbative (red) solution for $\Delta \rho_{33}$ as a function of time, in the case with no expansion of the Universe. The two curves are indistinguishable (left plot) until large 
times (right plot): the vertical line on the lower plot corresponds to $(\theta^2_{31} \Gamma_1)^{-1}$, while those on the upper one correspond to $\Gamma_2^{-1}$ and $\Gamma_1^{-1}$ respectively. }
\end{center}
\end{figure}

One can easily compute the solution of the eq.~(\ref{eq:nthorder}) up to order $n=3$, which is the first order that gives a non-vanishing result, as expected from general 
considerations on CP invariants. The result  at finite $t$ is not particularly illuminating but the limit $t\rightarrow \infty$ is rather  simple:
\begin{eqnarray}
\lim_{t\rightarrow \infty} {\Delta \rho_{33}\over \rho_{\rm eq}}\equiv \lim_{t\rightarrow \infty} {\rho_{N33} -\bar{\rho}_{N33}\over \rho_{\rm eq}} &=& 2 J_W   {(\Gamma_1-\Gamma_2) \Delta_{12} \Delta_{13} (\Delta_{12} - \Delta_{13})\over \left[\Delta_{13}^2 + {\Gamma_1^2\over 4} \right]  \left[(\Delta_{12}-\Delta_{13})^2 + {\Gamma_2^2\over 4} \right] }.
\label{eq:rho33nohubble}
\end{eqnarray}
A few comments are in order. 
We have not assumed any expansion in $\Gamma_i$ in this expression, only in the mixing angles. According to general theorems the equations should reach a stationary 
solution if all the eigenvalues of the matrix $A_0+ A_1+A_2+...$  are real and negative.   However, because $\Gamma_3=0$, one of the eigenvalues of $A_0$ vanishes and it is lifted only at second order in perturbation theory, 
$\sim \theta^2_{i3} \Gamma_{i}$, therefore we expect the  perturbative expansion should break down for $t \sim {1 \over \theta^2_{i3} \Gamma_{i}}$, which is the time scale of equilibration of the third state. On the other hand, if $\theta$ is small, the perturbative solution should be accurate for times $t \geq \Gamma_{1(2)}^{-1}$. Indeed this is precisely what we find comparing the perturbative and numerical solutions in figure ~\ref{fig:nohubble}.

The result is proportional to  $J_W$ which is the only CP rephasing invariant that can appear in this case. The result vanishes if any two of the masses or the yukawa's are degenerate, since the CP phase would be unphysical in this case.  
 
\subsubsection{Expanding Universe}
Let us turn now to the realistic case of an expanding Universe. As usual, we will consider the evolution as a function of the scale factor $x\equiv a$, in such a way 
that the Raffelt-Sigl equation  becomes
\begin{eqnarray}
\left.{d \over d t} \rightarrow x H_u(x) {\partial\over \partial x} \rho(x,y)\right|_{\rm fixed ~y} = - i[H(x,y), \rho(x,y)]  -{1 \over 2} \{\Gamma(x), \rho(x,y)-\rho_{eq}(y)\}, \nonumber\\
\end{eqnarray}
where $H_u(x)$ is the Hubble parameter, $H_u = \sqrt{4 \pi^3 g_*(T) \over 45 }{T^2\over M_{\rm Planck}}$, and 
$y\equiv {p\over T}$.  Assuming for simplicity a radiation dominated Universe with constant number of degrees of 
freedom, during the sterile evolution time we can assume $x T $= constant that we can fix to be one.  Therefore 
 the scaling of the different terms is
 \begin{eqnarray}
 H(x,y) \equiv x W {\Delta M^2\over 2 y} W^\dagger, ~\Gamma_i(x) \equiv {c_i\over x},  ~ x H_u(x) \equiv  {1\over M_{\rm P}^* x},
 \end{eqnarray}
 where $M_P^* \equiv M_{\rm Planck} \sqrt{45 \over 4 \pi^3 g_*(T_0)}$ and $g_*(T_0)$ is the number of relativistic degrees of freedom in the plasma during the sterile evolution. 
  
 Therefore the equation  as function of $x$ is:
  \begin{eqnarray}
 \dot{\rho} = - i x^2 [W \Delta W^\dagger, \rho]  -{1 \over 2} \{ \gamma, \rho-\rho_{eq}\} ,
\end{eqnarray}
where we have defined
\begin{eqnarray}
\Delta_{ij}  \equiv {\Delta M^2_{ij} \over 2 y} M_P^*, \;\;\; \gamma_i \equiv c_i M_P^*.  
\label{eq:reef}
\end{eqnarray}
The perturbative expansion works as in section~\ref{static}, but now all the $A_n(x)$ are $x$-dependent: $A_n(x)$ with $n\geq 1$ scale like the Hamiltonian, ie. $x^2$, while 
$A_0(x)$ contains terms that scale with $x^2$ and others that do not depend on $x$. Fortunately, there is an important simplification in that 
$A_0(x)$ can be diagonalised  by an $x$-independent matrix, therefore the path-ordered exponential can be easily evaluated.  The result can be written in the same form 
of eq.~(\ref{eq:nthorderU}), with the evolution operator given by
\begin{eqnarray}
U_0(t,r) =e^{\int_0^t A_0(x) dx } e^{-\int_0^r A_0(y) dy }.
\end{eqnarray}

  At third order  in the mixings, after algebraic simplifications and  partial integrations, the result can be given in terms of integrals of the form
\begin{eqnarray}
J_{n}(\alpha_1,\beta_1,..,\alpha_n,\beta_n,t) \equiv \int_0^t dx_1  e^{i \alpha_1   {x_1^3\over 3} + \beta_1 x_1} \int_0^{x_1}  dx_2 e^{i \alpha_2 {x_2^3\over 3} + \beta_2 x_2}..
 \int_0^{x_{n-1}} d x_n e^{i \alpha_n {x_n^3\over 3} + \beta_n x_n},\nonumber\\
 \end{eqnarray}
 where $\alpha_i$ are combinations of $\Delta_{ij}$ and $\beta_i$ are combinations of $\gamma_i$. Up to third order in the perturbative expansion only integrals with  $n\leq 3$ appear.

 Since we are in a regime where $\gamma_i \ll |\Delta_{ij}|^{(1/3)}$, the integrands are highly oscillatory and hard to deal with numerically. To evaluate the integrals, we  separate the integration interval  $[0, t] = [0, t_0] + [t_0,t]$  with $t_0$ such that  $t_0 |\Delta_{ij}|^{1/3} \gg 1$ and $t_0 \gamma_i \ll 1$: 
 \begin{eqnarray}
J_{n}(\alpha_1,\beta_1,..,\alpha_n,\beta_n,t) =J_{n}(\alpha_1, \beta_1,...\alpha_n,\beta_n,t_0) + \Delta J_n (\alpha_1, \beta_1,...\alpha_n,\beta_n,t_0, t).
 \end{eqnarray}
To solve the integrals up to $t_0$ we can safely Taylor expand in $\beta_i$  (which results in an expansion in $\gamma_{1,2}/|\Delta_{ij}|$) and write the integrals in terms of 
simpler integrals of the form: 
 \begin{eqnarray}
J_{n{\mathbf k}}(\alpha_1,..,\alpha_n,t) \equiv \int_0^t dx_1 x_1^{k_1}~ e^{i {\alpha_1   x_1^3\over 3} } \int_0^{x_1}  dx_2 x_2^{k_2} ~e^{i {\alpha_2 x_2^3\over 3} }..
 \int_0^{x_{n-1}} d x_n x_n^{k_n}~�e^{i {\alpha_n x_n^3\over 3} },\nonumber\\
  \end{eqnarray}
  up to third order in the $\beta_i$ expansion we just need integrals with $n+\sum_i k_i \leq 3$. 
    We can use the relation
  \begin{eqnarray}
 {d \over d x} \left[ F_n(x)  \right] =  x^n e^{i {\alpha  x^3\over 3}  }
  \end{eqnarray}
  with 
  \begin{eqnarray}
  F_n(\alpha, x) = - 3^{{n-2\over 3}} (-i \alpha )^{-{1+n\over 3}}~\Gamma\left[{1+n\over 3}, -{1\over 3} i \alpha x^3\right],
  \end{eqnarray}
to evaluate immediately the one-dimensional integrals in terms of incomplete $\Gamma$ functions.
The integrals in the range $[t_0, t]$ can be approximated by the large $t$ behaviour of the $J_{1n}(\alpha,t)$ functions, after resumming  the Taylor series in $\beta_i$.   Further details are presented in appendix A.

The finite $t$ dependence of the asymmetry $\Delta \rho_{33}$ is rather complicated, but the 
asymptotic value is non-zero and rather simple:
\begin{eqnarray}
\lim_{t\rightarrow\infty} {\Delta \rho_{33}\over \rho_{\rm eq}} &=& - J_W  \gamma_1 \gamma_2 (\gamma_2-\gamma_1) \lim_{t\rightarrow \infty} {\rm Im}[J_{3{\mathbf 0}}(\Delta_{12}-\Delta_{13}, -\Delta_{12},\Delta_{13}, t)+ J_{3\mathbf{0}}(\Delta_{12}-\Delta_{13},\Delta_{13}, -\Delta_{12},t)\nonumber\\
&+&J_{3{\mathbf 0}}(\Delta_{13}, -\Delta_{12},\Delta_{12}-\Delta_{13},t)+J_{3{\mathbf 0}}(\Delta_{13}, \Delta_{12}-\Delta_{13},-\Delta_{12},t)].\nonumber\\
\end{eqnarray}
This can be simplified to 
\begin{eqnarray}
\lim_{t\rightarrow\infty} {\Delta \rho_{33}\over \rho_{\rm eq}} &=&  - J_W  {\gamma_1 \gamma_2 (\gamma_2-\gamma_1) \over (\Delta_{13}\Delta_{12}\Delta_{23})^{1/3} } {\rm Im}\left[I\left( {\Delta_{12}\over \Delta_{23}},-{\Delta_{13}\over\Delta_{23}}\right)+I\left(-{\Delta_{12}\over\Delta_{13}},-{\Delta_{23}\over \Delta_{13}}\right)\right], \nonumber\\
\label{eq:rho33hubble}
\end{eqnarray} 
 where 
 \begin{eqnarray}
 I\left({\Delta_2\over \Delta_1},{\Delta_3\over \Delta_1}\right) \equiv  (\Delta_1 \Delta_2 \Delta_3)^{1/3} \int_0^\infty ~dx  e^{i {\Delta_1 x^3\over 3} }~  J_{10}(\Delta_2,x)   J_{10}(\Delta_3,x).
 \end{eqnarray}

Comparing eq.~(\ref{eq:rho33hubble}) and eq.~(\ref{eq:rho33nohubble}) we see that in the expanding case the asymmetry is 
cubic in $\gamma_i$ and not linear. Note that the dependence on the yukawa's is precisely that expected from a 
flavour invariant CP asymmetry. In fact this is effectively the situation in the expanding case, because the asymmetry 
is generated at times $t \ll \gamma_i^{-1}$ and the dependence in the yukawa's in this regime is therefore perturbative. 
This is in contrast with the non-expanding case, where the asymmetry evolves all the way till $t \sim \gamma_i^{-1}$. 
 To understand the reason behind this different behavior, it is useful to recall the definition of  $\Delta_{ij}$ from eq.~(\ref{eq:reef}). Then, we see that 
$\Delta_{ij} x^3 \gg 1$ implies $\Delta M_{ij}^2 /(4T)  \gg T^2/M_P^* = H_u(T)$, 
therefore in this regime the sterile neutrino oscillations  are much faster than the Hubble 
parameter and no asymmetry is produced anymore, since oscillations are averaged out.
Thus in the expanding Universe 
the generation of the asymmetry occurs at $x \sim |\Delta_{ij}|^{-1/3} \ll \gamma_i^{-1}$.

 Until now we have neglected the matter potentials, however given the suppression in three powers of $\gamma$ of the leading result, there are corrections of same order 
 coming from the potentials, and in fact they are numerically more important. 

 The equation including the potentials in the basis with diagonal neutrino Yukawas is:
  \begin{eqnarray}
 \dot{\rho} = - i x^2 [W \Delta W^\dagger, \rho]  -i [v,\rho] - {1 \over 2} \{ \gamma, \rho-\rho_{eq}\}, 
\end{eqnarray}
where
 \begin{eqnarray}
 v_{ij }= {y_i^2\over 8} {M_P^* }\delta_{ij}\equiv v_i \delta_{ij}. 
 \end{eqnarray}
   The result for the asymmetry including the potentials is given by:
    \begin{eqnarray}
\lim_{t\rightarrow\infty} {\Delta \rho_{33}\over \rho_{\rm eq}} =  J_W &\lim_{t\rightarrow\infty}  {\rm Re}&\left[ z_1 J_{30}(\Delta_{12}-\Delta_{13}, -\Delta_{12},\Delta_{13},t)+ z_2 J_{30}(\Delta_{12}-\Delta_{13}, \Delta_{13},-\Delta_{12},t) \right.\nonumber\\
&& + \left. z_2 J_{30}(\Delta_{13},\Delta_{12}-\Delta_{13}, -\Delta_{12},t)+  z_3 J_{30}(\Delta_{13}, -\Delta_{12},\Delta_{12}-\Delta_{13},t)\right].\nonumber\\
\label{eq:rho33hubblepot}
\end{eqnarray} 
with 
\begin{eqnarray}
z_1 &\equiv& \gamma_1 \gamma_2 \Delta_v + \gamma_1 v_2 \Delta_\gamma + i \left({\gamma_1 \gamma_2 \Delta_\gamma \over 2} - 2 \gamma_1 v_2\Delta_v\right),\nonumber\\
z_2 &\equiv& \left[ \gamma_1 v_2 -\gamma_2 v_1 + i \left({\gamma_1 \gamma_2 \over 2} +2 v_1v_2\right)\right] \Delta_\gamma ,\nonumber\\
z_3 &\equiv& -\gamma_1 \gamma_2 \Delta_v -   \gamma_2 v_1  \Delta_\gamma + i \left({\gamma_1 \gamma_2 \Delta_\gamma \over 2} - 2 \gamma_2  v_1 \Delta_v \right).
\end{eqnarray}
and $\Delta_v \equiv v_2 -v_1$ and $\Delta_\gamma \equiv (\gamma_2-\gamma_1)$.

The leading terms ${\mathcal O}(v^2\gamma)$ at asymptotic times $t \gg \gamma_{1,2}^{-1}$ are:
    \begin{eqnarray}
\lim_{t\rightarrow\infty} {\Delta \rho_{33}\over \rho_{\rm eq}} =  {9 y_t^6  \over 2048 \pi^3}J_W {y_1^2 y_2^2 (y_2^2 -y_1^2) M_P ^{^*2} \over |\Delta M^2_{12} \Delta M^2_{13} \Delta M^2_{23}|^{1/3}} \kappa,
\label{eq:rho33hubblepotnum}
\end{eqnarray} 
where
\begin{eqnarray}
\kappa &\equiv& |\Delta_{12} \Delta_{13} \Delta_{23}|^{1/3} {\rm Im}\big[  J_{30}(\Delta_{12}-\Delta_{13}, -\Delta_{12},\Delta_{13},t)-J_{30}(\Delta_{12}-\Delta_{13}, \Delta_{13},-\Delta_{12},t) \nonumber\\
&& - J_{30}(\Delta_{13},\Delta_{12}-\Delta_{13}, -\Delta_{12},t)+   J_{30}(\Delta_{13}, -\Delta_{12},\Delta_{12}-\Delta_{13},t)\big]
\end{eqnarray}
depends only on the ratios of mass differences and/or the ordering of the states. 
This result is parametrically the same as  the result of \cite{Akhmedov:1998qx} if we neglect the 
dependence of $\kappa$ on the mass differences and has the dependence on the yukawas expected from
eq.~(\ref{deltaCPN}).

Considering the naive seesaw scaling $y_i^2 \sim 2 {m_\nu M_i \over v^2}$, for $m_\nu \sim 1$ eV and assuming no
big hierarchies or degeneracies, i.e. $M_i^2 \sim \Delta M^2_{ij} \sim M^2$,  leads to
    \begin{eqnarray}
\lim_{t\rightarrow\infty} {\Delta \rho_{33}\over \rho_{\rm eq}} 
 \sim 2 \times 10^{-7} J_W \left({m_\nu \over 1 \, {\rm eV}}\right)^3 \left({M \over 10\, {\rm GeV}}\right).
\label{eq:rho33hubblepotnum2}
\end{eqnarray} 
The asymmetry is highly sensitive to the light neutrino mass. Note that we have pushed the value to the limit, a light neutrino mass in the  less constrained $0.1$ eV range would imply three orders of magnitude suppression. The asymmetry grows linearly with the mass of the heavy steriles. However, for masses larger than $\sim$ 10 -100~GeV lepton number violating transitions via the Majorana mass could washout further the asymmetry, an effect that requires a refinement of the formulation to be taken into account.

\subsection{Lepton asymmetries in the active sector }
 
 The asymmetry generated ignoring the $\mu$ evolution  depends only on the Dirac-type phase, $\delta$, appearing in $W$ as we have seen. However when 
the evolution of the leptonic chemical potentials is included, other phases contribute to the total lepton asymmetry. We will perform a perturbative expansion
 to third order in the mixings of both $V$ and $W $ matrices. 
  
  The result at finite $t  \ll \theta_{i3} ^2(\bar{\theta}_{i3}^2) \gamma_i^{-1}$ can be written in the form:
 \begin{eqnarray}
  {\rm Tr}[\mu](t)&=& \sum_{I_{\rm CP}}  I_{\rm CP} A_{I_{\rm CP}}(t)
 \end{eqnarray} 
 where all the four CP  invariants appear, $I_{\rm CP}= \left\{J_W, I_1^{(2)}, I_1^{(3)},I_2^{(3)}\right\}$, given in eqs.~(\ref{eq:cpinvs}).

At finite $t$, the result for the functions $A_{I_{CP}}$ is well approximated by 
  \begin{eqnarray}
  A_{I_1^{(2)} } (t)&=&   y_1 y_2 (y_2^2- y_1^2) \left(1- {\gamma_N\over \bar{\gamma}_N}\right) \gamma_N^2  G_1(t),\nonumber\\
  A_{I_1^{(3)} } (t)&=& -y_1 y_2 (y_2^2- y_1^2) \left(1- {\gamma_N\over \bar{\gamma}_N}\right) \gamma_N^2 G_2(t),\nonumber\\
  A_{I^{(3)}_{2}}(t) &=&  y_1 y_2 \left(1-{\gamma_N\over {\bar \gamma}_N}\right) \gamma_N G_3(t),\nonumber\\
  A_{J_W}(t) &=&  \gamma_1 \gamma_2 \left(1-{{\gamma}_N\over\bar{ \gamma}_N}\right) G_{41}(t) - {\gamma_N \over 2 \bar{\gamma}_N} G_{42}(t).
  \label{eq:AI1}
  \end{eqnarray}
 where  $\gamma_N \equiv \gamma_N^a+ \gamma_N^b$ and $\bar{\gamma}_N \equiv {2 \gamma_N^a+ 3 \gamma_N^b\over 2}$, while  
 \begin{eqnarray}
G_1(t) &\equiv&   \left(e^{-\bar{\gamma}_2  t} - e^{-\bar{\gamma}_1 t }\right) {\rm Re}\left[i J_{20}(\Delta_{12},-\Delta_{12}, t)+2 \Delta_v J_{201}(\Delta_{12},-\Delta_{12}, t)\right]  \nonumber\\
&+&{1\over 2}  \sum_{k=1}^2 (-1)^k e^{-\bar{\gamma}_k t}  {\rm Re}\left[J_{210}(\Delta_{12},-\Delta_{12}, t) \left(-2 \Delta_v + i (2 \bar{\gamma}_k - \gamma_1-\gamma_2)\right) \right],   \nonumber\\
\label{eq:g1}
\end{eqnarray}
and
\begin{eqnarray}
G_2(t) =\left. G_1(t)\right|_{\bar{\gamma}_1=0},
\label{eq:g2}
\end{eqnarray}
where we have defined $\bar{\gamma}_i \equiv y_i^2 \bar{\gamma}_N$ and $\Delta_v \equiv v_2 - v_1$, and the result for $G_3(t), G_{41}(t), G_{42}(t)$ are lengthier and reported in the appendix B.  These results  would get modified for  $\gamma_i t \gg 1$  had we included the non-linear terms that modify the rate of thermalisation at large times.  In these equations there is an implicit expansion up to third order in $\gamma_i(v_i)/\Delta^{1/3}$ when 
$\Delta^{1/3} t \gg 1$, while the terms $\gamma_i(v_i) t$ are resumed.

\begin{figure}
 \begin{center}
\includegraphics[scale=0.57]{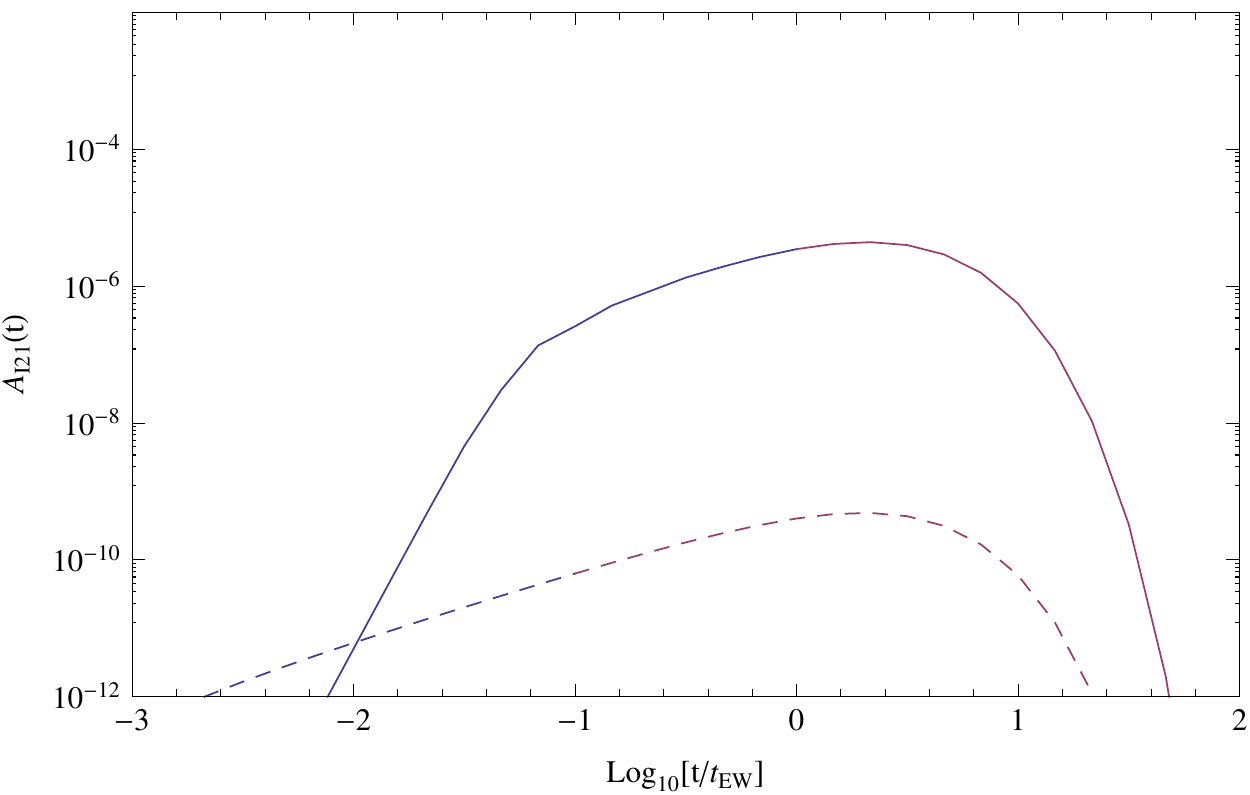} \includegraphics[scale=0.57]{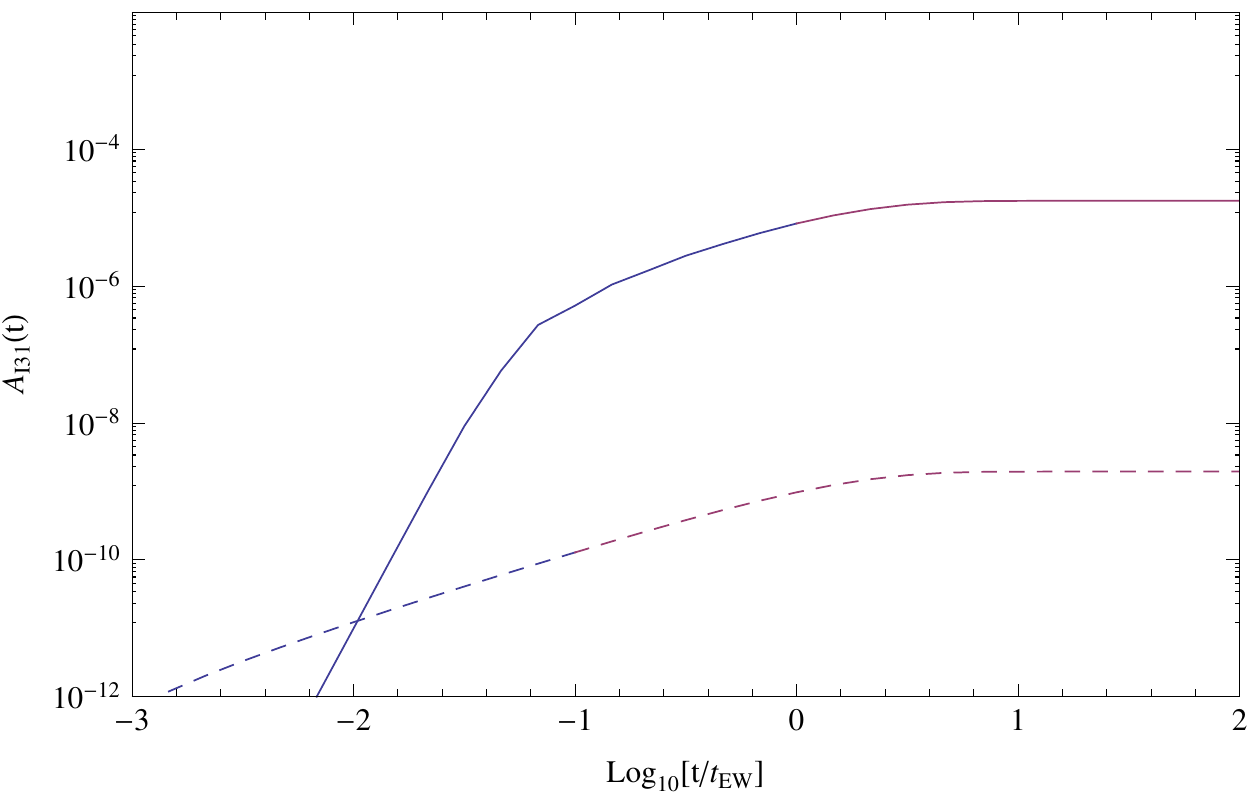}
\caption{\label{fig:AICP1} Functions $A_{I^{(2)}_1}(t)$ (left) and $A_{I^{(3)}_1}(t)$ (right)  assuming the rates are dominated by top quark scattering, and taking $y_2/\sqrt{2}= y_1= 10^{-7}$, for two choices of $\Delta M^2_{12} = 1 $GeV$^2$ (dashed) and $\Delta M^2_{12} = 10^{-6}$ GeV$^2$ (solid). $t_{EW}$ is the electroweak phase transition time, corresponding to $T_{EW}\simeq140$GeV. }
\end{center}
\end{figure}

In figure ~\ref{fig:AICP1} we plot the functions 
$A_{I^{(2)}_1}(t)$  and $A_{I^{(3)}_1}(t)$, which  depend only on one neutrino mass difference. We show  two physical situations: one with very degenerate neutrinos and the other with no strong degeneracies. 

These two invariants are the only ones relevant for the scenario that has been considered in most previous studies, 
where it has been assumed that only two sterile neutrinos have a role in generating the lepton asymmetry (see for 
instance \cite{Abada:2015rta} for a very recent analysis). This is the situation in the limit of  complete decoupling of $N_3$,  ensured by the condition  $\theta_{i3} = 0$, implying that only the invariants $I_1^{(2)}$ and $I_1^{(3)}$ survive.  In \cite{Asaka:2005pn} an approximate analytical solution was obtained, expanding in the yukawa's, under the assumption that $|\Delta_{12}|^{-1/3} \ll t_{EW} \ll \bar{\gamma}_i^{-1}$. In this limit,  the result of eqs.~(\ref{eq:AI1}) and (\ref{eq:g1}) can be simplified to 
\begin{eqnarray}
 {\rm Tr}[\mu](t_{EW})&\simeq& -\big((y_2^2-y_1^2\big)I_1^{(2)}  - y_2^2 I_1^{(3)}) y_1 y_2 (y_2^2- y_1^2) 
 \left(1- {\bar{\gamma}_N\over \gamma_N}\right) \gamma_N^3  { {\rm Im}[J_{20}(\Delta_{12}, -\Delta_{12}, \infty)] 
 \over T_{EW}}.\nonumber\\
 \end{eqnarray}
Comparing with eq.~(\ref{deltaCPN}), we see that the dependence on the yukawa's is again that expected from a 
flavour invariant CP asymmetry. Using 
 \begin{eqnarray}
 {\rm Im}[J_{20}(\Delta_{12}, -\Delta_{12}, \infty)] =- 2 \left({2\over 3}\right)^{1/3} {\pi^{3/2}\over  \Gamma[-1/6] }  {{\rm sign}(\Delta_{12})\over |\Delta_{12}|^{2/3}},
 \label{eq:imj2}
\end{eqnarray}
and $\bar{\gamma}_N = {4\over 3} \gamma_N$, and assuming the naive seesaw relations $y_1^2 = 2 {\sqrt{\Delta_{sol}} M_1\over v^2}$, $y_2^2=2 {\sqrt{\Delta_{atm}} M_2\over v^2}$  we find:
\begin{eqnarray}
 {\rm Tr}[\mu](t_{EW})&\simeq&  10^{-2}  (I_1^{(2)}  - I_1^{(3)})  {\sqrt{M_1 M_2^{7/3}}\over {\rm GeV}^{5/3}} \left( {M^2_2\over  |\Delta M^2_{12}|}\right)^{2/3}  , 
 \end{eqnarray}
 while for $y_1^2 = y_2^2/2 = 10^{-14}$ (that would correspond to light neutrino masses in the eV range and heavy ones in the GeV range)  we would have 
\begin{eqnarray}
{\rm Tr}[\mu](t_{EW})&\simeq&  7 \times 10^{-10} {I_1^{(2)}  - 2 I_1^{(3)}\over  |\Delta M^2_{12} ({\rm GeV}^2)|^{2/3}}  .
\end{eqnarray} 
Even if the CP  invariants are of ${\mathcal O}(1)$, the asymmetry is too small unless there is a significant degeneracy between the two states \cite{Asaka:2005pn}. 
  It is important however to realise that the naive seesaw scaling is too naive and a full exploration of parameter space is necessary. 

\begin{figure}
 \begin{center}
\includegraphics[scale=0.57]{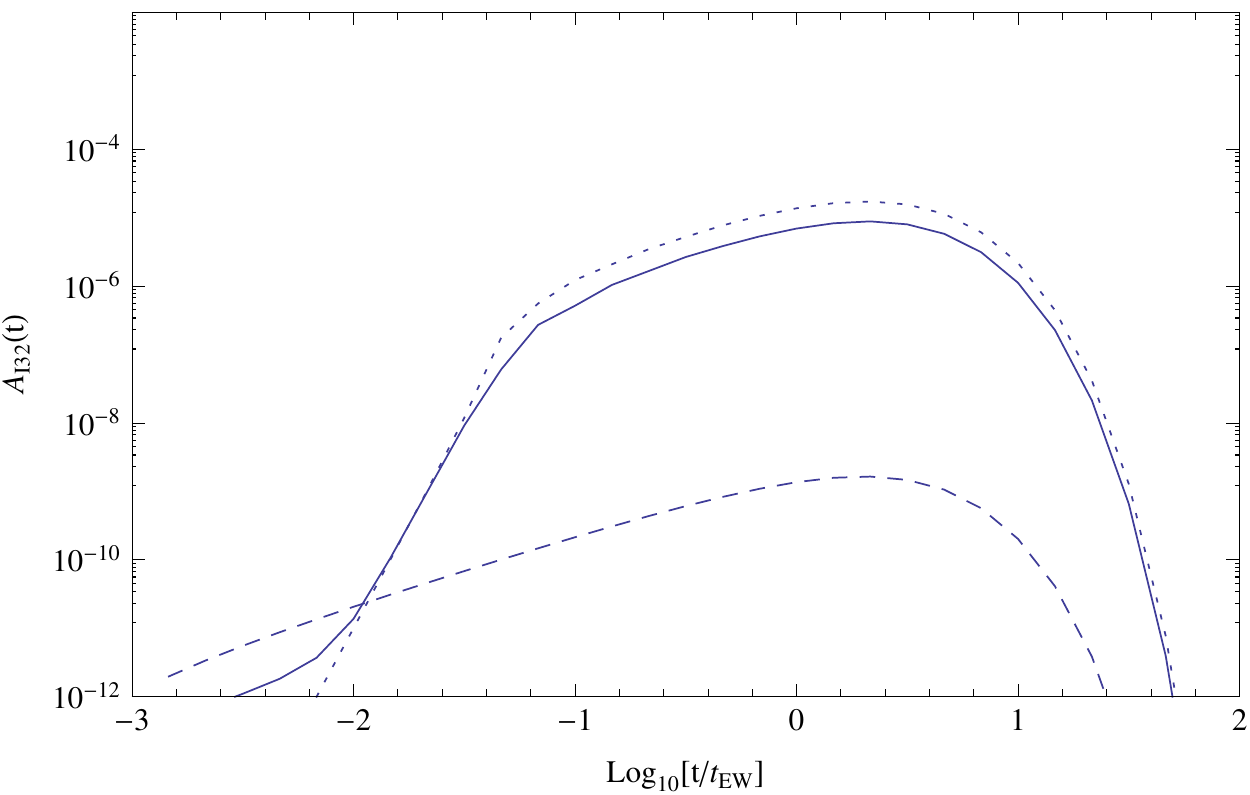} \includegraphics[scale=0.57]{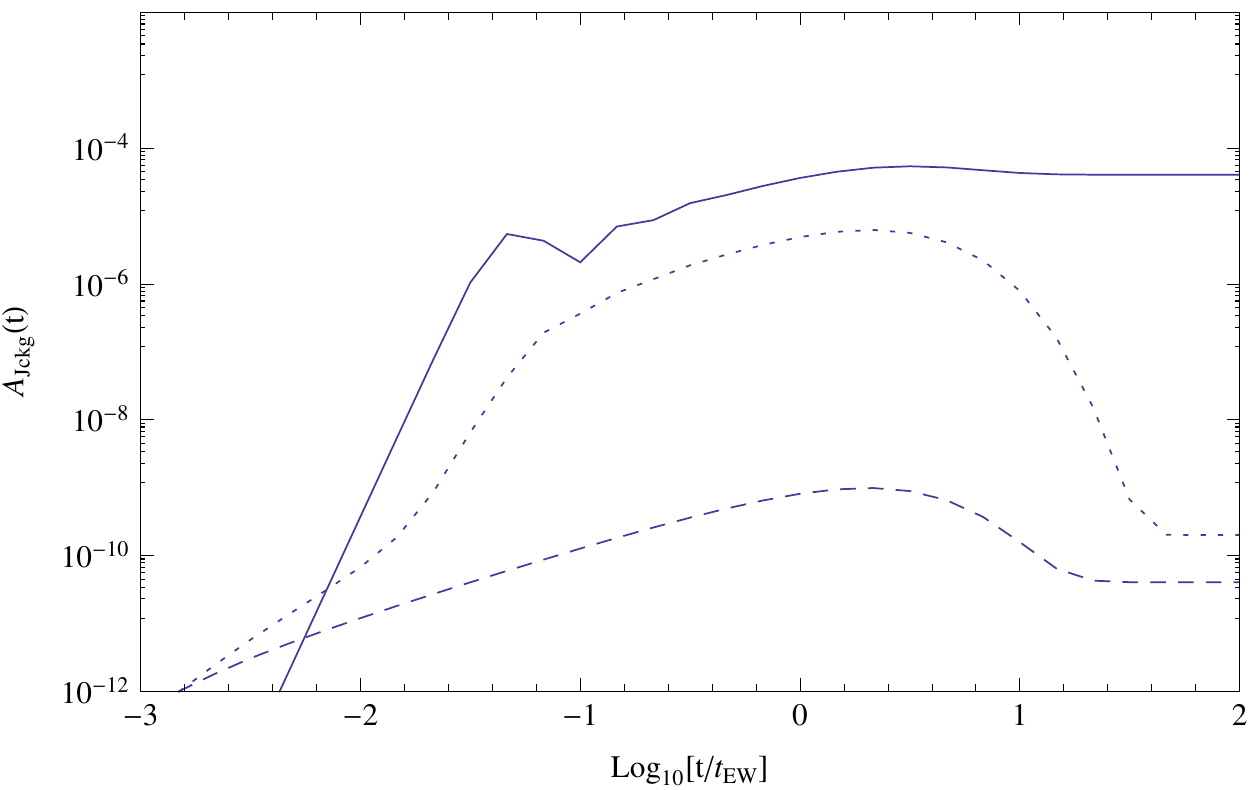}
\caption{\label{fig:AICP2} Functions $A_{I^{(3)}_2}(t)$ (left) and $A_{J_W}(t)$ (right)  assuming the rates are dominated by top quark scattering, and taking $y_2/\sqrt{2}= y_1= 10^{-7}$, for three choices of $[\Delta M^2_{12},\Delta M^2_{13}]  = [1,2], [10^{-6},2]$ and $[10^{-6}, 2\times 10^{-6}]$ in GeV$^2$ (dashed, dotted and  solid). $t_{EW}$ is the electroweak phase transition time, corresponding to $T_{EW}\simeq140$GeV. }
\end{center}
\end{figure}

 In figure~\ref{fig:AICP2}  we plot the functions $A_{I^{(3)}_2}(t)$  and $A_{J_W}(t)$. They depend on the two neutrino mass differences, so we show three examples here: one
 in which there are no degeneracies, one where there are two almost degenerate states, and the case where the three states are almost degenerate. As in the previous case
 we see a large enhancement when only one of the mass differences is small and a further enhancement when the two are small compared to the absolute scale. In the case
 of $A_{J_W}$ we find that there is a significant difference in the regime $\Delta_{ij}^{1/3} t \ll 1$ if we plot  $A_{J_W}(t)$ truncated to the terms of ${\mathcal  O}(y_i^6)$. As we will see in the next section, the latter is much closer to the numerical result. The reason for this difference is that 
 at small times,  $\Delta^{1/3} t \ll 1$,  only  some terms of order ${\mathcal O}(y^8_i)$ are kept in eqs.~(\ref{eq:AI1}), while there is a strong cancellation if all had  been included. Note however that this effect is only important at times where the asymmetry is suppressed and seems to affect only $A_{J_W}$. 

 It is interesting to note that even though the dependence on the yukawas of the functions $A_{I_{CP}}(t)$ is different 
(fourth or sixth order), the maxima for all cases are roughly of the  same order of magnitude.  
Note, however, that in the limit $t\gg \gamma_i^{-1}$, only the contribution of two invariants, $J_W$ and $I_1^{(3)}$, survive:
\begin{eqnarray}
 \lim_{t\rightarrow \infty} {\rm Tr}[\mu](t)&\simeq&-{\gamma_N \over 2 \bar{ \gamma}_N}  \lim_{t\rightarrow \infty}  \left[{ \Delta\rho_{33}(t)\over \rho_{\rm eq}} \right]_{\rm eq.~(3.41)} 
 \nonumber\\
 &-& {2^{4/3} \pi^{3/2}\over 3^{1/3}  \Gamma\left[-{1\over 6}\right]}  I^{(3)}_1 {y_1 y_2 (y_2^2- y_1^2)\over |\Delta_{12}|^{2/3}} \left(1- {\gamma_N\over \bar{\gamma}_N}\right) \gamma_N^2,
 \end{eqnarray} 
 where  we kept only the leading terms ${\mathcal O}(y^4)$ proportional to $I_1^{(3)}$ and
 we have used the result of eq.~(\ref{eq:imj2}).

The first term in this expression corresponds to the expectation of \cite{Akhmedov:1998qx}, ie. the final asymmetry is proportional to that stored in the third sterile state, eq.~(\ref{eq:rho33hubblepot}), while the second term was missing in the simplified treatment of \cite{Akhmedov:1998qx}. Note that they depend on different CP invariants. 

\section{Numerical solution}

In order to check the accuracy of the analytical solutions presented in the previous section, we have solved the differential equations numerically. As shown in \cite{Asaka:2011wq}, the momentum dependence does not change significantly the results so we will consider the average-momentum approximation.

\begin{figure}
 \begin{center}
\includegraphics[scale=0.165]{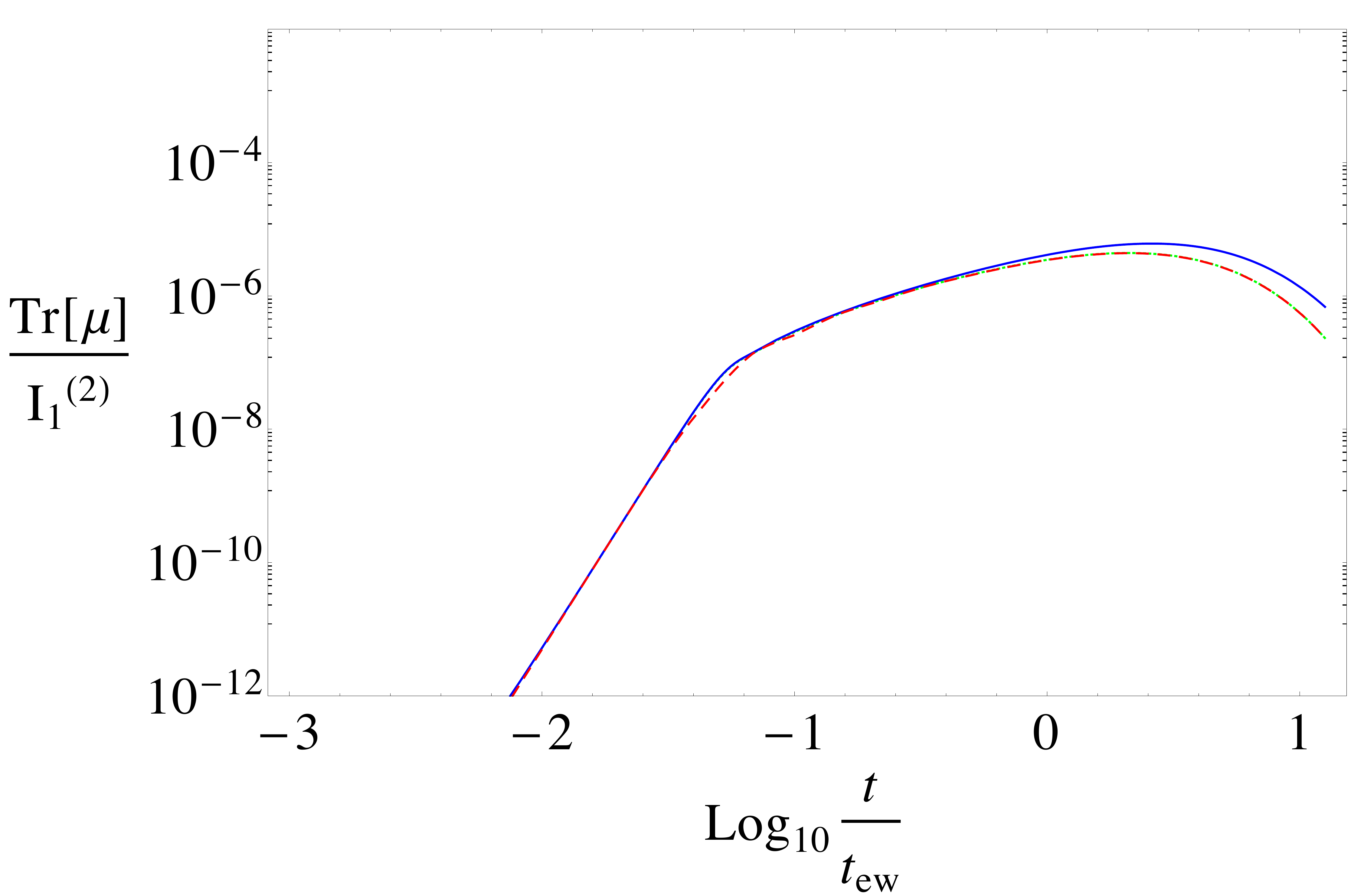} \includegraphics[scale=0.165]{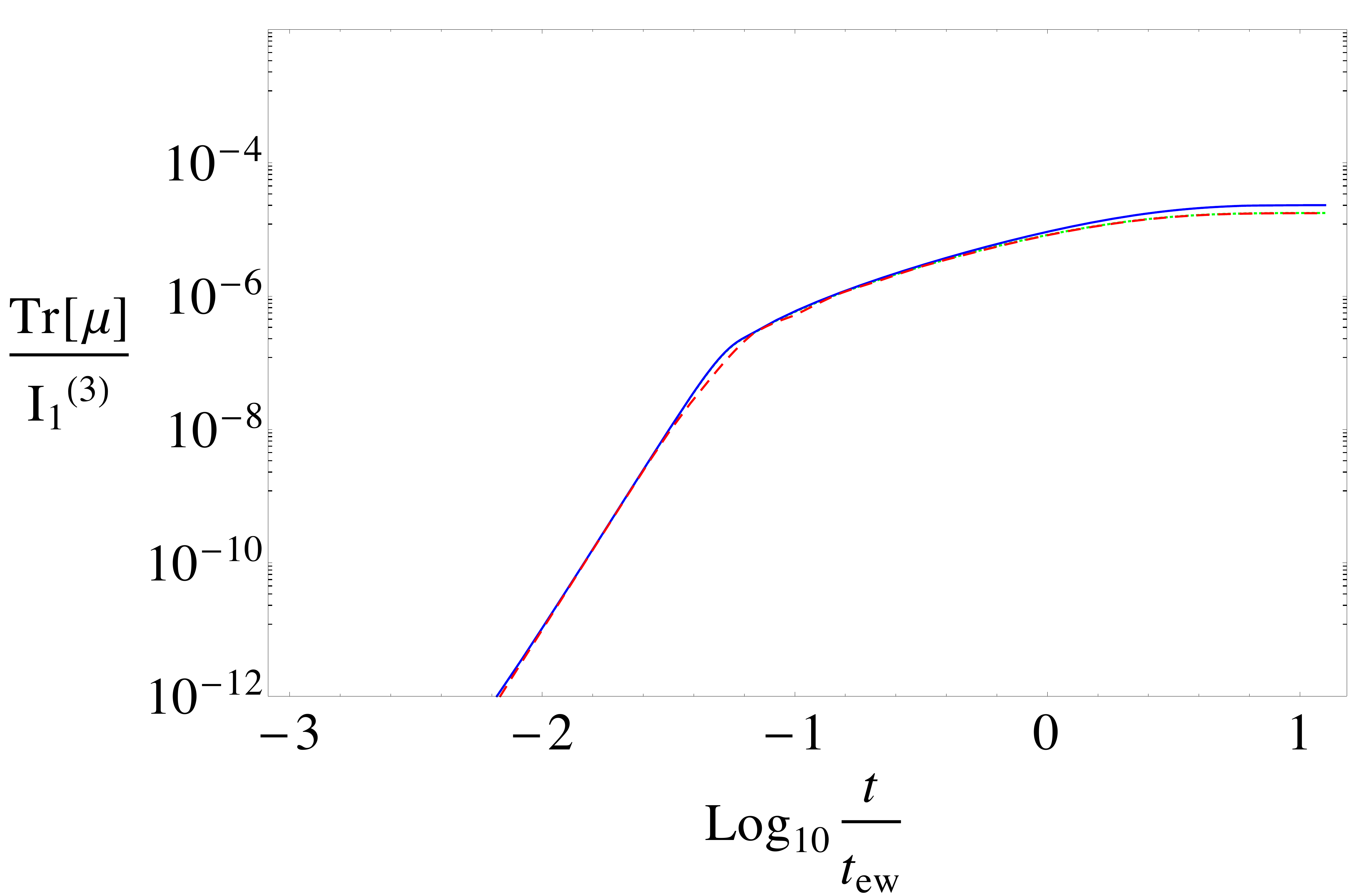}
\caption{\label{fig:AICPnum1} Left: full numerical solution (solid blue) and numerical solution neglecting non-linear terms (dotted green) for case 1, normalised to the invariant $I^{(2)}_1$,  compared with the prediction, $A_{I^{(2)}_1}(t)$ (dashed red). Right: same for case 2 normalised to the invariant $I^{(3)}_1$ compared to $A_{I^{(3)}_1}(t)$.  The parameters are the same as in figure~\protect\ref{fig:AICP1} for the degenerate case. }
\end{center}
\end{figure}

\begin{figure}
 \begin{center}
\includegraphics[scale=0.165]{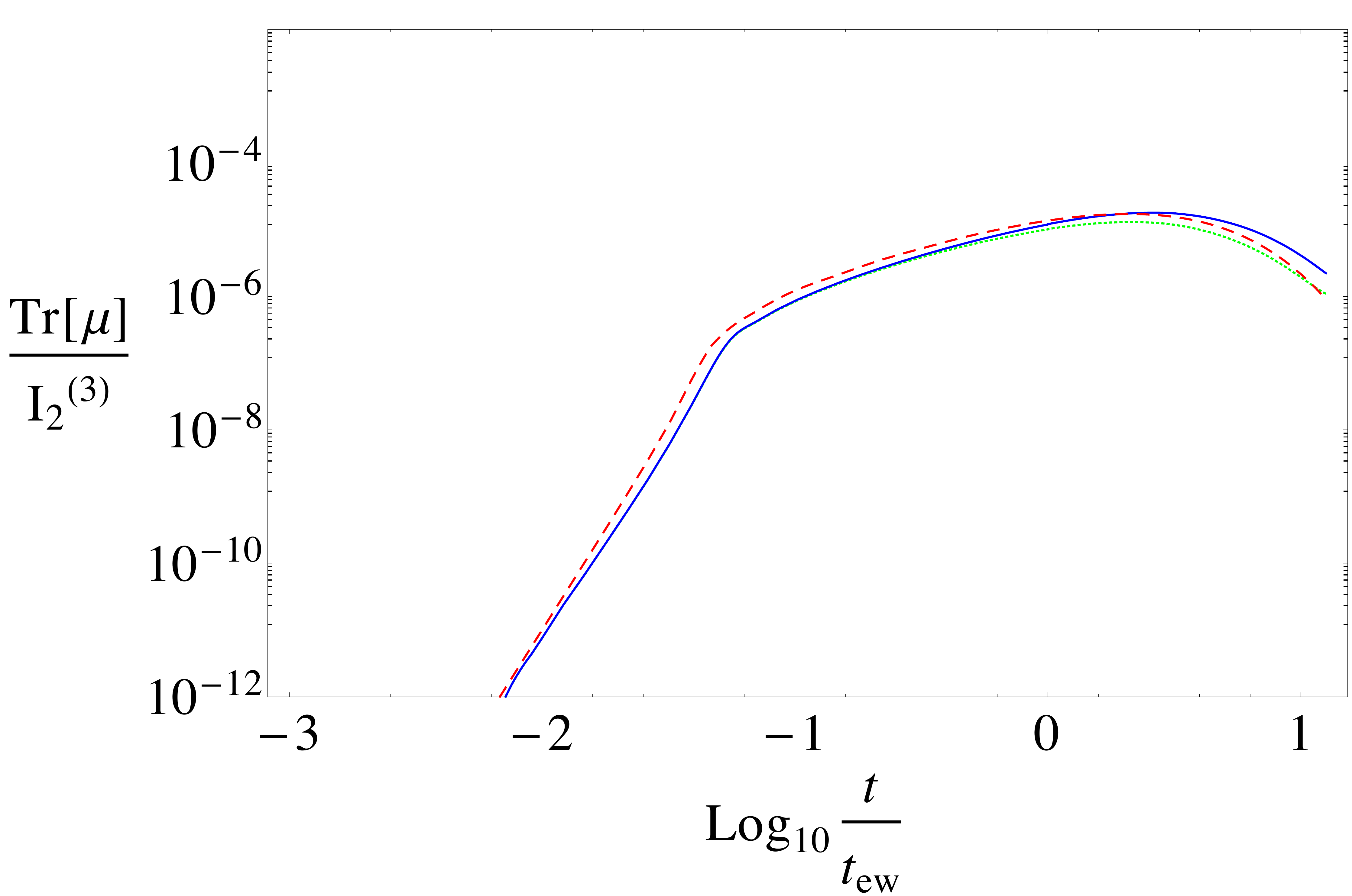} \includegraphics[scale=0.165]{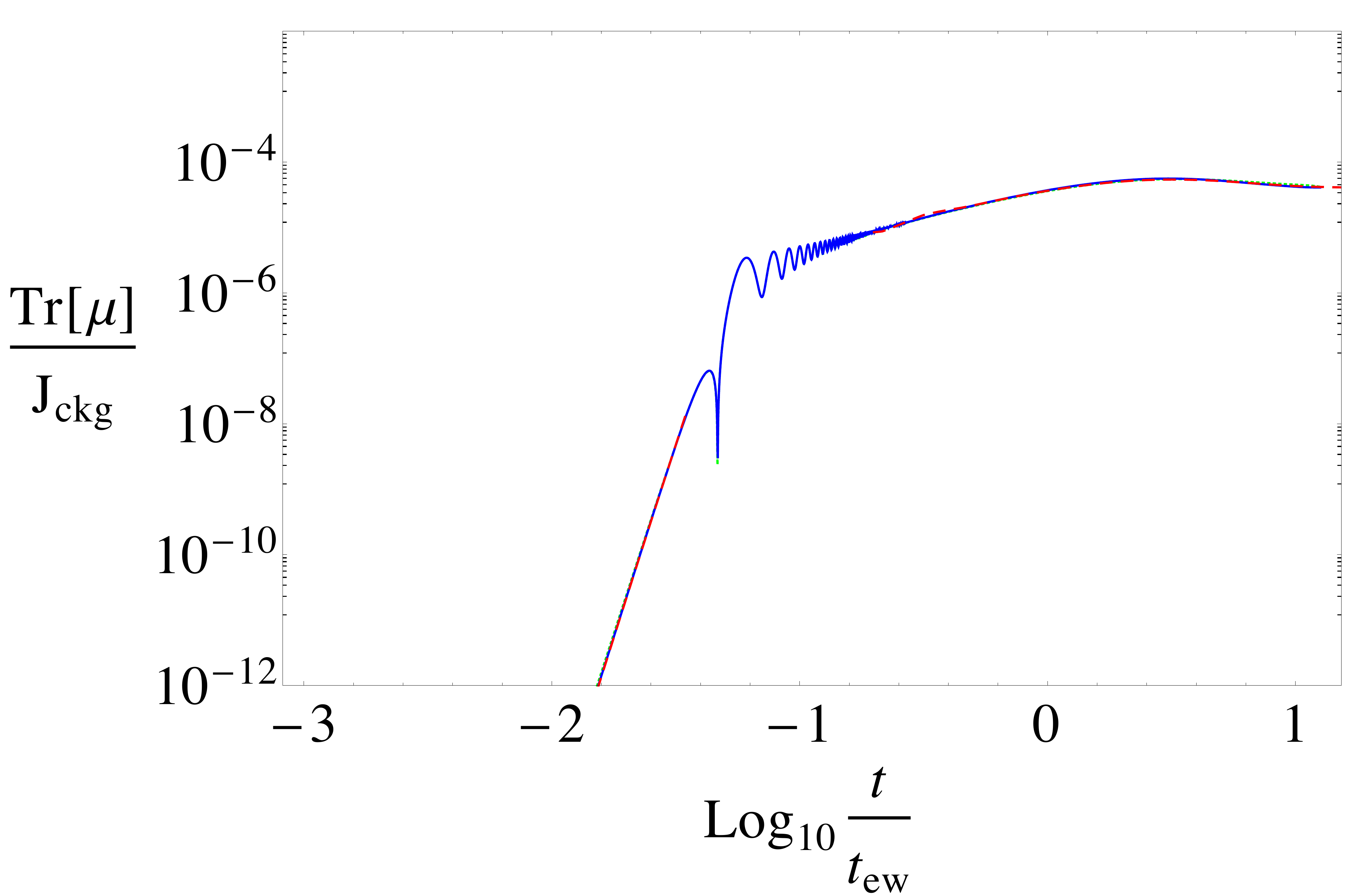}
\caption{\label{fig:AICPnum2} Left: full numerical solution (solid blue) and numerical solution neglecting non-linear terms (dotted green) for case 3, normalised to the invariant $I^{(3)}_2$,  compared with the prediction, $A_{I^{(3)}_2}(t)$ (dashed red). Right: same for case 4 normalised to the invariant $J_W$ compared to $A_{J_W}(t)$.  The parameters are the same as in figure~\protect\ref{fig:AICP2} for the double degenerate case. }
\end{center}
\end{figure}

In figures~\ref{fig:AICPnum1}-\ref{fig:AICPnum2} we compare the analytical and numerical solutions for the functions $A_{I_{CP}}(t)$ in the highly degenerate case (the values 
of the mixing angles are of ${\mathcal O}(10^{-2})$)
. 
In order to isolate the appropriate invariant we make the following choices:
\begin{itemize}
\item Case 1: $\theta_{i3} = \bar{\theta}_{i3}  = 0$ isolates $I_{1}^{(2)}$, 
\item Case 2: $\theta_{i3}= \bar{\theta}_{12} = 0$ isolates $I_{1}^{(3)}$, 
\item Case 3: $\theta_{12} = \bar{\theta}_{i3} = 0$ isolates $I_{2}^{(3)}$, 
\item Case 4: $\bar{\theta}_{ij} =0$ isolates $I_{J_W}$. 
\end{itemize}
The numerical results normalised by the corresponding CP invariant are shown together with the predictions of the previous section. In the case of $J_W$, we plotted the function $A_{J_W}$ keeping only the terms of ${\mathcal O}(y^6)$ that is more accurate at small $t$ and the full function at large $t$.  
 The agreement in all cases is quite good. The differences observed at large $t$ come from the non-linear terms in the equations. We also show the numerical results of the equations without them and find a very good agreement also at large $t$. Note that the approximation works well in the regime $\gamma t \gg 1$, that is in the strong washout regime of the fast modes.

Numerically it is very hard to go to regimes where the ratios $\gamma/|\Delta|^{1/3}$ become very  small, since the system becomes stiff. On the other hand, there is no reason 
why the perturbative solution is not accurate in such regime. We will therefore assume this to be the case in the following section and use the perturbative solution to perform 
a scan of parameter space.

\section{Baryon asymmetry}

The observed baryon asymmetry is usually quoted in terms of the abundance, which is the number-density asymmetry of baryons normalised by the entropy. After Planck  this quantity is known to per cent precision \cite{Ade:2013zuv}:
\begin{eqnarray}
Y_B^{\rm exp} \simeq 8.6(1) \times 10^{-11}.
\end{eqnarray}

The lepton asymmetries in the left-handed (LH) leptons generated in the production of the sterile neutrinos are efficiently transferred via sphaleron processes 
\cite{Kuzmin:1985mm} to the baryons.  The baryon asymmetry is given by
\begin{eqnarray}
Y_B = {28 \over 79} Y_{B-L}.
\end{eqnarray}
Since we have neglected spectator processes in the transport equations, the $B-L$ asymmetry is  related to the chemical potentials computed in the previous sections by the relation
\begin{eqnarray}
Y_{B-L} = -{90 \over \pi^4 g_*} {\rm Tr}[\mu],
\end{eqnarray}
where   $g_* = 106.75$  (which ignores the contribution to the entropy of the sterile states). Our estimate for the baryon asymmetry is therefore
\begin{eqnarray}
Y_B \simeq 3 \times 10^{-3} \left. {\rm Tr}[\mu(t)]\right|_{t_{EW}}.
\end{eqnarray}

We have performed a first scan of the full parameter space of the model. Given the theoretical uncertainties mentioned in different sections of the paper, we have considered as interesting the points that can explain the baryon asymmetry within a factor of  5. For this we have used the analytical solutions, even though 
in some regions of parameter space they will not be precise, since they are based on a perturbative expansion on the mixing angles of the matrices $V$ and $W$. We have 
considered however a few cases where the angles are not small and we find that the analytical solutions differ from the numerical ones only in some global numerical factor of a few
, but the time dependence is very similar. 

Even with an analytical expression the exploration of the large parameter space is a challenge. We have used the package Multinest \cite{Feroz:2007kg,Feroz:2008xx} to perform a scan on the Casas-Ibarra parameters \cite{Casas:2001sr}, where the Yukawa matrix is written as
\begin{eqnarray}
Y= -i U^*_{\rm PMNS} \sqrt{m_{\rm light}} R(z_{ij})^T \sqrt{M} {\sqrt{2}\over v}.
\end{eqnarray}
$m_{\rm light}$ is a diagonal matrix of the light neutrino masses and $R$ is a complex orthogonal matrix that depends on three complex angles $z_{ij}$. 
 We fix the light neutrino masses and mixings to the present best fit points 
in the global analysis of neutrino oscillation data of ref.~ \cite{Gonzalez-Garcia:2014bfa} and leave as free parameters: three complex angles, the three  phases of the PMNS matrix, the lightest neutrino mass as well
as the heavy Majorana masses that are allowed to vary in the range $M_i \in [0.1,100]$~GeV. In total thirteen free parameters. 

The scan searches for minima of the quantity $|\log_{10} |Y_B(t_{\rm EW})/Y^{\rm exp}_B||$ (in the range $\leq 1.5$) and the MultiNest algorithm is optimised to sample properly when there are several maxima. For the determination of $Y_B$ we use the analytical results of the previous sections, for which the CP invariants are computed directly from the matrix elements of the $V, W$ matrices that can be easily calculated by diagonalising the Yukawa mass matrix obtained in the Casas-Ibarra parametrization. Since the mechanism to work requires that at least one of the modes does not get to equilibrium before the electroweak phase transition we restrict the search to the range where one of the yukawa eigenvalues, $y_3$, is much smaller than the others and the following conditions are satisfied
\begin{eqnarray}
y_3 \leq 0.01 {\rm Min}[y_1,y_2],\;\;\; \sum_{i=1,2} \Gamma_i \left(|V_{i3}|^2+ |W_{i3}|^2\right) \leq 0.01 H_u(T_{\rm EW}).
\label{eq:cons}
\end{eqnarray}
 Furthermore, since the kinetic equations neglect lepton number violating effects  in the rates, we impose additionally the constraint
\begin{eqnarray}
\left({M_i\over T_{EW}}\right)^2 \Gamma_i \ll {H_u(T_{EW})}. 
\end{eqnarray}

We first consider a case where one of the sterile neutrinos is effectively decoupled from baryon number generation, that we can assume to be $N_3$. This can be achieved with the choice of parameters:
\begin{eqnarray}
m_{3(1)}=0, z_{i3} =0, ~R(z_{ij}) \rightarrow R(z_{ij}) (P) ,
\label{eq:dec}
\end{eqnarray}
for the IH(NH), where $P$ is the $123\rightarrow 312$ permutation matrix (only necessary for the NH). 
With this choice, only the terms corresponding to the CP invariants $I_1^{(2)}$ and $I_1^{(3)}$ contribute. 
This case is the one that has been considered in most previous works on the subject \cite{Asaka:2005pn,Shaposhnikov:2008pf,Canetti:2010aw,Asaka:2011wq,Shuve:2014zua}, where the number of parameters is reduced to six: only one complex angle, two PMNS CP phases and two Majorana neutrino masses are relevant.  

\begin{figure}
 \begin{center}
\includegraphics[scale=0.25]{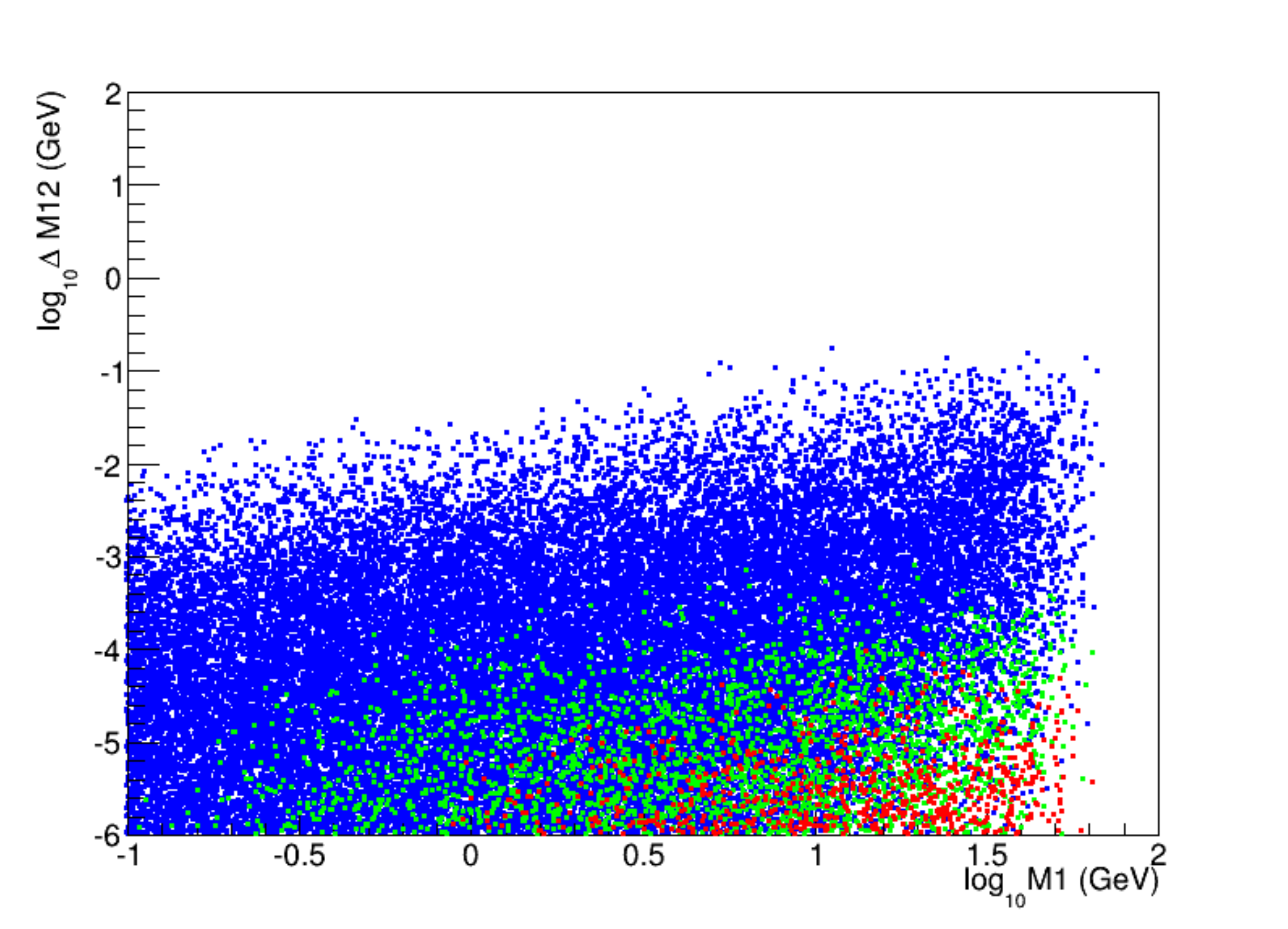} \includegraphics[scale=0.25]{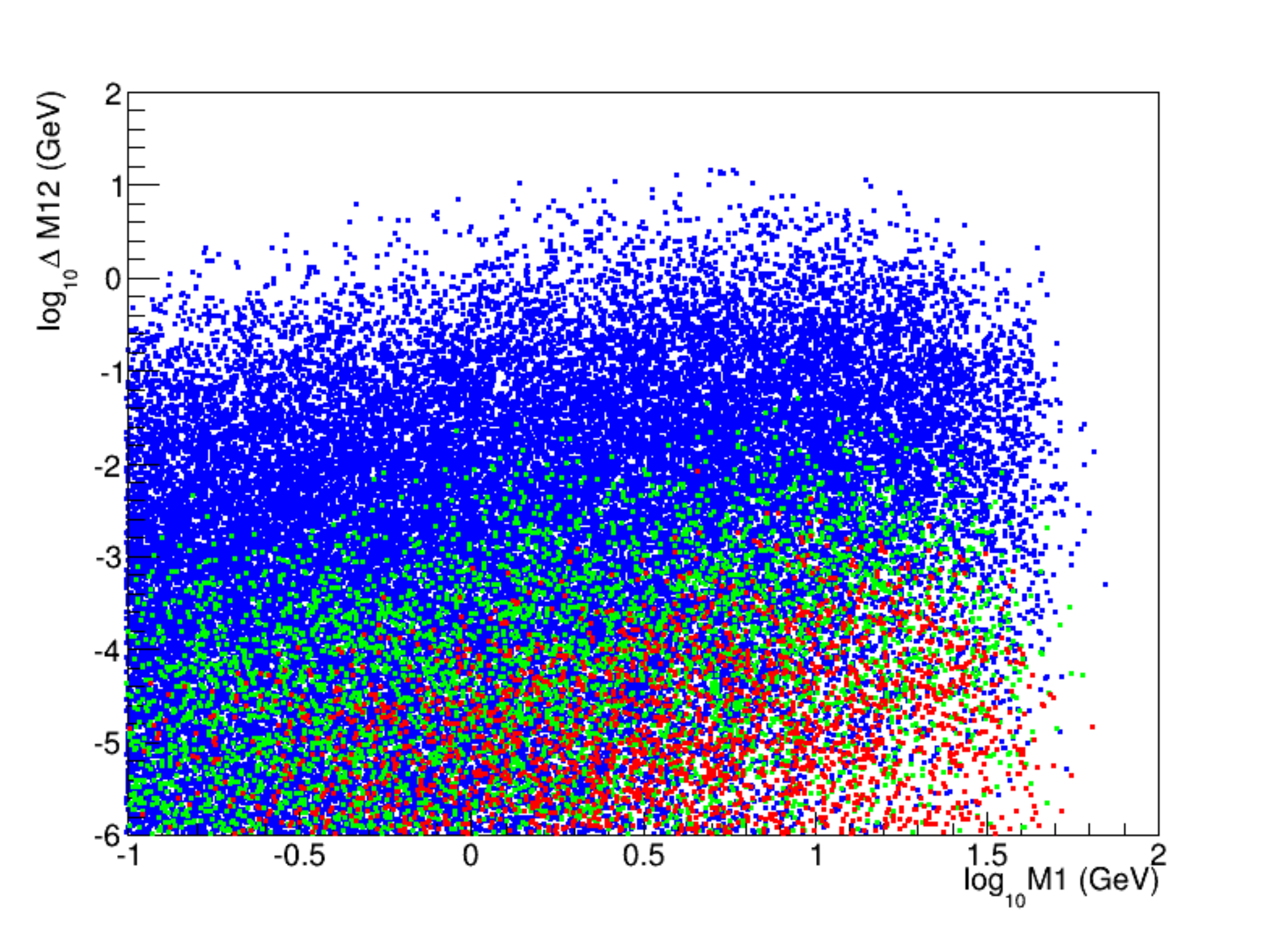} 
\caption{\label{fig:dmvsm_shapos} Points on the plane $\Delta M=M_2-M_1$ versus $M_1$ for which   $Y_B > 1/5 \times Y_B^{\rm exp}$ (blue),  $Y_B> Y_B^{\rm exp}$ (green) and $Y_B > 5 \times Y_B^{\rm exp}$ (red) for NH (left) and IH (right), with only two sterile neutrino species.}
\end{center}
\end{figure}

\begin{figure}
 \begin{center}
\includegraphics[scale=0.25]{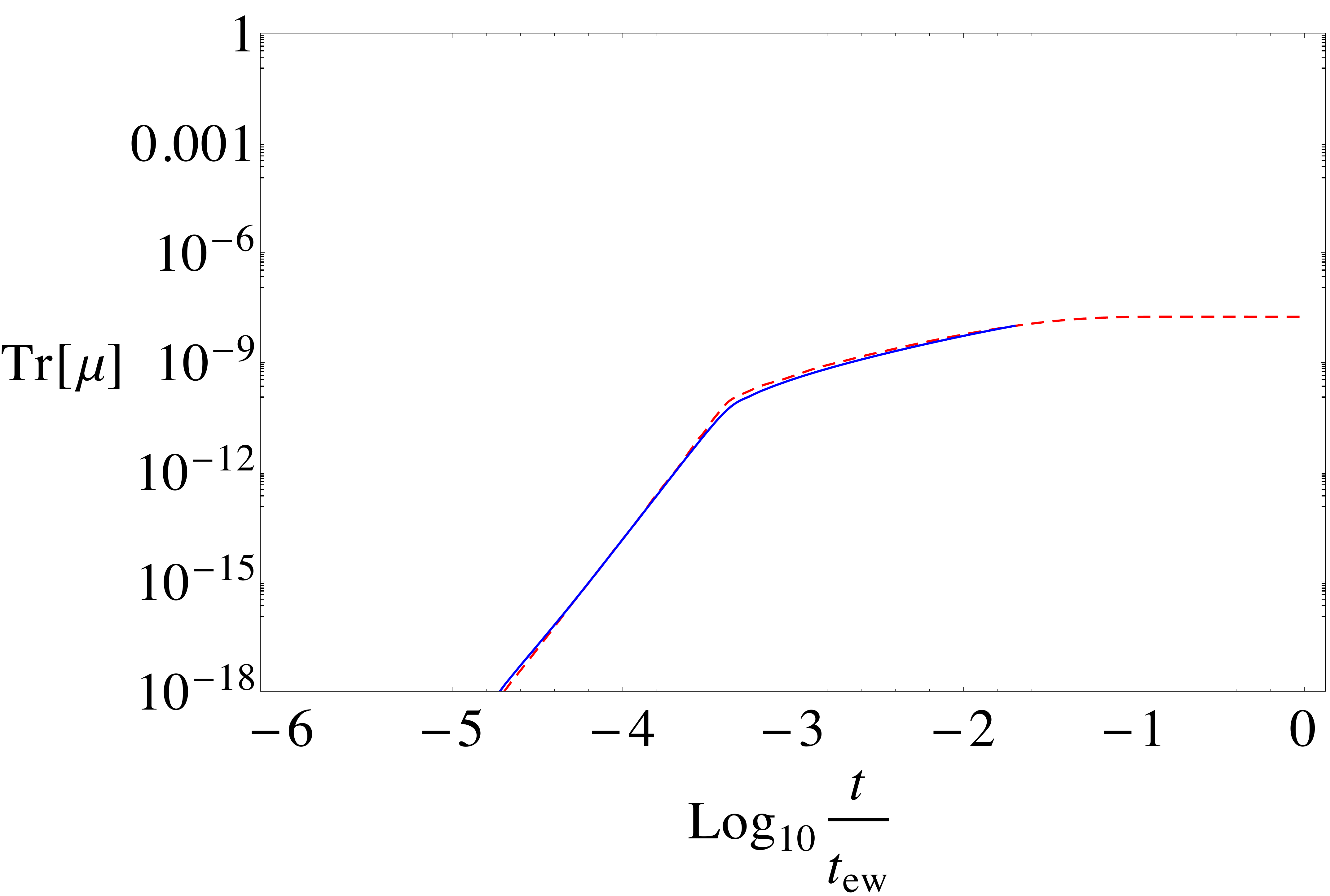} 
\caption{\label{fig:sample_shap} Comparison of the analytical (red-dashed) and numerical (blue-solid) solution for one of the points with mild degeneracy and $Y_B \geq Y_B^{\rm exp}$, corresponding to 
$\log_{10}(M_1(GeV)) = 0.9$ and $\log_{10}(\Delta M(GeV)) = -0.92$ and yukawa couplings $y_1=1.3 \times 10^{-6}, y_2 = 9.8 \times 10^{-9}$. }
\end{center}
\end{figure}

It is believed that  a large degeneracy of the two  sterile neutrinos is needed to obtain the correct baryon asymmetry.  
In figure~\ref{fig:dmvsm_shapos} we show the result of the scan under the conditions of eq.~(\ref{eq:dec}) on the plane $\Delta M_{12} \equiv M_2 -M_1$ versus $M_1$ for normal and inverted orderings of the light neutrinos. The  different colours correspond to values of $Y_B >   1/5,1, 5 \times Y_B^{\rm exp}$ (blue,green,red). Successful leptogenesis is possible in a larger range of parameter space for IH than for NH.  
 In the range shown our results agree reasonably well with those in ref.~\cite{Asaka:2013jfa} for the IH, while the range for NH looks a bit smaller. We see that there are a significant number of points for which the degeneracy is mild for the IH. We have analysed more carefully some of these points by solving the full numerical equations. We find that even though these points correspond to cases where the angles in $V, W$ are not small, the analytical and numerical solution agree very well and have the same $t$ dependence  as shown in figure~\ref{fig:sample_shap}. Note that the numerical solution is difficult at large times for non-degenerate solutions and the standard methods that we use fail. An optimised numerical method is needed to solve the stiffness problem and this will be studied elsewhere.  
It is very interesting to correlate the baryon asymmetry with observables that could be in principle measured such as the Dirac CP phase of the PMNS matrix, the 
amplitude of neutrinoless double beta decay or the active-sterile mixings that control the probability for the heavy sterile states to be observed in accelerators or in rare
decays of heavy mesons. The effective mass entering the $0\nu\beta\beta$ decay is given by
\begin{equation}
m_{\beta\beta}= \sum_{i=1}^3 U_{ei}^{2} m_i +
\sum_{i=1}^3 U_{e(i+3)}^2 M_{i} \frac{\mathcal{M}^{0\nu\beta\beta}(M_{i})}{\mathcal{M}^{0\nu\beta\beta}(0)},
\label{mbb}
\end{equation}
where  $\mathcal{M}^{0\nu\beta\beta}$ are the Nuclear Matrix Elements (NMEs) defined in \cite{Blennow:2010th}\footnote{The results for the NMEs computation in the interacting shell model~\cite{Caurier:2004gf,Caurier:2007wq} are available in Appendix A
of \cite{Blennow:2010th}}. The first term corresponds to the standard light neutrino contribution and the second is the contribution from the heavy
states. $U_{ei}$ with $i\geq4$ is the active-sterile neutrino mixing.

\begin{figure}
 \begin{center}
\includegraphics[scale=0.18]{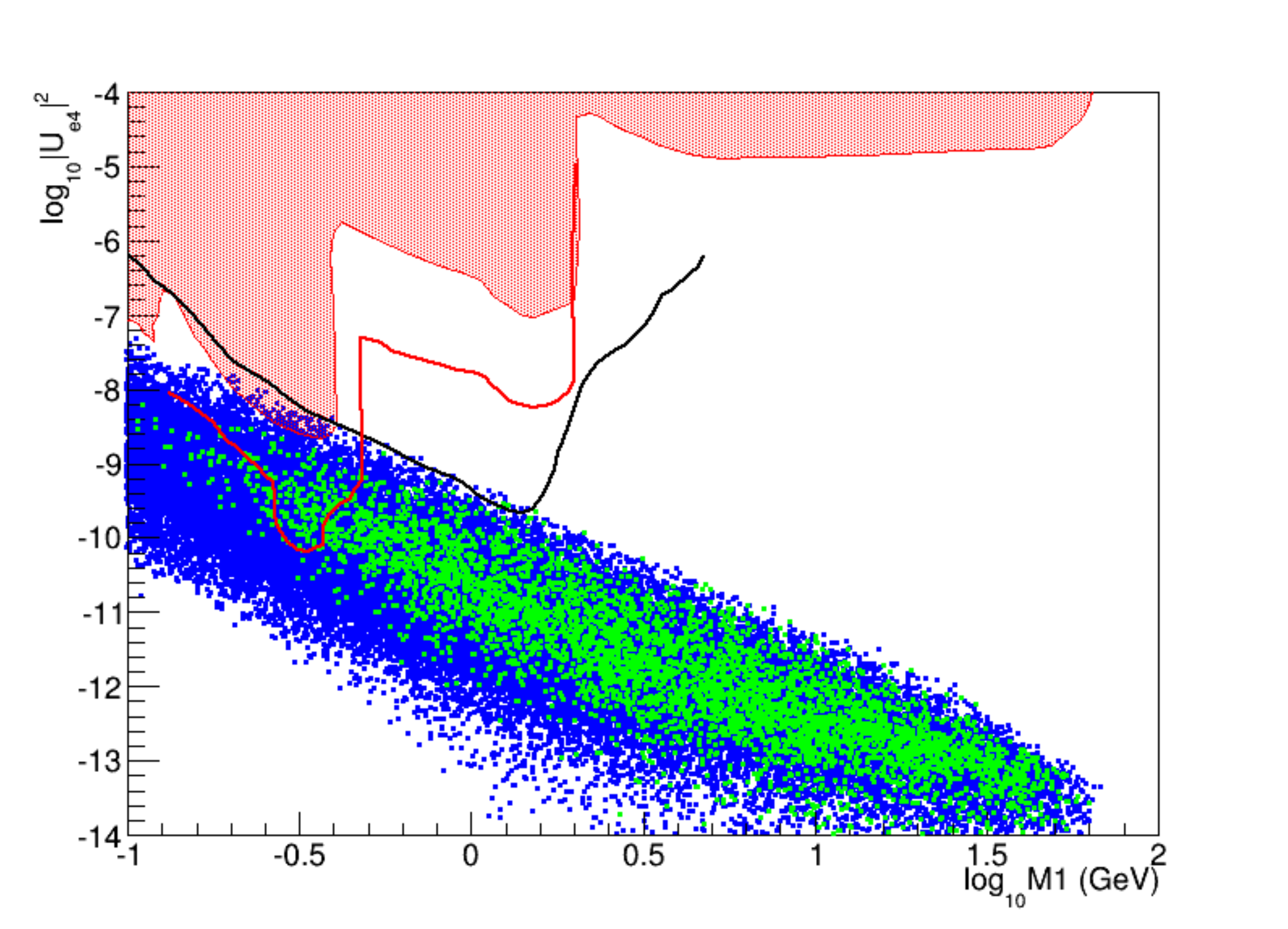}\includegraphics[scale=0.18]{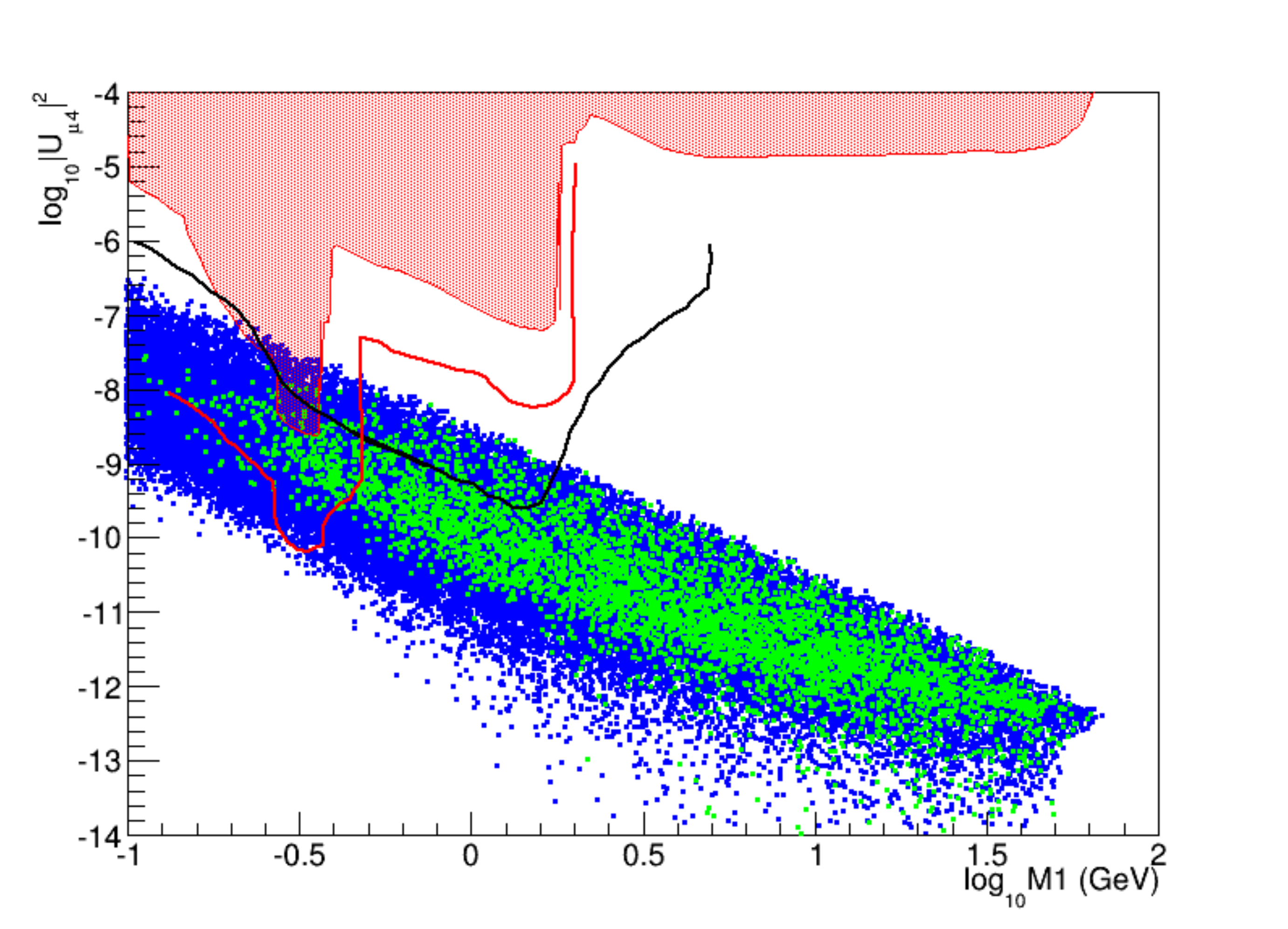}\includegraphics[scale=0.18]{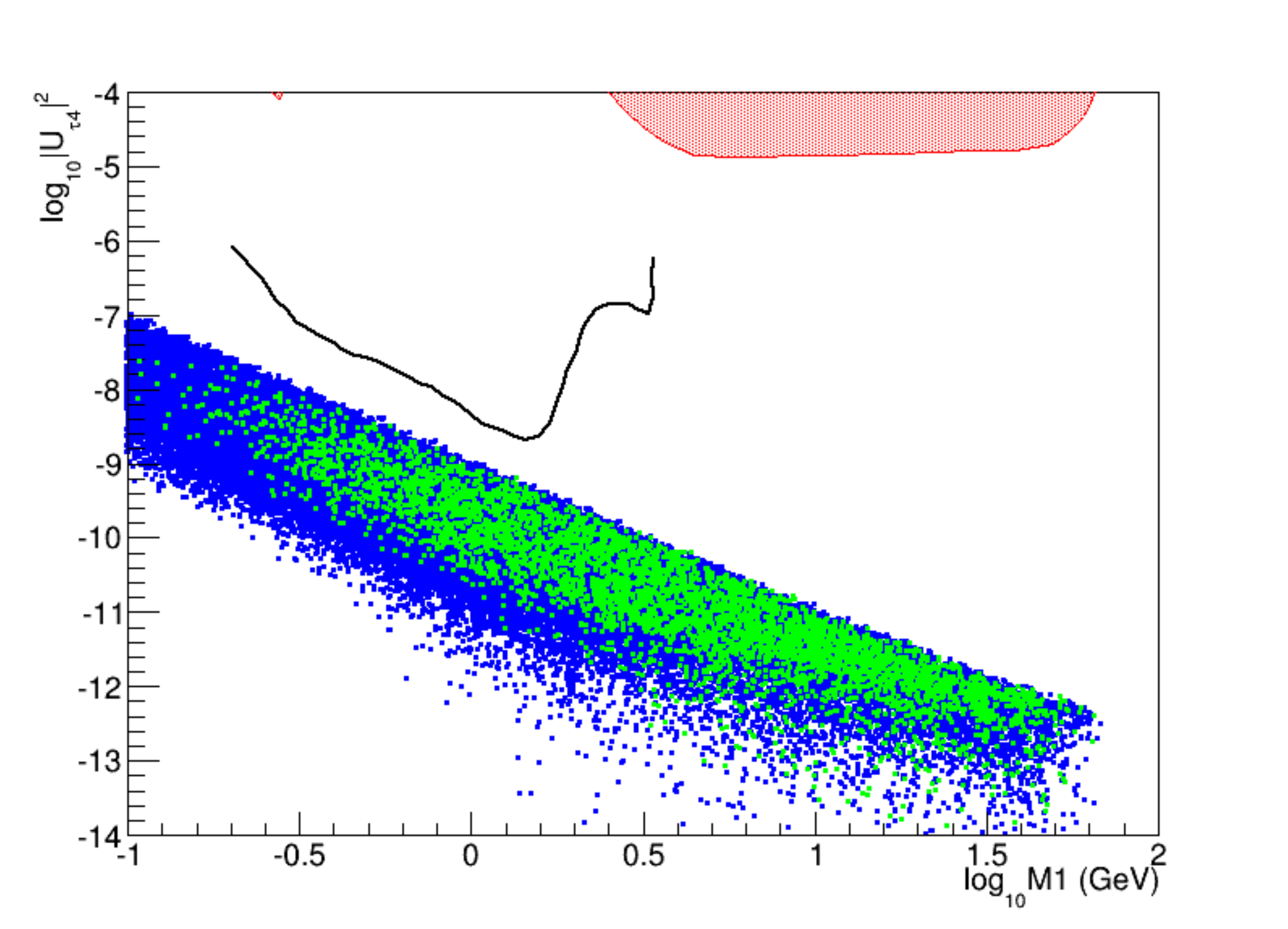}
\includegraphics[scale=0.18]{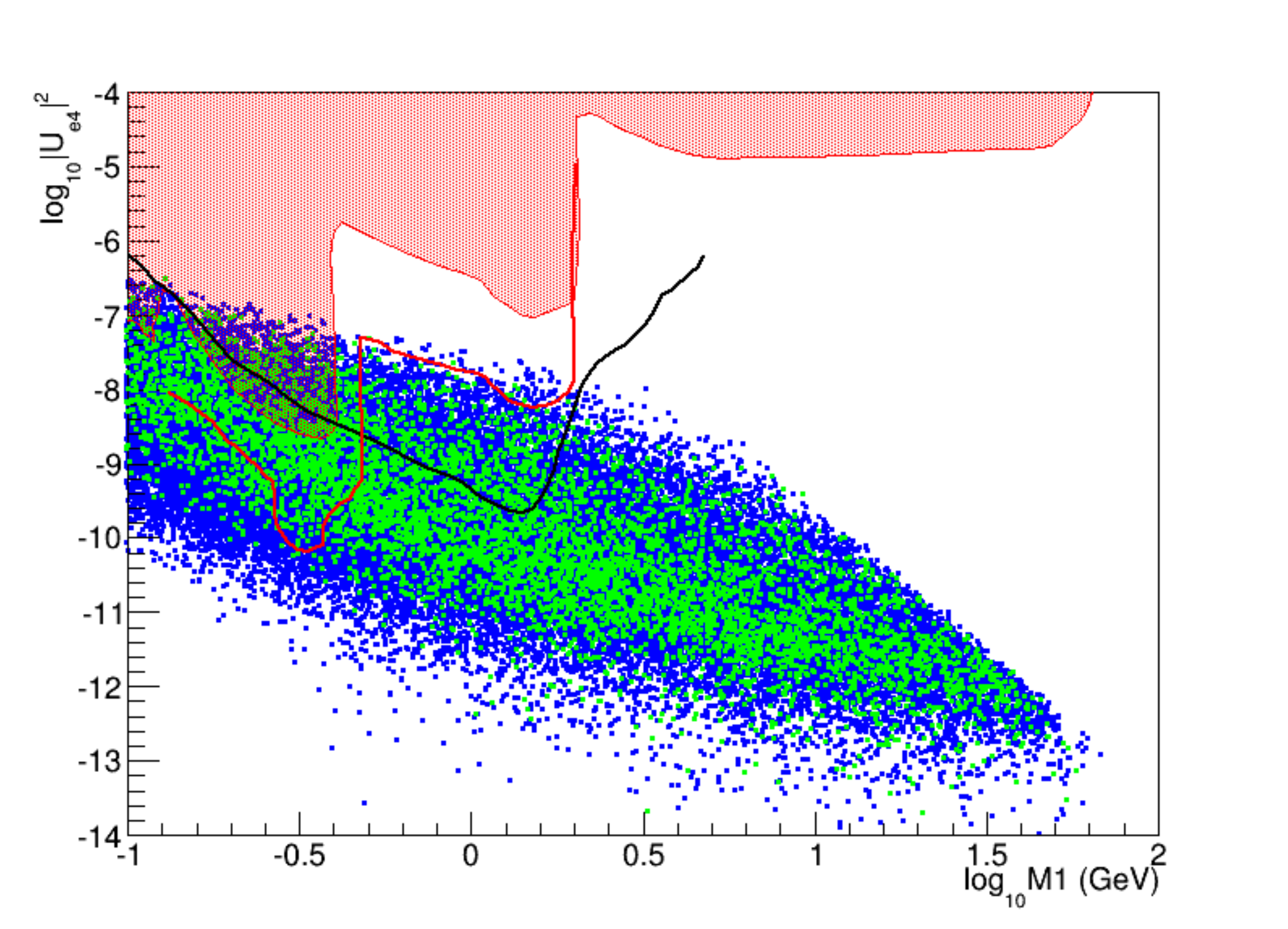}\includegraphics[scale=0.18]{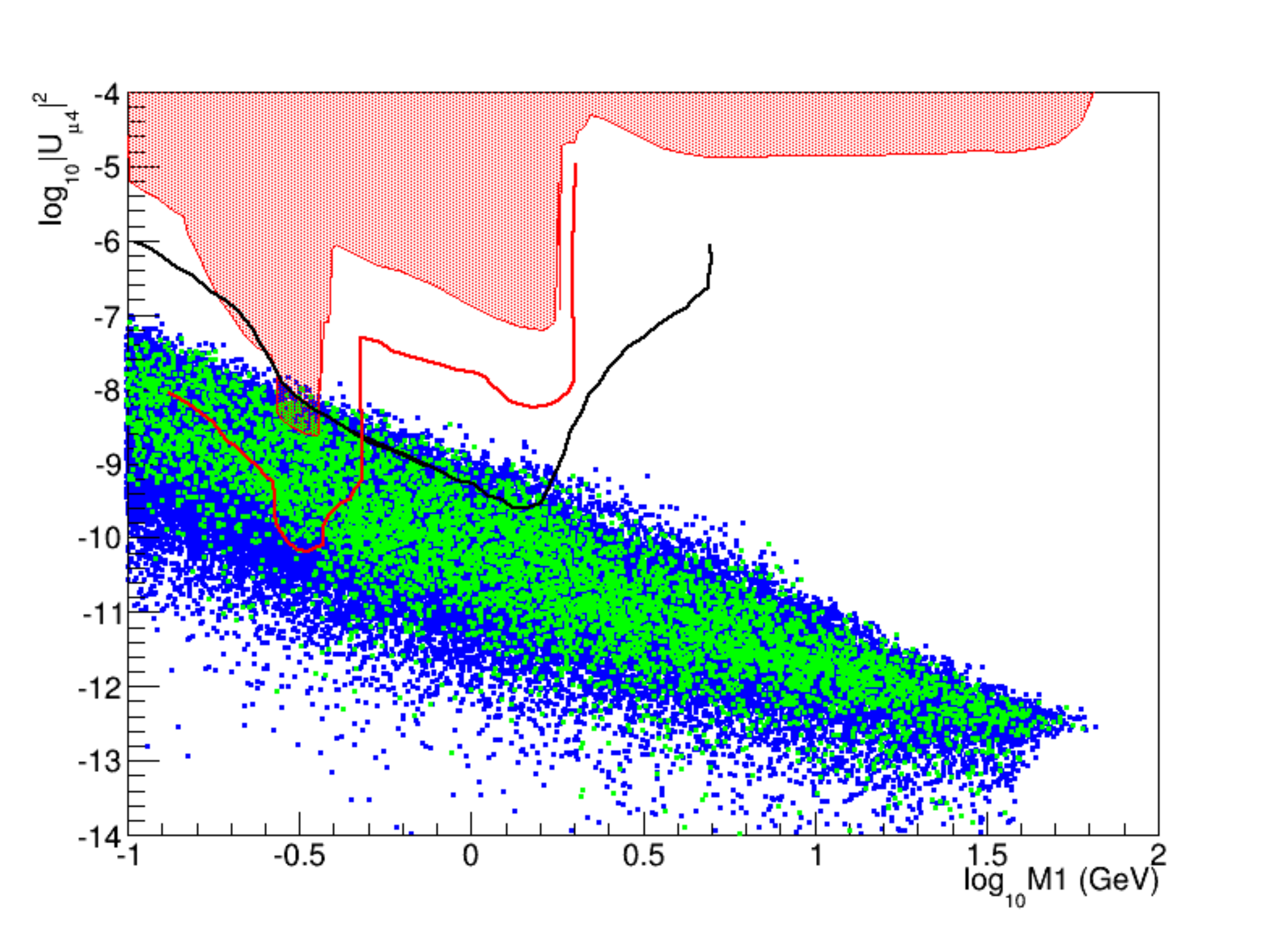}\includegraphics[scale=0.18]{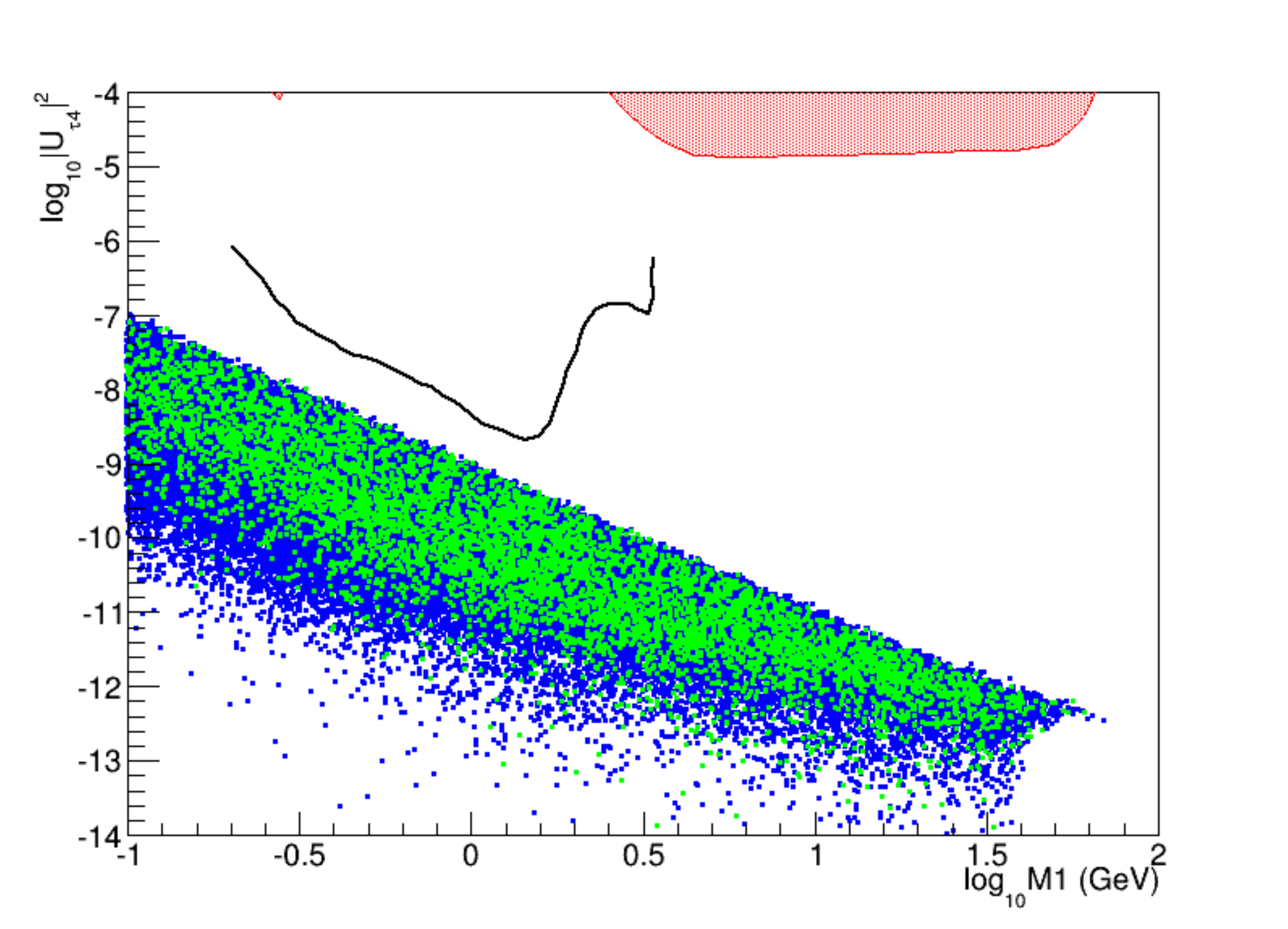}
\caption{\label{fig:uea_shapos} Points on the plane $|U_{e4}|^2$(left),  $|U_{\mu4}|^2$(middle), $|U_{\tau4}|^2$(right) versus $M_1$ for which $Y_B$  is in the range $[1/5-1]\times Y_B^{\rm exp}$ (blue) and $[1-5] \times Y_B^{\rm exp}$ (green)  for NH (up) and IH (down), with only two sterile neutrino species. The red bands are the present constraints \cite{Atre:2009rg}, the solid black line shows the reach of the SHiP experiment \cite{Alekhin:2015byh} and the solid red line is the reach of LBNE near detector \cite{Adams:2013qkq}. }
\end{center}
\end{figure}

\begin{figure}
 \begin{center}
 \includegraphics[scale=0.3]{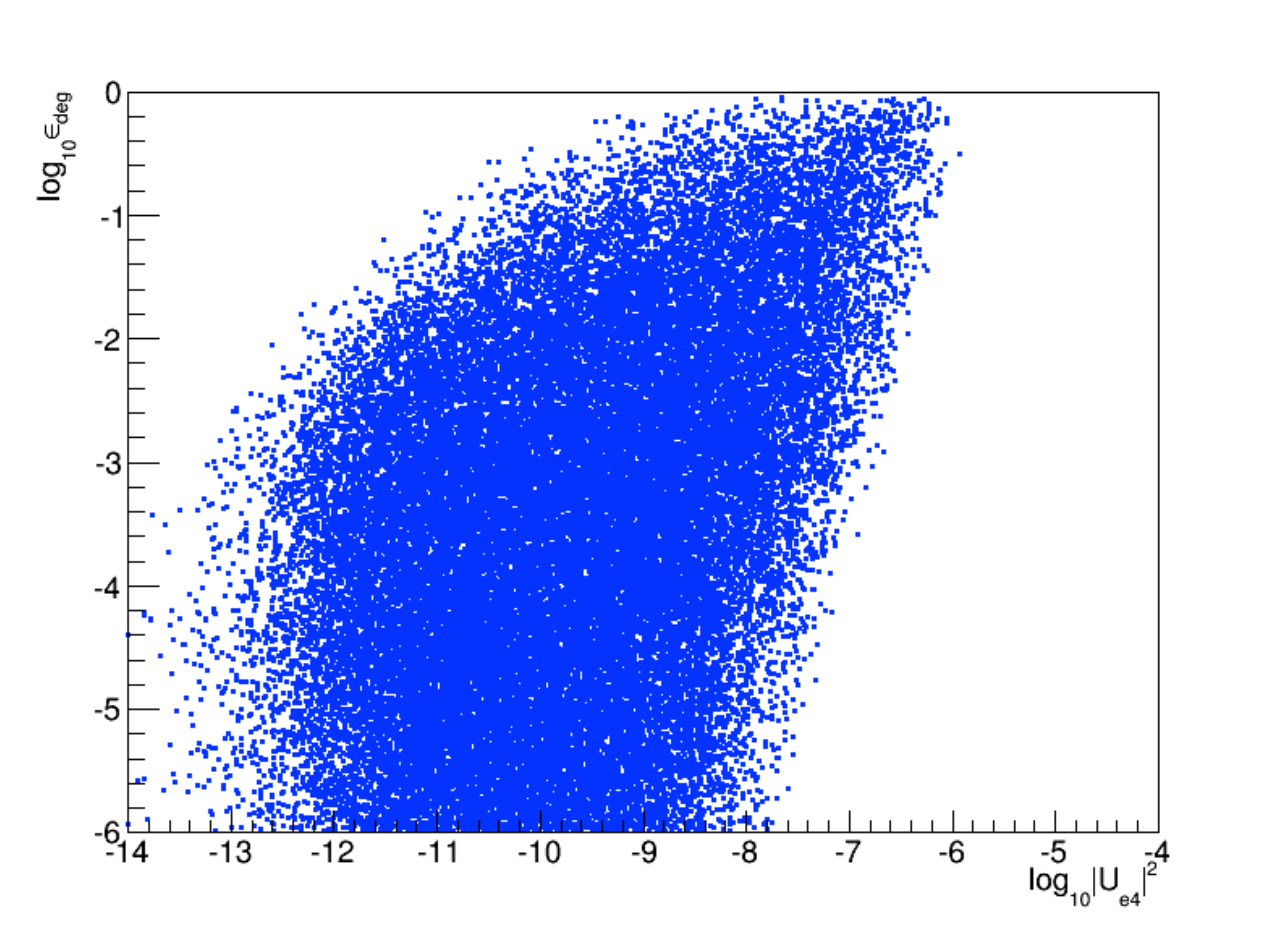} 
\caption{\label{fig:corrdegmix} Points on the plane $\epsilon_{deg}={|M_2-M_1|\over M_2+M_1}$ versus $|U_{e4}|^2$ for which the asymmetry is in the range $[1/5,5] \times Y_B^{\rm exp}$ in the range explored for  IH.}
\end{center}
\end{figure}

In figure~\ref{fig:uea_shapos} we show the results for the active-sterile mixing as function of the sterile mass and compare them with present direct bounds and the prospects of SHiP\cite{Alekhin:2015byh} and LBNE near detector \cite{Adams:2013qkq}. We show the result for $M_1$ but the one for $M_2$ is almost identical. 
We see that most of the parameter space for successfull baryogenesis is not excluded by present constraints and that the active-sterile mixings tend to be larger for the IH.  A sizeable region in the range of the GeV could be explored in the future experiment SHiP in the case of the IH and by LBNE near detectors. 
It is interesting to note that the less degenerate solutions 
 can not have very small active-sterile mixing, 
as shown in figure~\ref{fig:corrdegmix}, where we plot the points on the plane $\epsilon_{deg} \equiv |M_2-M_1|/(M_2+M_1)$ versus the active-sterile mixing in the electron flavour. 
The degeneracy can be lifted to some extent at the expense of larger yukawa couplings which also imply larger mixings. 

We have looked for direct correlations of the baryon asymmetry with the phases of the PMNS matrix. We have found that the distribution on the Dirac phase and the Majorana phase are flat. This is due to the fact that the complex angle can provide the necessary CP violation, even if the PMNS phases would vanish. The same is true for the effective mass of neutrinoless double beta decay, which depends on the Majorana phase. A dedicated scan is needed to quantify how the putative measurement of various observables could  constrain the lepton asymmetry. This will be done elsewhere. 

\begin{figure}
 \begin{center}
\includegraphics[scale=0.25]{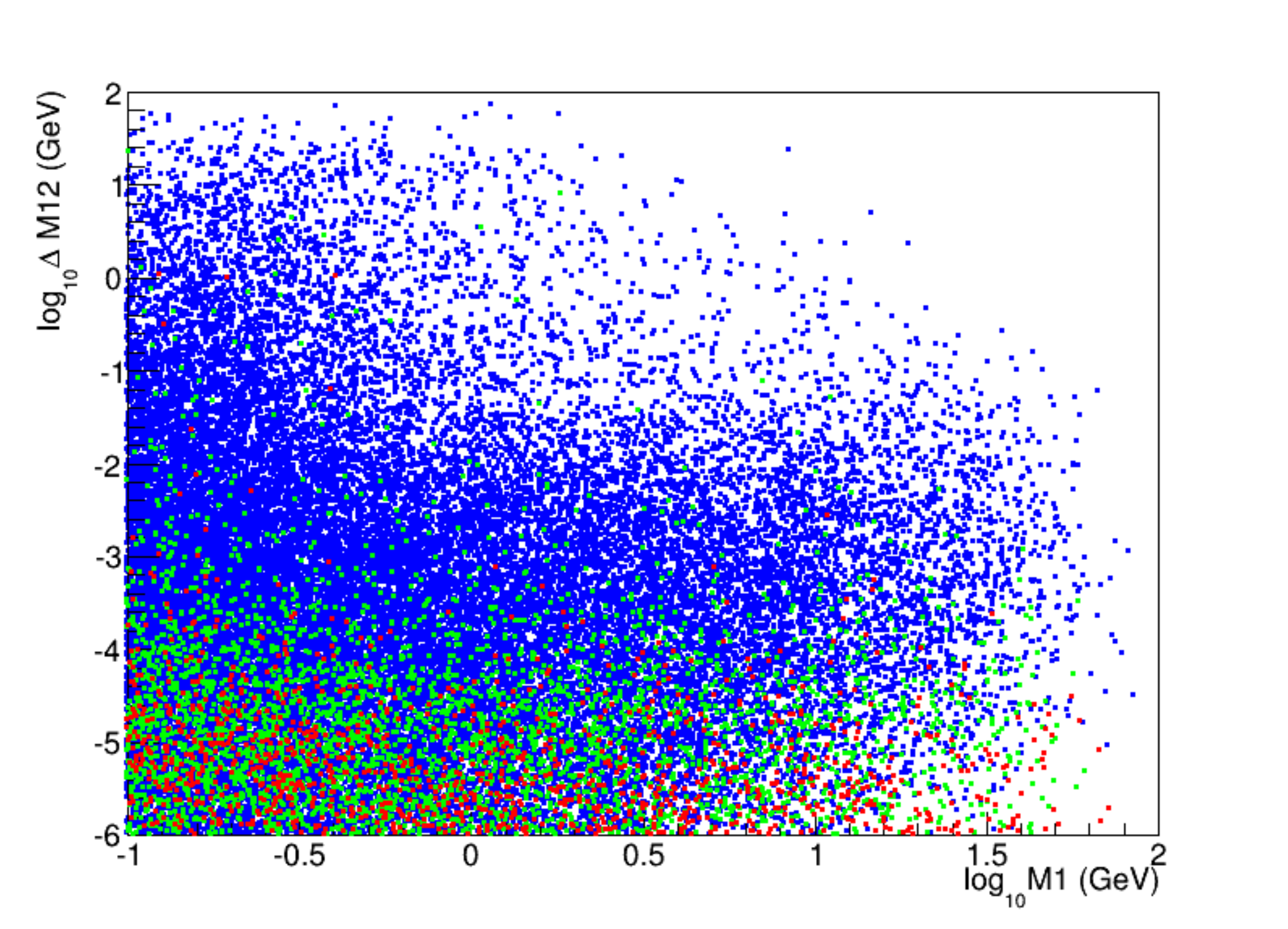} \includegraphics[scale=0.25]{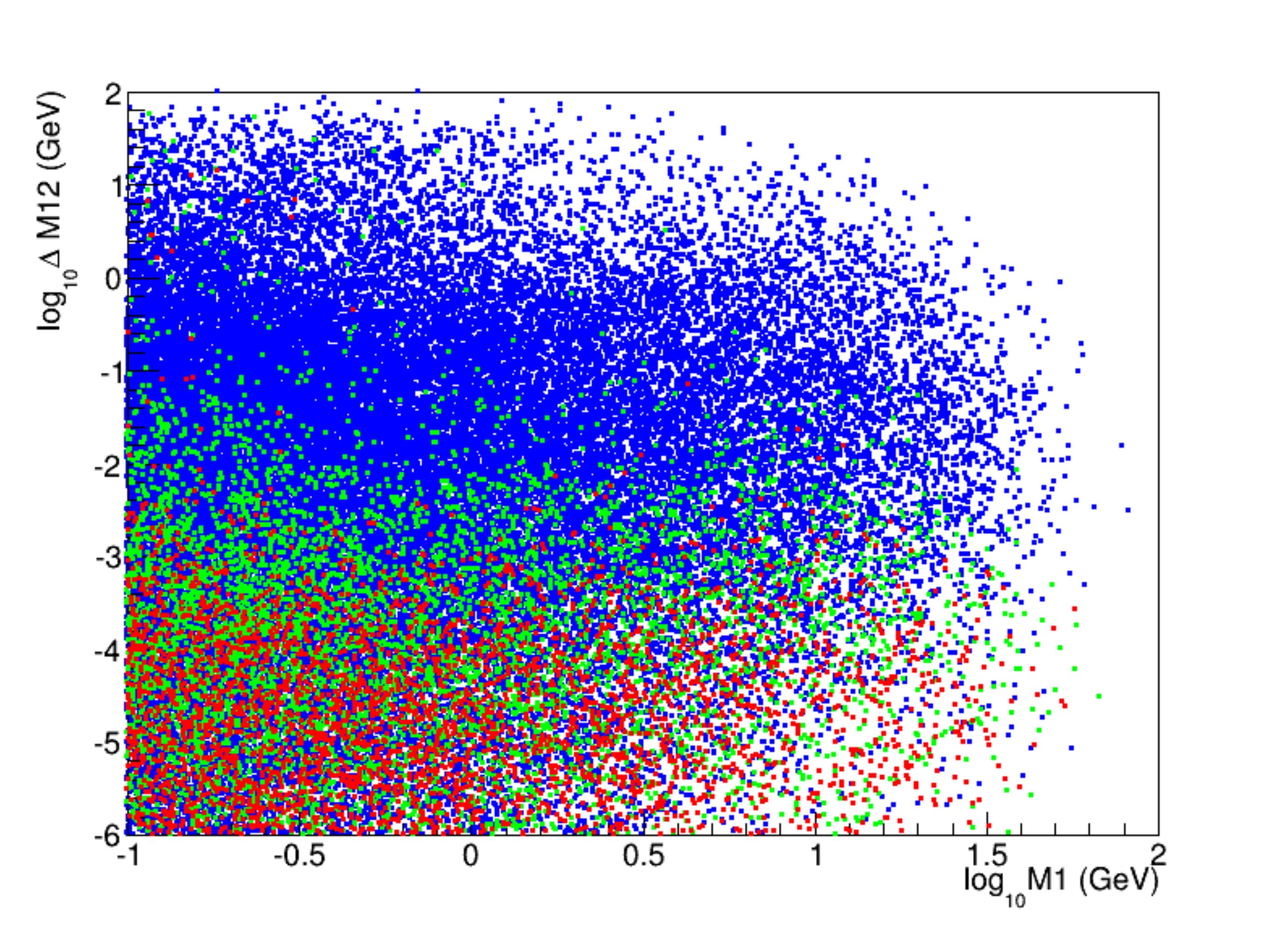} 
\caption{\label{fig:dmvsm} Points on the plane $\Delta M=M_2-M_1$ versus $M_1$ for which   $Y_B > 1/5 \times Y_B^{\rm exp}$ (blue),  $Y_B> Y_B^{\rm exp}$ (green) and $Y_B > 5 \times Y_B^{\rm exp}$ (red) for NH (left) and IH (right), in the general case with three neutrinos.}
\end{center}
\end{figure}

\begin{figure}
 \begin{center}
\includegraphics[scale=0.18]{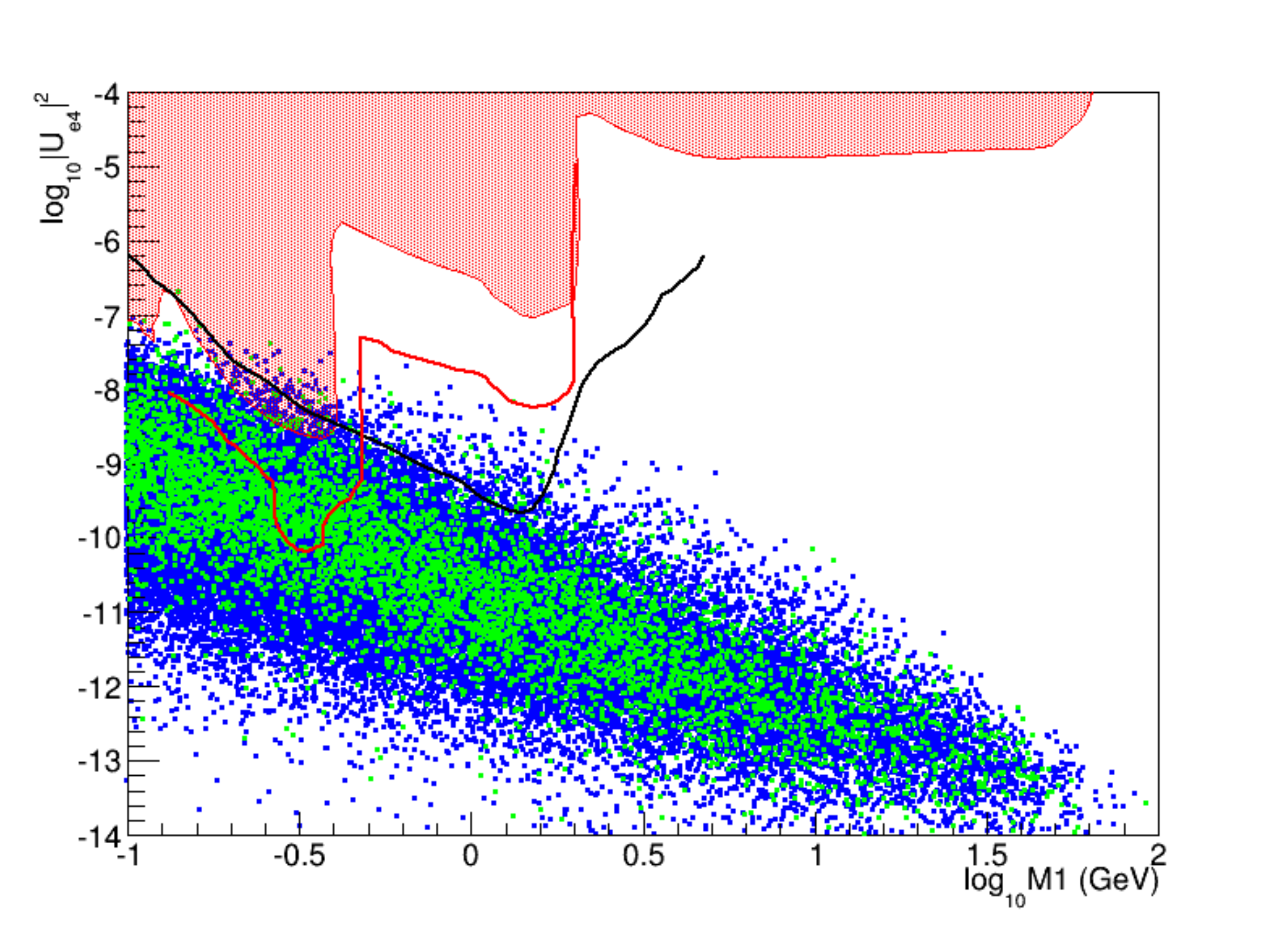}\includegraphics[scale=0.18]{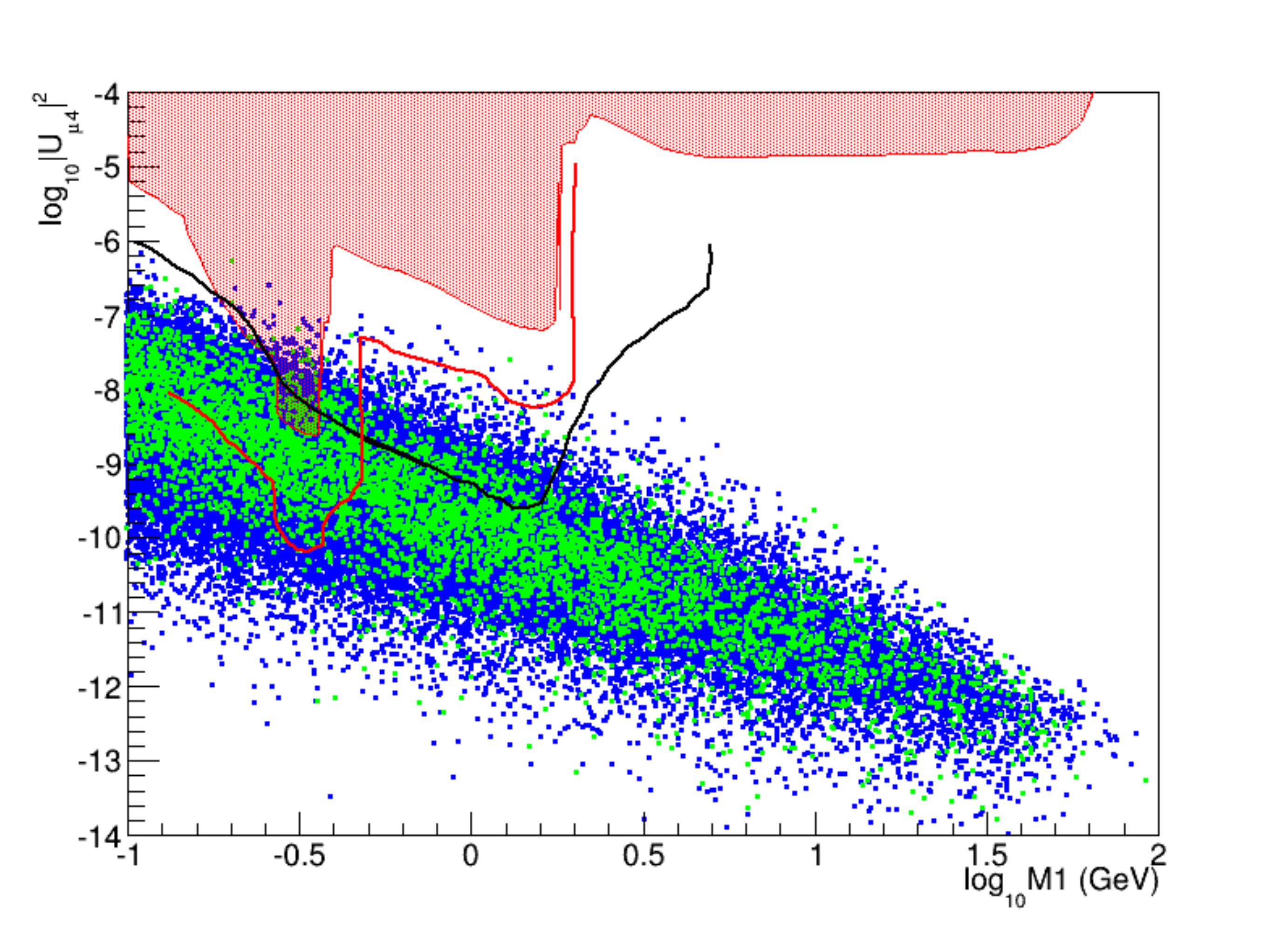}\includegraphics[scale=0.18]{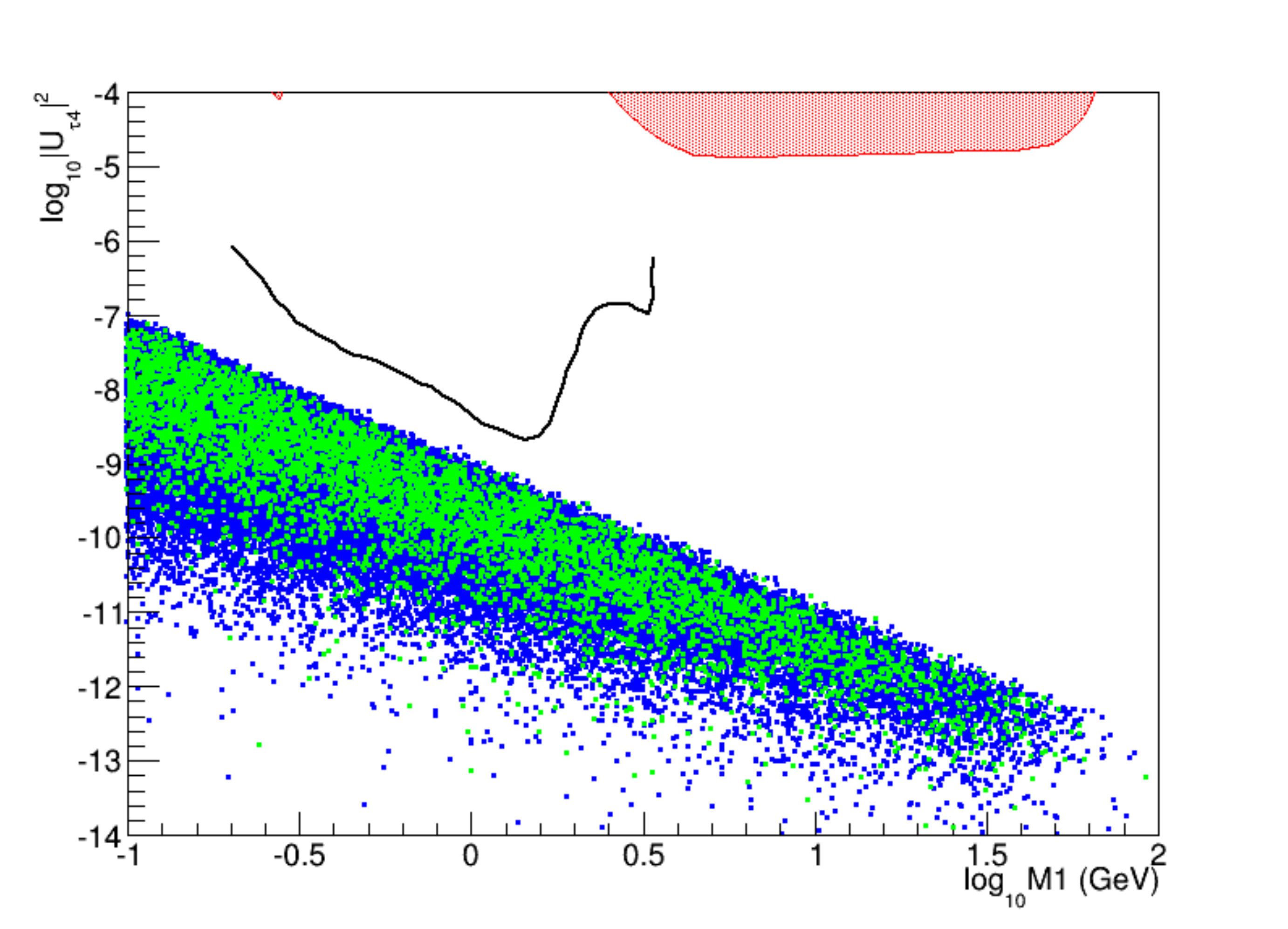}
\includegraphics[scale=0.18]{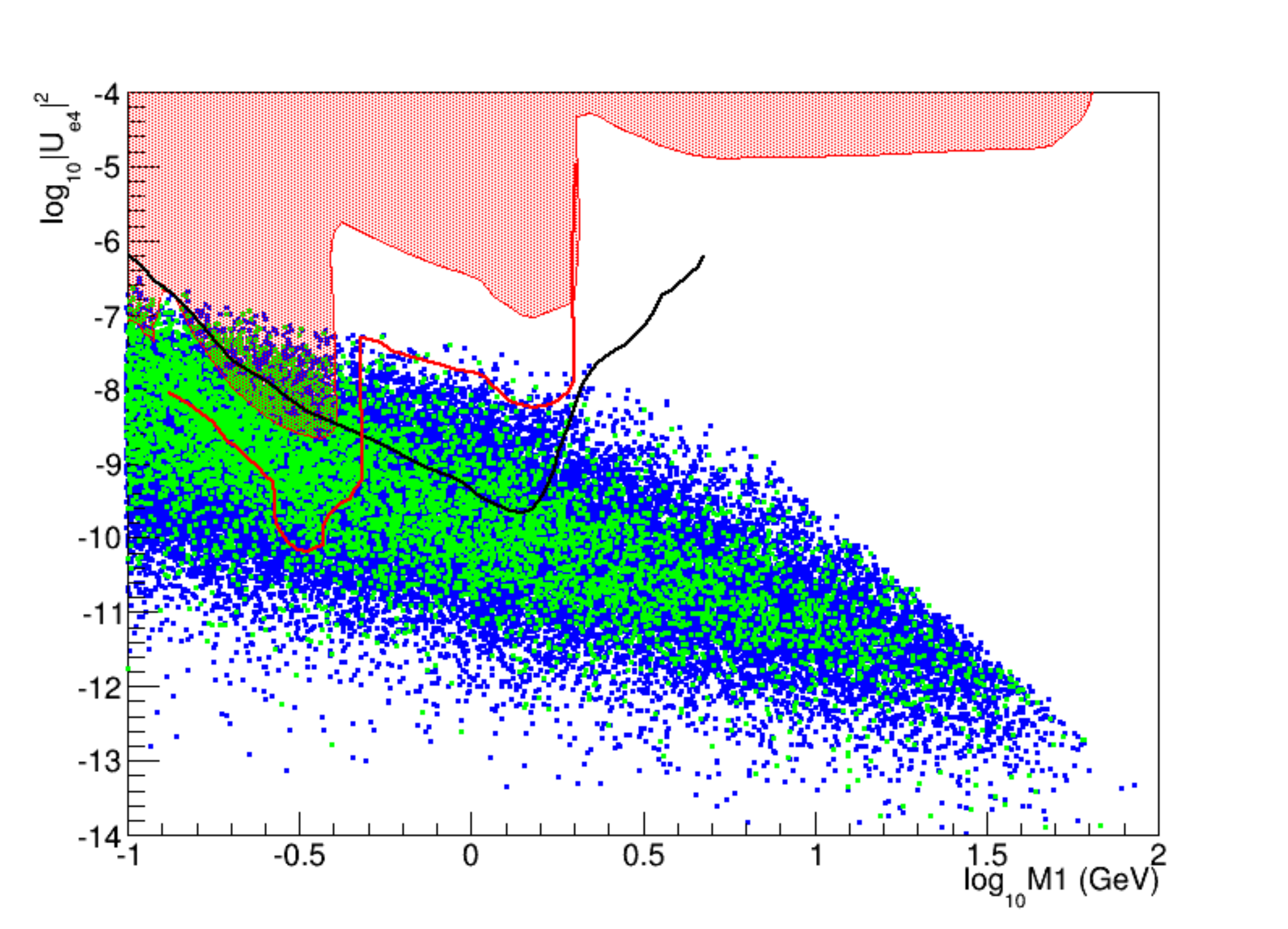}\includegraphics[scale=0.18]{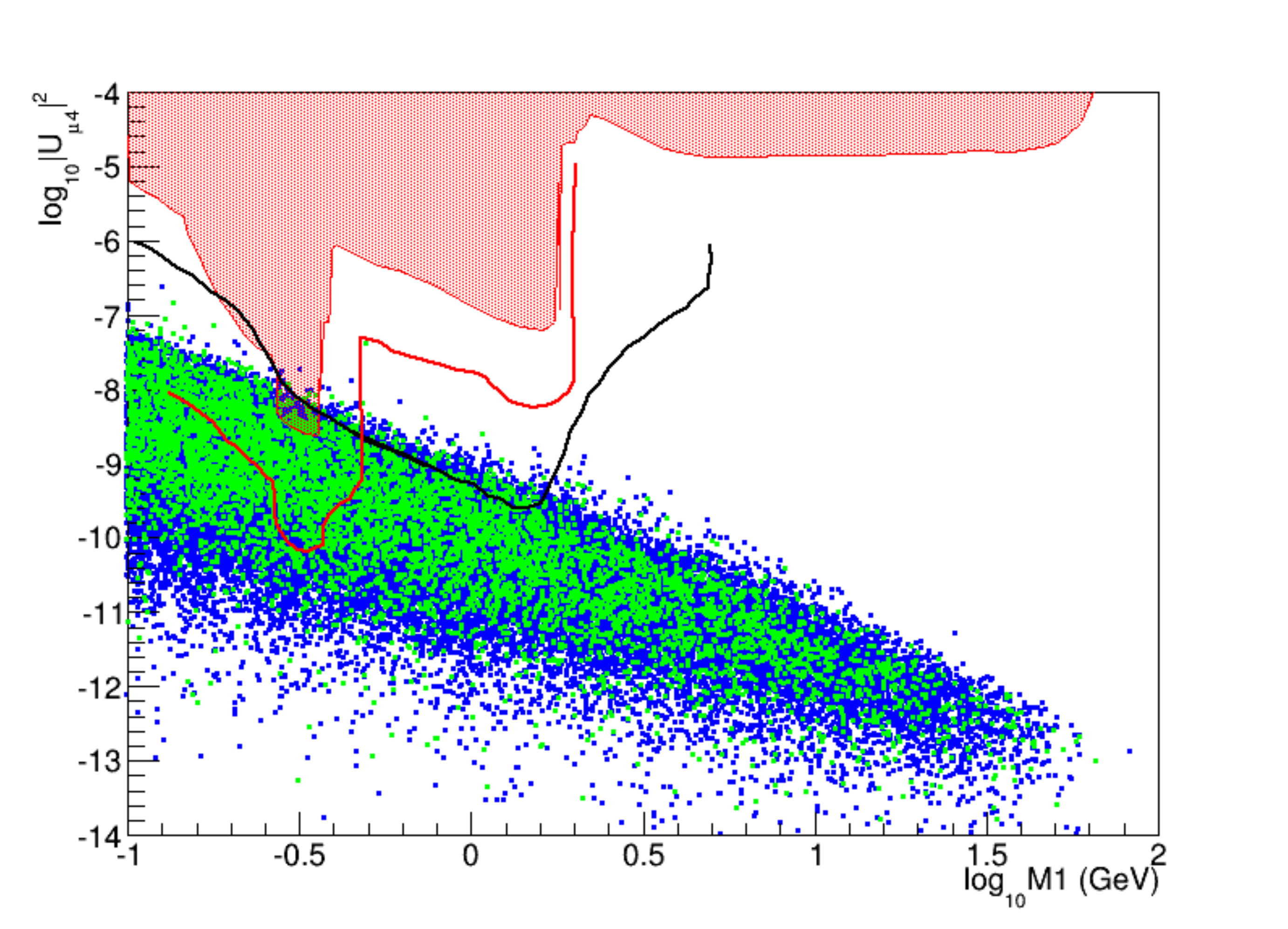}\includegraphics[scale=0.18]{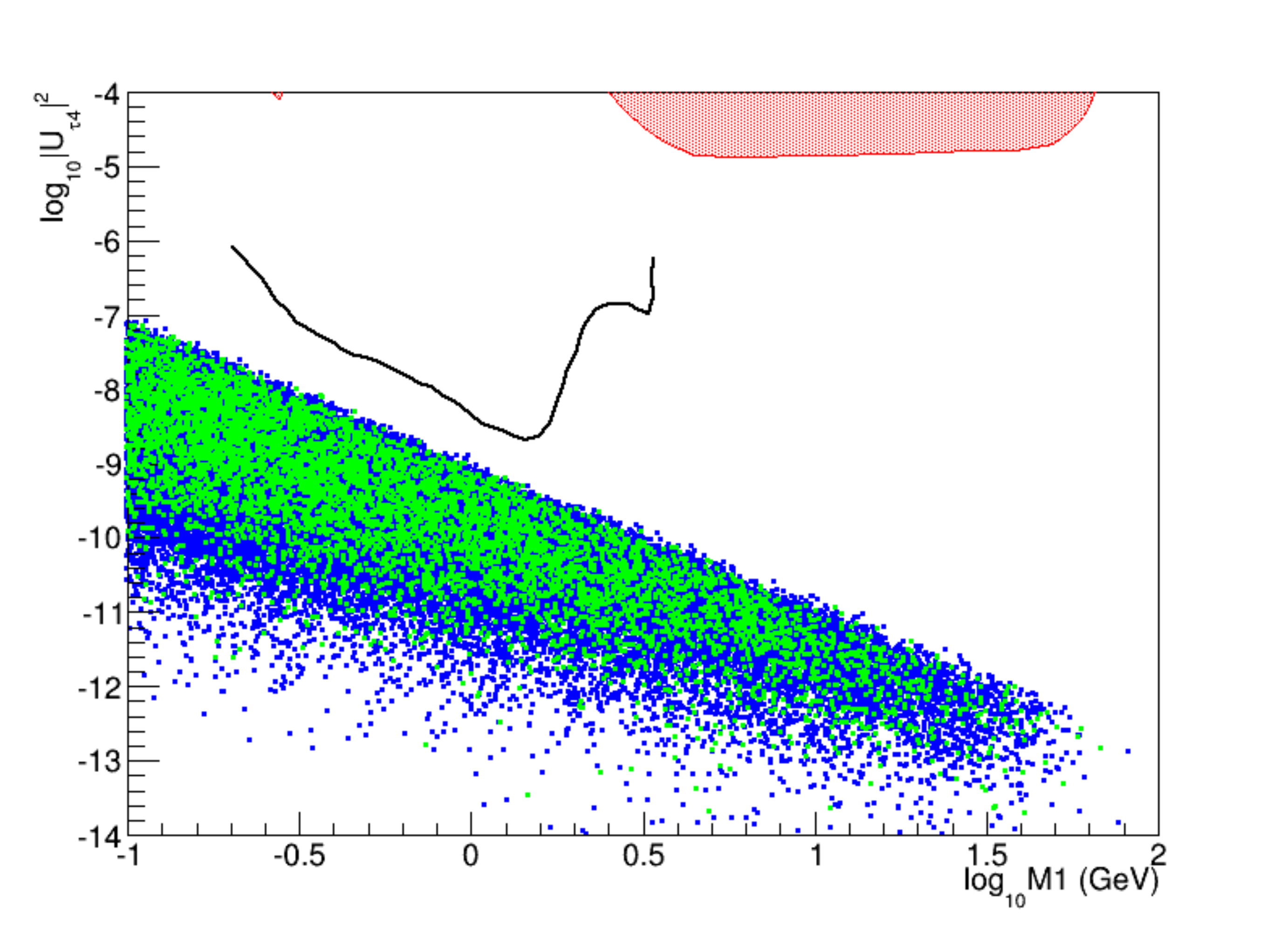}
\caption{\label{fig:uea_tot} Points on the plane $|U_{e4}|^2$(left),  $|U_{\mu4}|^2$(middle), $|U_{\tau4}|^2$(right) versus $M_1$ for which the $Y_B$ is in the range $[1/5-1]\times Y_B^{\rm exp}$ (blue) and $[1-5]\times Y_B^{\rm exp}$ (green) for  NH (up) and IH (down), with three sterile species. The red bands are the present constraints, the solid black line shows the reach of the SHiP experiment \cite{Alekhin:2015byh} and the solid red line is the reach of LBNE near detector \cite{Adams:2013qkq}. }
\end{center}
\end{figure}

In the general case, $N_3$ is also relevant and the main difference with respect to the previous 
situation is that there is a significantly enlarged parameter space where degeneracy is not necessary. 
This was already found in refs. ~\cite{Drewes:2012ma} for some points of parameter space.  
In figure~\ref{fig:dmvsm} we show the points on the plane $(\Delta M_{12}, M_1)$ for the general case. 
 The active-sterile mixings 
are shown in figure~\ref{fig:uea_tot}. These mixings can be larger in this case, specially in the case of the NH. 
The SHiP prospects are therefore more promising in this context. As in the $N=2$ case there is no direct connection between the asymmetry 
and the PMNS CP phases. On the other hand, the lightest neutrino mass is non-zero in this case, but the requirement that one yukawa needs to be
significantly smaller than the others, eq.~(\ref{eq:cons}), implies that the lightest neutrino mass must be small. In figure~\ref{fig:m1} we show the distribution of this quantity for those points that satisfy $Y_B \geq Y_{B}^{\rm exp}$ in the case of NH (the IH being very similar).  
\begin{figure}
 \begin{center}
\includegraphics[scale=0.45]{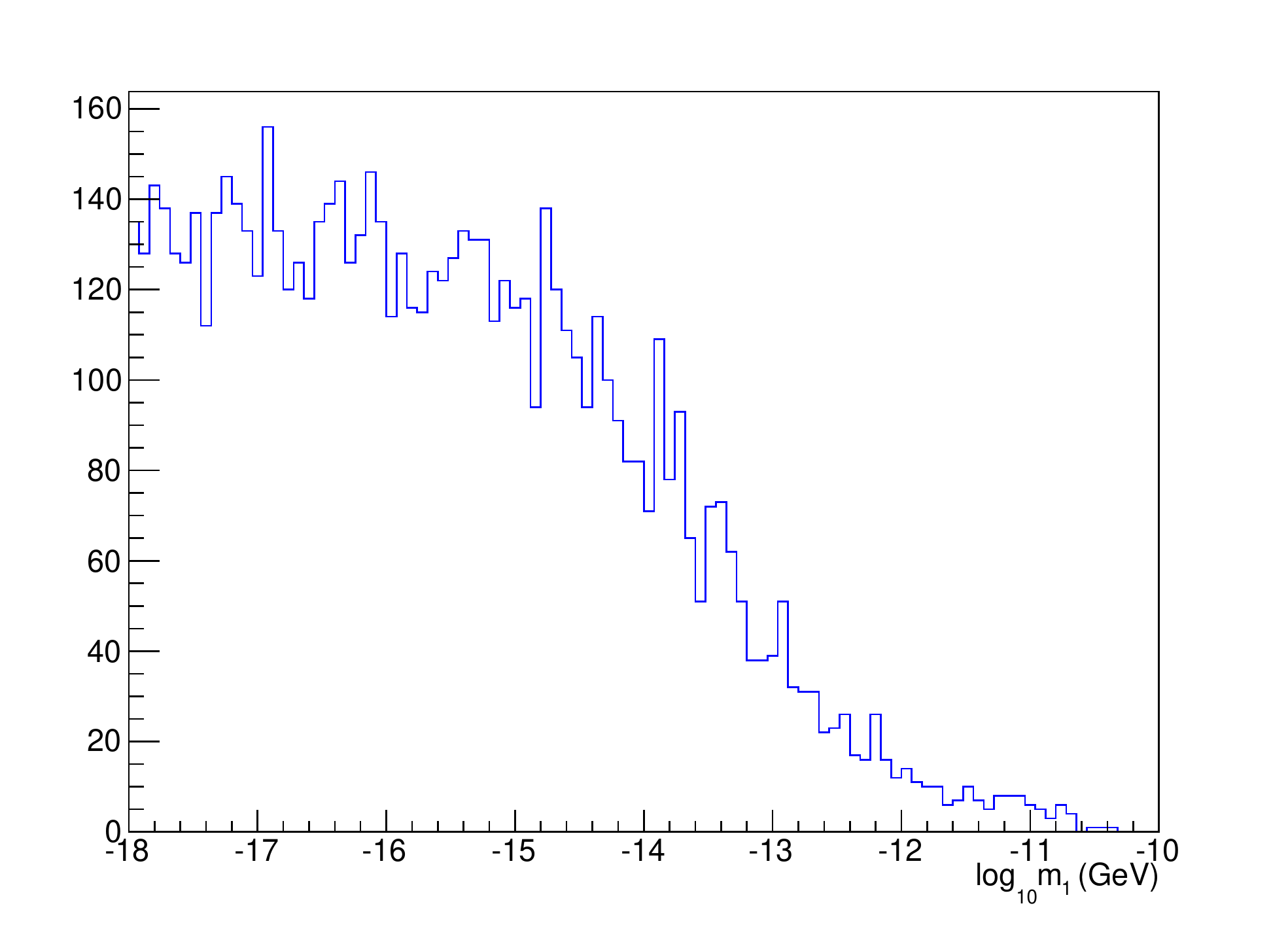} 
\caption{\label{fig:m1} Distribution of $m_1$ for points that satisfy $Y_B > Y_B^{\rm exp}$ for the NH. }
\label{fig:m1}
\end{center}
\end{figure}

\section{Conclusions}

We have studied the mechanism of leptogenesis in a low-scale seesaw model that is arguably the simplest extension of the Standard Model that can account for neutrino masses. 
For Majorana neutrino masses in the GeV range, sizeable lepton asymmetries can be generated in the production of these 
states some of which never reach thermal equilibrium before the electroweak phase transition. Lepton asymmetries are 
efficiently transferred to baryons via sphaleron processes. This mechanism  was proposed in \cite{Akhmedov:1998qx,Asaka:2005pn} and studied in many works, but a full exploration of parameter space in the general case of three neutrinos is lacking. To this aim we have developed an accurate 
analytical approximation to the quantum kinetic equations which works both in the weak and 
strong washout regimes of the fast modes (there is always a slow mode that does not reach thermal equilibrium before the EW phase transition). It relies on a perturbative 
expansion in the mixing angles of the two unitary matrices that diagonalise the Yukawa matrix. This analytical approximation 
allows us to identify the relevant CP invariants, and explore with confidence the regime of non-degenerate neutrino masses which is 
very challenging from the numerical point of view. We have used this analytical solution to scan the full parameter space using the MultiNest package to identify the regions where the baryon asymmetry is within an order of magnitude of the experimental value. We have performed first a scan in the simpler setting where one of the sterile neutrino decouples, which reduces the parameter space, and is the approximation that has been considered in most previous works on the subject, for example in the so-called $\nu$MSM. Although baryon asymmetries
tend to be larger in the case of highly degenerate neutrinos, we find solutions with a very mild degeneracy that also correlate with a larger active-sterile mixing. These non-degenerate solutions appear for an inverted ordering of the light neutrinos. On the other hand we do not observe a direct correlation with other observables, such as the PMNS CP phases nor the neutrinoless double beta decay amplitude.  

We have also performed a scan in the full parameter space, with the only requirement that one of the yukawa matrix eigenvalues is very small, and that one mode 
will not reach equilibrium before the electroweak transition, for the washout not to be complete. The main difference with the simpler case of two neutrinos is that the  parameter space with successfull baryogenesis is significantly enlarged, in particular as regards non-degenerate spectra. Also the active-sterile mixings can reach larger values, particularly in the normal hierarchy case, improving the chances of future experiments such as SHiP or LBNE to find the GeV sterile neutrinos. There is much less difference in this case between normal and inverted neutrino orderings and also no direct correlation with the PMNS phases. On the other hand, the requirement of a small yukawa eigenvalue implies that the lightest neutrino mass cannot be large.

A number of refinements are needed to improve the precision of the determination of the baryon asymmetry. First a more precise determination of the scattering rates of the sterile neutrinos is required. Most previous studies, and this one, have included only top-quark scatterings, but it has been pointed out recently that gauge scatterings are also very important. A correct treatment of these processes in the kinetic equations is necessary. Also the kinetic equations neglect effects of ${\mathcal O}((M_i/T)^2)$. Such effects are not so small for masses in the GeV near the electroweak phase transition and their effect should be quantified. Finally,  spectator processes and the asymmetries of fields other than the sterile neutrinos and LH leptons have not been taken into account in the kinetic equations.
A proper treatment could easily bring corrections of ${\mathcal O}(1)$. Finally, a more ambitious scan of parameter space should define more accurately the limits of eq.~(\ref{eq:cons}) for successfull baryogenesis. These effects will be studied in the future.

 \begin{acknowledgments}
We wish to thank  R.~ Ruiz de Austri, J.~ Mart\'{\i}n-Albo and J.M.~Mart\'{\i} for their help with the used software, 
and D. Bodeker, V.~Domcke, M.~Drewes and M.~Laine for useful discussions.  
This work was partially supported by grants FPA2011-29678, FPA2014-57816-P, PROMETEOII/2014/050, CUP (CSD2008-00037), 
ITN INVISIBLES (Marie Curie Actions, PITN-GA-2011-289442), the INFN program on Theoretical Astroparticle Physics (TASP) 
and the grant  2012CPPYP7 ({\it  Theoretical Astroparticle Physics}) under the program  PRIN 2012 funded by 
the Italian Ministry of Education, University and Research (MIUR). PH acknowledges the support of the Aspen Center
for Physics (National Science Foundation grant PHY-1066293), where this work was completed. MK thanks Fermilab for hosting her while part of this work was done.
\end{acknowledgments}

\appendix
   \section {Results for the perturbative integrals}
  
  \subsection {One dimensional integrals}
  We just need them up to ${\mathcal O}(\beta/\alpha)^2$:
\begin{eqnarray}
J_1(\alpha_1,\beta_1,t) &\simeq& J_{10}(\alpha_1,t_0) + \beta_1 J_{11}(\alpha_1,t_0) + {\beta_1^2 \over 2} J_{12}(\alpha_1,t_0)+\Delta J_1(\alpha_1,\beta_1,t,t_0) ,\nonumber\\
\end{eqnarray}  
with 
\begin{eqnarray}
\Delta J_1(\alpha_1,\beta_1,t,t_0) &=& \sum_n {\beta_1^n\over n!} J_{1n}(\alpha_1, t,t_0) \simeq i \sum_n {\beta_1^n \over n!} \left( {e^{i {\alpha_1 t_0^3\over 3} }\over\alpha_1 t_0^{2-n}}- {e^{i {\alpha_1 t^3 \over 3}} \over \alpha_1 t^{2-n}}\right)+ {\mathcal O}(t^{-4}, t_0^{-4}) \nonumber\\
&=& i \left({e^{i {\alpha_1 t_0^3\over 3}  + \beta_1 t_0}\over \alpha_1 t_0^2}-  {e^{i {\alpha_1 t^3 \over 3} + \beta_1 t}\over \alpha_1 t^2}\right).
\end{eqnarray}
We can factor out the $\alpha$ dependence and define:
\begin{eqnarray}
J_{10}(\alpha,t) = {1 \over |\alpha|^{1/3}} \Big( {\rm Re}\big[J_{10}(1,t |\alpha|^{1/3})\big] + i {\rm sign}(\alpha) {\rm Im}\big[J_{10}(1,t |\alpha|^{1/3})\big] \Big).
\end{eqnarray}
 \subsection {Two dimensional integrals}
   We just need them up to ${\mathcal O}(\beta/\alpha)$:
\begin{eqnarray}
J_{2}(\alpha_1,\beta_1,\alpha_2,\beta_2,t) &\simeq& J_{200}(\alpha_1,\alpha_2,t_0) + \beta_1 J_{210}(\alpha_1,\alpha_2,t_0) +\beta_2 J_{201}(\alpha_1,\alpha_2,t_0) \nonumber\\
&+&\Delta J_{2}(\alpha_1,\beta_1,\alpha_2,\beta_2,t,t_0) ,
\end{eqnarray}  
where if $\sum_i \alpha_i\neq 0$: 
\begin{eqnarray}
\Delta J_2(\alpha_1,\beta_1,\alpha_2,\beta_2,t,t_0) &=&\left(J_{1}(\alpha_2,\beta_2,t_0) +  i {e^{i {\alpha_2 t_0^3\over 3}  + \beta_2 t_0}\over \alpha_2 t_0^2}\right)\Delta J_1(\alpha_1,\beta_1,t,t_0) \nonumber\\
&-&i \left( i {e^{i {\sum_i \alpha_i t_0^3\over 3}  + \sum_i \beta_i t_0}\over \alpha_2 \sum_i \alpha_i t_0^4}- i {e^{i {\sum_i \alpha_i t^3\over 3}  + \sum_i \beta_i t}\over \alpha_2 \sum_i \alpha_i t^4}\right),
\end{eqnarray}
and for those terms where $\sum_i \alpha_i =0$ 
\begin{eqnarray}
\Delta J_2(\alpha_1,\beta_1,\alpha_2,\beta_2,t,t_0) &=&\left(J_{1}(\alpha_2,\beta_2,t_0) +  i {e^{i {\alpha_2 t_0^3\over 3}  + \beta_2 t_0}\over \alpha_2 t_0^2}\right)\Delta J_1(\alpha_1,\beta_1,t,t_0) \nonumber\\
&-&{i\over \alpha_2} \left( \int_{t_0}^t {e^{\sum_i \beta_i x}\over x^2}\right).
\end{eqnarray}

We can factorize the $\alpha$-dependence:
\begin{eqnarray}
J_{200}(-\alpha,\alpha,t) &=& {1�\over |\alpha|^{2/3}}  \Big( {\rm Re}\big[J_{200}(-1,1,t |\alpha|^{1/3})\big] + i ~{\rm sign}(\alpha) {\rm Im}\big[J_{200}(-1,1,t |\alpha|^{1/3})\big]\Big),\nonumber\\
J_{201}(-\alpha,\alpha,t) &=& {1�\over |\alpha|}  \Big({\rm Re}\big[J_{201}(-1,1,t |\alpha|^{1/3})\big] + i ~{\rm sign}(\alpha) {\rm Im}\big[J_{201}(-1,1,t |\alpha|^{1/3})\big]\Big),\nonumber\\
J_{210}(-\alpha,\alpha,t) &=& {1�\over |\alpha|} \Big( {\rm Re}\big[J_{210}(-1,1,t |\alpha|^{1/3})\big] + i {\rm sign}(\alpha) {\rm Im}\big[J_{210}(-1,1,t |\alpha|^{1/3})\big]\Big),\nonumber\\
\end{eqnarray}
and reduce the integrals to the basic ones.
\subsection {Three dimensional integrals}
We need the integrals up to ${\mathcal O}(\beta/\alpha)^0$ in this case. We can use the relation:
\begin{eqnarray}
J_{3{\mathbf 0}}(\alpha_1,\alpha_2,\alpha_3,t) =J_{10}(\alpha_1,t) J_{200}(\alpha_2,\alpha_3,t)  - \int_0^t dx ~e^{{i \alpha_2 x^3\over 3}} J_{10}(\alpha_1,x) J_{10}(\alpha_3,x).\nonumber\\
\end{eqnarray}
Since $\sum_i \alpha_i =0$ for the cases of interest, we can rewrite the result in terms of some basic integrals, $I_1$ and $I_2$:
\begin{eqnarray}
J_{200}(\alpha_1,\alpha_2,t) =  {1�\over |\alpha_1 \alpha_2|^{1/3}}  I_1\big(|\alpha_2/\alpha_1|,{\rm sign}(\alpha_2), {\rm sign}(\alpha_1), t |\alpha_1|^{1/3}\big)
\end{eqnarray}
and
\begin{eqnarray}
\int_0^t dx ~e^{{i \alpha_2 x^3\over 3}} J_{10}(\alpha_1,x) J_{10}(\alpha_3,x) &=& {I_2\big(|\alpha_1/\alpha_2|,{\rm sign}(\alpha_1),{\rm sign}(\alpha_3),{\rm sign}(\alpha_2),t |\alpha_2|^{1/3}\big)�\over |\alpha_1 \alpha_2 \alpha_3|^{1/3}}\nonumber\\
\end{eqnarray}
where
\begin{eqnarray}
I_1(r,s_1,s_2,t) \equiv \int_0^t d x~ e^{i s_2 x^3/3} J_{10}(s_1, r^{1/3} x) 
\end{eqnarray}
\begin{eqnarray}
I_2(r,s_1,s_2,s_3,t) \equiv \int_0^t d x~ e^{i s_3 x^3/3} J_{10}(s_1, r^{1/3} t) J_{10}(s_2, (-s_3/s_2-s_1/s_2 r)^{1/3} t)
\end{eqnarray}

\section{Perturbative result for the invariants $J_W$ and $I^{(3)}_2$ }

The finite $t$ perturbative results proportional to the invariants $J_W$  are given by the following expressions (we have used the property $\gamma_i \propto v_i$  to simplify them):
     \begin{eqnarray}
A_{J_W}(t) =  \gamma_1 \gamma_2 \left(1-{{\gamma}_N\over\bar{ \gamma}_N}\right) G_{41}(t) - {\gamma_N \over 2 \bar{\gamma}_N} G_{42}(t) ,
\end{eqnarray}
where
\begin{eqnarray}
 G_{41}(t) &\equiv&   \sum_{k=1}^2 (-1)^k~e^{-\bar{\gamma}_k t} \Bigg\{ \nonumber\\
&&  \sum_{i<j} {\rm Re}\left[a_{ij}   J_{20}(\Delta_{ij},-\Delta_{ij}, t_0) + b_{ij} J_{201}(\Delta_{ij},-\Delta_{ij}, t_0)+ c^{(k)}_{ij}(t) J_{210}(\Delta_{ij},-\Delta_{ij}, t_0)\right]\Bigg\},\nonumber\\
\end{eqnarray}
with 
\begin{center}
\begin{tabular}{lll}
$a_{12} = i$ ,~~& $b_{12} = 2 \Delta_v,$ ~~& $c^{(k)} _{12} = -  \Delta_v + {i\over 2} (2 {\bar \gamma}_k  - \gamma_1-\gamma_2),$ \\
$a_{13}= -i,$ ~~& $b_{13} = 2 v_1,$  ~~& $c^{(k)}_{13} = -  v_1 -  {i\over 2} (2 {\bar\gamma}_k  - \gamma_1),$\\
$a_{23} = i,$ ~~& $b_{23} = -2 v_2,$ ~~& $c^{(k)}_{23} = v_2 + {i\over 2} (2 {\bar\gamma}_k  - \gamma_2),$
\end{tabular}
\end{center}
and  $\Delta_v \equiv v_2 -v_1$, $\Delta_\gamma \equiv (\gamma_2-\gamma_1)$. 

\begin{eqnarray}
G_{42}(t) &\equiv& {\rm Re}\Big[ d_1 J_{30}(\Delta_{12}-\Delta_{13}, -\Delta_{12},\Delta_{13},t)+ d_2 J_{30}(\Delta_{12}-\Delta_{13}, \Delta_{13},-\Delta_{12},t) \nonumber\\
&&\phantom{+{\rm Re}} +  d_3 J_{30}(\Delta_{13},\Delta_{12}-\Delta_{13}, -\Delta_{12},t)+  d_4 J_{30}(\Delta_{13}, -\Delta_{12},\Delta_{12}-\Delta_{13},t)\nonumber\\
&& \phantom{+{\rm Re}}+  d_5 J_{30}(\Delta_{12},-\Delta_{12}+\Delta_{13}, -\Delta_{13},t)+  d_6 J_{30}(\Delta_{12}, -\Delta_{13},-\Delta_{12}+\Delta_{13},t)\Big],\nonumber\\
\label{eq:muhubblepotmu}
\end{eqnarray} 
with
\begin{eqnarray}
d_1 & =& z_1 + i {\gamma_1\over 2} \big[2 \Delta_v + i \Delta_\gamma\big]\big[2 v_2 -i(2 \bar{\gamma}_2-\gamma_2)\big]   e^{-\bar{\gamma}_2 t},   \nonumber\\
d_2 & =& z_2 + i {\Delta_\gamma\over 2} \big[-2 v_1 + i \gamma_1\big]\big[2 v_2 -i(2 \bar{\gamma}_2-\gamma_2)\big]  e^{-\bar{\gamma}_2 t},   \nonumber\\
d_3 & =& z_2 - i {\Delta_\gamma\over 2} \big[2 v_2 + i \gamma_2\big] \big[2 v_1+ i \big( 2 \bar{\gamma}_1 - \gamma_1\big)\big] e^{-\bar{\gamma}_1 t} ,\nonumber\\
d_4 &=& z_3 +i {\gamma_2\over 2}  \big[2 \Delta_v - i \Delta_\gamma\big] \big[2 v_1 + i (2 \bar{\gamma}_1 - \gamma_1)\big] e^{-\bar{\gamma}_1 t}, \nonumber\\
d_5 &=& {\gamma_1\over 2} \big( 2 v_2 + i \gamma_2\big) \Big[e^{- \bar{\gamma_1} t} (2 i \Delta_v +2 \bar{\gamma}_1- \gamma_1- \gamma_2 )  - e^{-\bar{\gamma}_2 t} (2 i \Delta_v + 2 \bar{\gamma}_2-\gamma_1 -\gamma_2 )\Big],\nonumber\\
d_6 &=&  -{\gamma_2\over 2} \big( 2 v_1 - i \gamma_1\big) \Big[e^{- \bar{\gamma_1} t} (2 i \Delta_v +2 \bar{\gamma}_1- \gamma_1- \gamma_2 )  -e^{-\bar{\gamma}_2 t} (2 i \Delta_v +  2 \bar{\gamma}_2 -\gamma_1- \gamma_2) \Big].\nonumber\\
 \end{eqnarray}
On the other hand, for the invariant  $I^{(3)}_2$ 
 \begin{eqnarray}
 A_{I^{(3)}_{2}}(t) = & y_1 y_2 \left(1-{\gamma_N\over {\bar \gamma}_N}\right) \gamma_N G_3(t),
 \end{eqnarray}
 \begin{eqnarray}
 G_3(t) &\equiv&   \sum_{k=1}^2 (-1)^k~e^{-\bar{\gamma}_k t} \Bigg\{\nonumber\\
 && \sum_{i<j}  {\rm Re} \Big [a' _{ij}   J_{20}(\Delta_{ij},-\Delta_{ij}, t_0) + b'_{ij}(t) J_{201}(\Delta_{ij},-\Delta_{ij}, t_0)+ c'^{(k)}_{ij}(t) J_{210}(\Delta_{ij},-\Delta_{ij}, t_0)\Big] \nonumber\\
 & &+ {\rm Re} \Big[w_1 J_{30}(\Delta_{12},-\Delta_{12}+\Delta_{13}, -\Delta_{13},t)+  w_2 J_{30}(\Delta_{12}, -\Delta_{13},-\Delta_{12}+\Delta_{13},t)\Big]\Bigg\},\nonumber\\
\label{eq:muhubblepotmu}
\end{eqnarray} 
with 
\begin{eqnarray}
w_1 = {1 \over 2} \left[2  v_2 \gamma_1 + i \gamma_1 \gamma_2 \right], \;\;\;w_2 = {1 \over 2} \left[-2 v_1 \gamma_2 + i \gamma_1 \gamma_2 \right] , 
 \end{eqnarray}
and
\begin{center}
\begin{tabular}{lll}
$a'_{12} = i \gamma_2,$ \;\;\; & $b'_{12} =  2 \gamma_2 v_2 - v_1 \gamma_2 -v_2 \gamma_1,$\;\;&$c'^{(k)}_{12} =  {1\over 2}\gamma_2  \Big(-2  \Delta_v +i(2 \bar{\gamma}_k - \gamma_2 -\gamma_1) \Big),$  \\
$a'_{13} = -i \gamma_1,$\;\; &$b'_{13} = 2 \gamma_1 v_1, $\;\;\; &$c'^{(k)}_{13} = -{1\over 2}\gamma_1  \Big(2 v_1 + i (2 \bar{\gamma}_k -\gamma_1) \Big),$\\
$a'_{23} = i \gamma_2, $&$b'_{23} = -2 \gamma_2 v_2,$ &$c'^{(k)}_{23} = {1\over 2}\gamma_2  \Big(2 v_2 + i (2 \bar{\gamma}_k -\gamma_2 )\Big).$
\end{tabular}
\end{center}
\bibliographystyle{JHEP}
\bibliography{biblio}

\providecommand{\href}[2]{#2}\begingroup\raggedright\begin{thebibliography}{10}

\bibitem{Fukugita:1986hr}
M.~Fukugita and T.~Yanagida, \emph{{Baryogenesis Without Grand Unification}},
  \href{http://dx.doi.org/10.1016/0370-2693(86)91126-3}{\emph{Phys.Lett.} {\bf
  B174} (1986) 45}.

\bibitem{davidson08}
S.~Davidson, E.~Nardi and Y.~Nir, \emph{{Leptogenesis}},
  \href{http://dx.doi.org/10.1016/j.physrep.2008.06.002}{\emph{Phys. Rept.}
  {\bf 466} (2008) 105--177},
  [\href{http://arxiv.org/abs/hep-ph/0802.2962}{{\tt hep-ph/0802.2962}}].

\bibitem{davidson02}
S.~Davidson and A.~Ibarra, \emph{{A lower bound on the right-handed neutrino
  mass from leptogenesis}},
  \href{http://dx.doi.org/10.1016/S0370-2693(02)01735-5}{\emph{Phys. Lett.}
  {\bf B535} (2002) 25}, [\href{http://arxiv.org/abs/hep-ph/0202239}{{\tt
  hep-ph/0202239}}].

\bibitem{hambye03}
T.~Hambye, Y.~Lin, A.~Notari, M.~Papucci and A.~Strumia, \emph{{Constraints on
  neutrino masses from leptogenesis models}},
  \href{http://dx.doi.org/10.1016/j.nuclphysb.2004.06.027}{\emph{Nucl. Phys.}
  {\bf B695} (2004) 169}, [\href{http://arxiv.org/abs/hep-ph/0312203}{{\tt
  hep-ph/0312203}}].

\bibitem{racker12}
J.~Racker, M.~Pe\~na and N.~Rius, \emph{{Leptogenesis with small violation of
  B-L}}, \href{http://dx.doi.org/10.1088/1475-7516/2012/07/030}{\emph{JCAP}
  {\bf 1207} (2012) 030}, [\href{http://arxiv.org/abs/arXiv:1205.1948}{{\tt
  arXiv:1205.1948}}].

\bibitem{pilaftsis03}
A.~Pilaftsis and T.~E. Underwood, \emph{{Resonant leptogenesis}},
  \href{http://dx.doi.org/10.1016/j.nuclphysb.2004.05.029}{\emph{Nucl.Phys.}
  {\bf B692} (2004) 303--345}, [\href{http://arxiv.org/abs/hep-ph/0309342}{{\tt
  hep-ph/0309342}}].

\bibitem{Akhmedov:1998qx}
E.~K. Akhmedov, V.~Rubakov and A.~Y. Smirnov, \emph{{Baryogenesis via neutrino
  oscillations}},
  \href{http://dx.doi.org/10.1103/PhysRevLett.81.1359}{\emph{Phys.Rev.Lett.}
  {\bf 81} (1998) 1359--1362}, [\href{http://arxiv.org/abs/hep-ph/9803255}{{\tt
  hep-ph/9803255}}].

\bibitem{Asaka:2005pn}
T.~Asaka and M.~Shaposhnikov, \emph{{The nuMSM, dark matter and baryon
  asymmetry of the universe}},
  \href{http://dx.doi.org/10.1016/j.physletb.2005.06.020}{\emph{Phys.Lett.}
  {\bf B620} (2005) 17--26}, [\href{http://arxiv.org/abs/hep-ph/0505013}{{\tt
  hep-ph/0505013}}].

\bibitem{Shaposhnikov:2008pf}
M.~Shaposhnikov, \emph{{The nuMSM, leptonic asymmetries, and properties of
  singlet fermions}},
  \href{http://dx.doi.org/10.1088/1126-6708/2008/08/008}{\emph{JHEP} {\bf 0808}
  (2008) 008}, [\href{http://arxiv.org/abs/arXiv:0804.4542}{{\tt
  arXiv:0804.4542}}].

\bibitem{Canetti:2012kh}
L.~Canetti, M.~Drewes, T.~Frossard and M.~Shaposhnikov, \emph{{Dark Matter,
  Baryogenesis and Neutrino Oscillations from Right Handed Neutrinos}},
  \href{http://dx.doi.org/10.1103/PhysRevD.87.093006}{\emph{Phys.Rev.} {\bf
  D87} (2013) 093006}, [\href{http://arxiv.org/abs/arXiv:1208.4607}{{\tt
  arXiv:1208.4607}}].

\bibitem{Feroz:2007kg}
F.~Feroz and M.~Hobson, \emph{{Multimodal nested sampling: an efficient and
  robust alternative to MCMC methods for astronomical data analysis}},
  \href{http://dx.doi.org/10.1111/j.1365-2966.2007.12353.x}{\emph{Mon.Not.Roy.Astron.Soc.}
  {\bf 384} (2008) 449}, [\href{http://arxiv.org/abs/arXiv:0704.3704}{{\tt
  arXiv:0704.3704}}].

\bibitem{Feroz:2008xx}
F.~Feroz, M.~Hobson and M.~Bridges, \emph{{MultiNest: an efficient and robust
  Bayesian inference tool for cosmology and particle physics}},
  \href{http://dx.doi.org/10.1111/j.1365-2966.2009.14548.x}{\emph{Mon.Not.Roy.Astron.Soc.}
  {\bf 398} (2009) 1601--1614},
  [\href{http://arxiv.org/abs/arXiv:0809.3437}{{\tt arXiv:0809.3437}}].

\bibitem{deGouvea:2009fp}
A.~de~Gouvea, W.-C. Huang and J.~Jenkins, \emph{{Pseudo-Dirac Neutrinos in the
  New Standard Model}},
  \href{http://dx.doi.org/10.1103/PhysRevD.80.073007}{\emph{Phys.Rev.} {\bf
  D80} (2009) 073007}, [\href{http://arxiv.org/abs/arXiv:0906.1611}{{\tt
  arXiv:0906.1611}}].

\bibitem{deGouvea:2011zz}
A.~de~Gouvea and W.-C. Huang, \emph{{Constraining the (Low-Energy) Type-I
  Seesaw}},
  \href{http://dx.doi.org/10.1103/PhysRevD.85.053006}{\emph{Phys.Rev.} {\bf
  D85} (2012) 053006}, [\href{http://arxiv.org/abs/arXiv:1110.6122}{{\tt
  arXiv:1110.6122}}].

\bibitem{Donini:2011jh}
A.~Donini, P.~Hernandez, J.~Lopez-Pavon and M.~Maltoni, \emph{{Minimal models
  with light sterile neutrinos}},
  \href{http://dx.doi.org/10.1007/JHEP07(2011)105}{\emph{JHEP} {\bf 1107}
  (2011) 105}, [\href{http://arxiv.org/abs/arXiv:1106.0064}{{\tt
  arXiv:1106.0064}}].

\bibitem{Donini:2012tt}
A.~Donini, P.~Hernandez, J.~Lopez-Pavon, M.~Maltoni and T.~Schwetz, \emph{{The
  minimal 3+2 neutrino model versus oscillation anomalies}},
  \href{http://dx.doi.org/10.1007/JHEP07(2012)161}{\emph{JHEP} {\bf 1207}
  (2012) 161}, [\href{http://arxiv.org/abs/arXiv:1205.5230}{{\tt
  arXiv:1205.5230}}].

\bibitem{Dolgov:2000pj}
A.~D. Dolgov, S.~H. Hansen, G.~Raffelt and D.~V. Semikoz, \emph{{Cosmological
  and astrophysical bounds on a heavy sterile neutrino and the KARMEN
  anomaly}}, \href{http://dx.doi.org/10.1016/S0550-3213(00)00203-0}{\emph{Nucl.
  Phys.} {\bf B580} (2000) 331--351},
  [\href{http://arxiv.org/abs/hep-ph/0002223}{{\tt hep-ph/0002223}}].

\bibitem{Dolgov:2000jw}
A.~D. Dolgov, S.~H. Hansen, G.~Raffelt and D.~V. Semikoz, \emph{{Heavy sterile
  neutrinos: Bounds from big bang nucleosynthesis and SN1987A}},
  \href{http://dx.doi.org/10.1016/S0550-3213(00)00566-6}{\emph{Nucl. Phys.}
  {\bf B590} (2000) 562--574}, [\href{http://arxiv.org/abs/hep-ph/0008138}{{\tt
  hep-ph/0008138}}].

\bibitem{Ruchayskiy:2012si}
O.~Ruchayskiy and A.~Ivashko, \emph{{Restrictions on the lifetime of sterile
  neutrinos from primordial nucleosynthesis}},
  \href{http://dx.doi.org/10.1088/1475-7516/2012/10/014}{\emph{JCAP} {\bf 1210}
  (2012) 014}, [\href{http://arxiv.org/abs/arXiv:1202.2841}{{\tt
  arXiv:1202.2841}}].

\bibitem{Hernandez:2013lza}
P.~Hernandez, M.~Kekic and J.~Lopez-Pavon, \emph{{Low-scale seesaw models
  versus $N_{eff}$}},
  \href{http://dx.doi.org/10.1103/PhysRevD.89.073009}{\emph{Phys.Rev.} {\bf
  D89} (2014) 073009}, [\href{http://arxiv.org/abs/arXiv:1311.2614}{{\tt
  arXiv:1311.2614}}].

\bibitem{Hernandez:2014fha}
P.~Hernandez, M.~Kekic and J.~Lopez-Pavon, \emph{{$N_{\rm eff}$ in low-scale
  seesaw models versus the lightest neutrino mass}},
  \href{http://dx.doi.org/10.1103/PhysRevD.90.065033}{\emph{Phys.Rev.} {\bf
  D90} (2014) 065033}, [\href{http://arxiv.org/abs/arXiv:1406.2961}{{\tt
  arXiv:1406.2961}}].

\bibitem{Drewes:2015iva}
M.~Drewes and B.~Garbrecht, \emph{{Experimental and cosmological constraints on
  heavy neutrinos}},  \href{http://arxiv.org/abs/arXiv:1502.00477}{{\tt
  arXiv:1502.00477}}.

\bibitem{Branco:2001pq}
G.~C. Branco, T.~Morozumi, B.~Nobre and M.~Rebelo, \emph{{A Bridge between CP
  violation at low-energies and leptogenesis}},
  \href{http://dx.doi.org/10.1016/S0550-3213(01)00425-4}{\emph{Nucl.Phys.} {\bf
  B617} (2001) 475--492}, [\href{http://arxiv.org/abs/hep-ph/0107164}{{\tt
  hep-ph/0107164}}].

\bibitem{Jenkins:2007ip}
E.~E. Jenkins and A.~V. Manohar, \emph{{Rephasing Invariants of Quark and
  Lepton Mixing Matrices}},
  \href{http://dx.doi.org/10.1016/j.nuclphysb.2007.09.031}{\emph{Nucl.Phys.}
  {\bf B792} (2008) 187--205},
  [\href{http://arxiv.org/abs/arXiv:0706.4313}{{\tt arXiv:0706.4313}}].

\bibitem{Covi:1996wh}
L.~Covi, E.~Roulet and F.~Vissani, \emph{{CP violating decays in leptogenesis
  scenarios}},
  \href{http://dx.doi.org/10.1016/0370-2693(96)00817-9}{\emph{Phys. Lett.} {\bf
  B384} (1996) 169--174}, [\href{http://arxiv.org/abs/hep-ph/9605319}{{\tt
  hep-ph/9605319}}].

\bibitem{Sigl:1992fn}
G.~Sigl and G.~Raffelt, \emph{{General kinetic description of relativistic
  mixed neutrinos}},
  \href{http://dx.doi.org/10.1016/0550-3213(93)90175-O}{\emph{Nucl.Phys.} {\bf
  B406} (1993) 423--451}.

\bibitem{Nardi:2005hs}
E.~Nardi, Y.~Nir, J.~Racker and E.~Roulet, \emph{{On Higgs and sphaleron
  effects during the leptogenesis era}},
  \href{http://dx.doi.org/10.1088/1126-6708/2006/01/068}{\emph{JHEP} {\bf 01}
  (2006) 068}, [\href{http://arxiv.org/abs/hep-ph/0512052}{{\tt
  hep-ph/0512052}}].

\bibitem{Asaka:2011wq}
T.~Asaka, S.~Eijima and H.~Ishida, \emph{{Kinetic Equations for Baryogenesis
  via Sterile Neutrino Oscillation}},
  \href{http://dx.doi.org/10.1088/1475-7516/2012/02/021}{\emph{JCAP} {\bf 1202}
  (2012) 021}, [\href{http://arxiv.org/abs/arXiv:1112.5565}{{\tt
  arXiv:1112.5565}}].

\bibitem{Luty:1992un}
M.~Luty, \emph{{Baryogenesis via leptogenesis}},
  \href{http://dx.doi.org/10.1103/PhysRevD.45.455}{\emph{Phys.Rev.} {\bf D45}
  (1992) 455--465}.

\bibitem{Besak:2012qm}
D.~Besak and D.~Bodeker, \emph{{Thermal production of ultrarelativistic
  right-handed neutrinos: Complete leading-order results}},
  \href{http://dx.doi.org/10.1088/1475-7516/2012/03/029}{\emph{JCAP} {\bf 1203}
  (2012) 029}, [\href{http://arxiv.org/abs/arXiv:1202.1288}{{\tt
  arXiv:1202.1288}}].

\bibitem{Shuve:2014zua}
B.~Shuve and I.~Yavin, \emph{{Baryogenesis through Neutrino Oscillations: A
  Unified Perspective}},
  \href{http://dx.doi.org/10.1103/PhysRevD.89.075014}{\emph{Phys.Rev.} {\bf
  D89} (2014) 075014}, [\href{http://arxiv.org/abs/arXiv:1401.2459}{{\tt
  arXiv:1401.2459}}].

\bibitem{Ghisoiu:2014ena}
I.~Ghisoiu and M.~Laine, \emph{{Right-handed neutrino production rate at T >
  160 GeV}}, \href{http://dx.doi.org/10.1088/1475-7516/2014/12/032}{\emph{JCAP}
  {\bf 1412} (2014) 032}, [\href{http://arxiv.org/abs/arXiv:1411.1765}{{\tt
  arXiv:1411.1765}}].

\bibitem{Abada:2015rta}
A.~Abada, G.~Arcadi, V.~Domcke and M.~Lucente, \emph{{Lepton number violation
  as a key to low-scale leptogenesis}},
  \href{http://arxiv.org/abs/arXiv:1507.06215}{{\tt arXiv:1507.06215}}.

\bibitem{Ade:2013zuv}
{\scshape Planck} collaboration, P.~Ade et~al., \emph{{Planck 2013 results.
  XVI. Cosmological parameters}},
  \href{http://dx.doi.org/10.1051/0004-6361/201321591}{\emph{Astron.Astrophys.}
  {\bf 571} (2014) A16}, [\href{http://arxiv.org/abs/arXiv:1303.5076}{{\tt
  arXiv:1303.5076}}].

\bibitem{Kuzmin:1985mm}
V.~A. Kuzmin, V.~A. Rubakov and M.~E. Shaposhnikov, \emph{{On the Anomalous
  Electroweak Baryon Number Nonconservation in the Early Universe}},
  \href{http://dx.doi.org/10.1016/0370-2693(85)91028-7}{\emph{Phys. Lett.} {\bf
  B155} (1985) 36}.

\bibitem{Casas:2001sr}
J.~A. Casas and A.~Ibarra, \emph{{Oscillating neutrinos and muon ---> e,
  gamma}}, \href{http://dx.doi.org/10.1016/S0550-3213(01)00475-8}{\emph{Nucl.
  Phys.} {\bf B618} (2001) 171--204},
  [\href{http://arxiv.org/abs/hep-ph/0103065}{{\tt hep-ph/0103065}}].

\bibitem{Gonzalez-Garcia:2014bfa}
M.~C. Gonzalez-Garcia, M.~Maltoni and T.~Schwetz, \emph{{Updated fit to three
  neutrino mixing: status of leptonic CP violation}},
  \href{http://dx.doi.org/10.1007/JHEP11(2014)052}{\emph{JHEP} {\bf 11} (2014)
  052}, [\href{http://arxiv.org/abs/arXiv:1409.5439}{{\tt arXiv:1409.5439}}].

\bibitem{Canetti:2010aw}
L.~Canetti and M.~Shaposhnikov, \emph{{Baryon Asymmetry of the Universe in the
  NuMSM}}, \href{http://dx.doi.org/10.1088/1475-7516/2010/09/001}{\emph{JCAP}
  {\bf 1009} (2010) 001}, [\href{http://arxiv.org/abs/arXiv:1006.0133}{{\tt
  arXiv:1006.0133}}].

\bibitem{Asaka:2013jfa}
T.~Asaka and S.~Eijima, \emph{{Direct Search for Right-handed Neutrinos and
  Neutrinoless Double Beta Decay}},
  \href{http://dx.doi.org/10.1093/ptep/ptt094}{\emph{PTEP} {\bf 2013} (2013)
  113B02}, [\href{http://arxiv.org/abs/arXiv:1308.3550}{{\tt
  arXiv:1308.3550}}].

\bibitem{Blennow:2010th}
M.~Blennow, E.~Fernandez-Martinez, J.~Lopez-Pavon and J.~Menendez,
  \emph{{Neutrinoless double beta decay in seesaw models}},
  \href{http://dx.doi.org/10.1007/JHEP07(2010)096}{\emph{JHEP} {\bf 07} (2010)
  096}, [\href{http://arxiv.org/abs/arXiv:1005.3240}{{\tt arXiv:1005.3240}}].

\bibitem{Caurier:2004gf}
E.~Caurier, G.~Martinez-Pinedo, F.~Nowacki, A.~Poves and A.~P. Zuker,
  \emph{{The Shell model as unified view of nuclear structure}},
  \href{http://dx.doi.org/10.1103/RevModPhys.77.427}{\emph{Rev. Mod. Phys.}
  {\bf 77} (2005) 427--488}, [\href{http://arxiv.org/abs/nucl-th/0402046}{{\tt
  nucl-th/0402046}}].

\bibitem{Caurier:2007wq}
E.~Caurier, J.~Menendez, F.~Nowacki and A.~Poves, \emph{{The Influence of
  pairing on the nuclear matrix elements of the neutrinoless beta beta
  decays}}, \href{http://dx.doi.org/10.1103/PhysRevLett.100.052503}{\emph{Phys.
  Rev. Lett.} {\bf 100} (2008) 052503},
  [\href{http://arxiv.org/abs/arXiv:0709.2137}{{\tt arXiv:0709.2137}}].

\bibitem{Alekhin:2015byh}
S.~Alekhin et~al., \emph{{A facility to Search for Hidden Particles at the CERN
  SPS: the SHiP physics case}},
  \href{http://arxiv.org/abs/arXiv:1504.04855}{{\tt arXiv:1504.04855}}.

\bibitem{Adams:2013qkq}
{\scshape LBNE} collaboration, C.~Adams et~al., \emph{{The Long-Baseline
  Neutrino Experiment: Exploring Fundamental Symmetries of the Universe}},
\newblock \href{http://arxiv.org/abs/arXiv:1307.7335}{{\tt arXiv:1307.7335}}.

\bibitem{Atre:2009rg}
A.~Atre, T.~Han, S.~Pascoli and B.~Zhang, \emph{{The Search for Heavy Majorana
  Neutrinos}},
  \href{http://dx.doi.org/10.1088/1126-6708/2009/05/030}{\emph{JHEP} {\bf 05}
  (2009) 030}, [\href{http://arxiv.org/abs/arXiv:0901.3589}{{\tt
  arXiv:0901.3589}}].

\bibitem{Drewes:2012ma}
M.~Drewes and B.~Garbrecht, \emph{{Leptogenesis from a GeV Seesaw without Mass
  Degeneracy}}, \href{http://dx.doi.org/10.1007/JHEP03(2013)096}{\emph{JHEP}
  {\bf 03} (2013) 096}, [\href{http://arxiv.org/abs/arXiv:1206.5537}{{\tt
  arXiv:1206.5537}}].

\end{thebibliography}\endgroup

\end{document}